\documentclass[11pt]{article}
\usepackage{mathtools,amssymb,amsfonts,graphicx}

\DeclareFontFamily{U}{bbold}{}
\DeclareFontShape{U}{bbold}{m}{n}
 {  <-5.5> s*[1.04] bbold5
    <5.5-6.5> s*[1.04] bbold6
    <6.5-7.5> s*[1.04] bbold7
    <7.5-8.5> s*[1.04] bbold8
    <8.5-9.5> s*[1.04] bbold9
    <9.5-11.5> s*[1.04] bbold10
    <11.5-16> s*[1.04] bbold12
    <16-> s*[1.04] bbold17
 }{}

\usepackage[bbgreekl]{mathbbol}
\DeclareSymbolFontAlphabet{\mathbbm}{bbold}
\DeclareSymbolFontAlphabet{\mathbb}{AMSb}%

\usepackage{hyperref}
\usepackage[nosort]{cite}
\usepackage{verbatim}
\usepackage{multicol,color,xcolor,longtable}

\DeclareFontFamily{U}{matha}{\hyphenchar\font45}
\DeclareFontShape{U}{matha}{m}{n}{
      <5> <6> <7> <8> <9> <10> gen * matha
      <10.95> matha10 <12> <14.4> <17.28> <20.74> <24.88> matha12
      }{}
\DeclareSymbolFont{matha}{U}{matha}{m}{n}

\DeclareMathSymbol{\oleft}{2}{matha}{"68}
\DeclareMathSymbol{\oright}{2}{matha}{"69}

\usepackage[small,bf,hang]{caption}
\usepackage{slashed}
\usepackage{latexsym,epsfig}

\definecolor{darkred}{rgb}{0.65,0.15,0}
\definecolor{darkgreen}{rgb}{.05,.5,.05}
\hypersetup{pdfborder={0 0 0},colorlinks=true,urlcolor=darkred,citecolor=blue,linkcolor=darkred,linktocpage=true}

\usepackage{mathrsfs}
\usepackage{dsfont}
\usepackage[Symbolsmallscale]{upgreek}

\makeatletter
\g@addto@macro\bfseries{\boldmath}
\makeatother

\newcommand{\vir}{\mathfrak{vir}}
\newcommand{\hevir}{\hat{\mathfrak{e}}_{8}\oleft\mathfrak{vir}}

\newcommand{\Virm}{\mathrm{Vir}^{-}}

\newcommand{\dgr}[1]{{\scalebox{0.5}{$(#1)$}}}

\newcommand{\nn}{\nonumber}

\newcommand{\ints}{\mathds{Z}}
\newcommand{\reals}{\mathds{R}}

\def\bea{\begin{eqnarray}}\def\eea{\end{eqnarray}}
\newcommand{\be}{\begin{equation}}
\newcommand{\ee}{\end{equation}}
\newcommand{\CR}{\nonumber \\*}

\def\cL{{\cal L}}

\def\cE{{\mathcal{E}}}

\newcommand{\Is}{{\tilde{I}}}
\newcommand{\Js}{{\tilde{J}}}
\newcommand{\Ks}{{\tilde{K}}}
\newcommand{\Ls}{{\tilde{L}}}

\newcommand{\ord}[1]{{\scriptscriptstyle (#1)}}

\newcommand{\dK}{{\mathsf{K}}}
\newcommand{\dd}{{\mathrm{d}}}

\newcommand{\lb}{\left[}
\newcommand{\rb}{\right]}

\newcommand{\mf}[1]{{\mathfrak{#1}}}

\makeatletter

\@addtoreset{equation}{section}
\makeatother

\newcommand{\ket}[1]{ |#1 \rangle}
\newcommand{\bra}[1]{ \langle #1 |}

\def\hE8{\widehat{E}_{8}}
\def\he8{\hat{\mathfrak{e}}_{8}}

\makeatletter
\def\sbm{\@ifstar\sbm@starred\sbm@unstarred}
\newcommand{\sbm@unstarred}[1]{%
\raisebox{1.1pt}{\scalebox{1.}[.88]{\ensuremath{[}}}%
\mathbbm{#1}\raisebox{1.1pt}{\scalebox{1.}[.88]{\ensuremath{]}}}%
}
\newcommand{\sbm@starred}[1]{%
\raisebox{1.1pt}{\scalebox{1.}[.88]{\ensuremath{[}}}%
{#1}\raisebox{1.1pt}{\scalebox{1.}[.88]{\ensuremath{]}}}%
}
\makeatother

\newcommand{\cS}{\mathcal{S}}
\newcommand{\M}{\mathsf{m}}
\newcommand{\T}{\mathsf{T}}
\newcommand{\LL}{\mathsf{L}}
\newcommand{\sS}{\mathsf{S}}
\newcommand{\cU}{\mathcal{U}}

\newcommand{\coup}{\mathsf{g}}
\newcommand{\Vpot}{V_{\rm gsugra}}
\newcommand{\V}{\mathsf{v}}
\newcommand{\sugrarho}{\varrho}
\newcommand{\sugrasigma}{\sigma}
\newcommand{\twistrho}{r}
\newcommand{\twistsigma}{s}

\newcommand{\sP}{\mathsf{P}}
\newcommand{\sQ}{\mathsf{Q}}

\makeatletter
\DeclareRobustCommand\fJ{\@ifnextchar({\@shiftfJ}{\mathfrak{J}}}
\def\@shiftfJ(#1){\mathfrak{J}^{(#1)}}
\DeclareRobustCommand\fP{\@ifnextchar({\@shiftfP}{\mathcal{P}}}
\def\@shiftfP(#1){\mathcal{P}^{(#1)}}
\makeatother

\def\y{{\zeta}}

\newcommand{\sugraupgamma}{{\hspace{.02em}\breve{\hspace{-.02em}\upgamma\hspace{.02em}}\hspace{-.02em}}}

\begin{document}

{\flushright {CPHT-RR062.092023}\\[5mm]}

\vspace{10mm}

\begin{center}
  {\LARGE \sc 
  Consistent truncation\\[4mm]  of
   eleven-dimensional
  supergravity
  on $S^8\times S^1$
   }
    \\[13mm]

{\large
Guillaume Bossard${}^{1}$, Franz Ciceri${}^2$,\\[1ex]Gianluca Inverso${}^{3}$ and Axel Kleinschmidt${}^{2,4}$}

\vspace{8mm}
${}^1${\it Centre de Physique Th\'eorique, CNRS, Institut Polytechnique de Paris, \\ FR-91128 Palaiseau cedex, France}
\vskip 1.2 ex
${}^2${\it Max-Planck-Institut f\"{u}r Gravitationsphysik (Albert-Einstein-Institut)\\
Am M\"{u}hlenberg 1, DE-14476 Potsdam, Germany}
\vskip 1.2 ex
${}^3${\it INFN, Sezione di Padova \\
    Via Marzolo 8, 35131 Padova, Italy}
\vskip 1.2 ex
${}^4${\it International Solvay Institutes\\
ULB-Campus Plaine CP231, BE-1050 Brussels, Belgium}

\end{center}

\vspace{5mm}

\begin{center} 
\hrule

\vspace{6mm}

\begin{tabular}{p{14cm}}
{\small%
Eleven-dimensional supergravity on $S^8\times S^1$ is conjectured to be dual to the M-theory matrix model.
We prove that the dynamics of a subset of   fluctuations around this background is consistently described by $D=2$ SO(9) gauged maximal supergravity.
We provide the full non-linear uplift formul\ae{} for all bosonic fields.
We also present a further truncation to the SO(3)$\times$SO(6) invariant sector and discuss its relation to the BMN matrix model at finite temperature.
The construction relies on the framework of generalised Scherk--Schwarz reductions,  established for E$_9$ exceptional field theory in a companion paper.
As a by-product, we severely constrain the most general gauge deformations in $D=2$ admitting an uplift to higher dimensions.
}
\end{tabular}
\vspace{5mm}
\hrule
\end{center}

\thispagestyle{empty}

\newpage
\setcounter{page}{1}


\setcounter{tocdepth}{2}
\tableofcontents

\vspace{5mm}
\noindent\rule{\textwidth}{1.5pt}

\section{Introduction and summary}
\label{sec:intro}

The SU($N$) matrix quantum mechanics, first introduced as a regularisation of the supermembrane~\cite{Hoppe,deWit:1988wri}, has been proposed as a non-perturbative definition of M-theory in the infinite momentum frame~\cite{Banks:1996vh}. 
A more recent perspective on this conjecture is provided by holography~\cite{Itzhaki:1998dd,Boonstra:1998mp}, where the strong coupling limit of the matrix model is described by eleven-dimensional supergravity on the SO(9)-invariant pp-wave solution~\cite{Hull:1984vh,Nicolai:2000zt}. The corresponding ten-dimensional description involves IIA supergravity on the near-horizon geometry of $N$ D0-branes, whose metric is conformal to AdS$_2 \times S^8$. The above holographic duality has been the subject of several studies, including numerical evaluations of some correlation functions, see for example \cite{Sekino:1999av,Sekino:2000mg,Hanada:2011fq,Ortiz:2014aja,Filev:2015hia}.

In order to apply holographic techniques such as holographic renormalisation  \cite{Bianchi:2001de,Bianchi:2001kw,Skenderis:2002wp}, it is generally very useful to have a consistent truncation to a lower-dimensional supergravity theory, capturing a subset of fluctuations in the asymptotically AdS space-time.
For the $S^8\times S^1$ pp-wave background, the natural candidate is SO(9) gauged maximal supergravity in $D=2$ space-time dimensions~\cite{Ortiz:2012ib}, in which the pp-wave is a  $1/2$-BPS domain wall solution with a running dilaton. A U(1)$^4$ axion-free subsector has been shown to consistently uplift to ten dimensions in \cite{Anabalon:2013zka}. Holographic renormalisation was used in this model to derive the two-point functions of quadratic and cubic operators \cite{Ortiz:2014aja}. 
In order to further probe the connection between SO(9) gauged supergravity and the M-theory matrix model, it is necessary to have at one's disposal a consistent embedding in eleven dimensions that captures all possible fluctuations. The consistent uplift of the entire two-dimensional theory, which was announced in~\cite{Bossard:2022wvi}, is the main result of this paper.

In a companion paper~\cite{gss}, we have described how generalised Scherk--Schwarz reductions~\cite{Grana:2008yw,Aldazabal:2011nj,Geissbuhler:2011mx,Grana:2012rr,Berman:2012uy,Musaev:2013rq,Aldazabal:2013mya,Berman:2013cli,Aldazabal:2013via,Lee:2014mla,Hohm:2014qga,Hohm:2017wtr} of E$_9$ exceptional field theory~\cite{Bossard:2018utw,Bossard:2021jix} can be used to obtain the complete bosonic dynamics of two-dimensional gauged maximal supergravity theories that admit a consistent uplift to maximal supergravity in $D=10$ or $D=11$ dimensions. The resulting theory was described uniformly by a pseudo-Lagrangian whose Euler--Lagrange equations need to be supplemented by a set of duality equations that reduce the number of propagating bosonic degrees of freedom to $128$ as required by  maximal supersymmetry. The pseudo-Lagrangian consists of a potential and a topological term.

In the present paper, we apply the general results obtained in~\cite{gss}  to the particular case of SO(9) gauged supergravity in $D=2$ dimensions. 
We recover the SO(9) gauged supergravity theory that was originally derived by Ortiz and Samtleben~\cite{Ortiz:2012ib} using supersymmetry, and provide moreover concrete formul\ae{} for the uplift of any two-dimensional configuration to $D=11$ supergravity. 
The complete form of the metric and the three-form gauge field in eleven dimensions is necessary to interpret holographically the solutions of SO(9) gauged supergravity.
We then focus on the SO(3)$\times $SO(6) invariant subsector of the theory, including the axion that was not captured in~\cite{Anabalon:2013zka}. This truncation is a priori relevant to the description of the BMN mass deformation of the BFSS matrix model~\cite{Berenstein:2002jq}. We will show that it includes a non-normalisable mode that triggers the BMN deformation at finite temperature~\cite{Costa:2014wya}.

The generalised Scherk--Schwarz reduction of E$_9$ exceptional field theory rests, as all such reductions, first and foremost on the identification of a twist matrix taking values in the hidden symmetry group and depending on the so-called internal coordinates of exceptional field theory. We recall that in exceptional field theory~\cite{Berman:2010is,Hohm:2013pua,Hohm:2013vpa,Hohm:2013uia,Hohm:2014fxa,Abzalov:2015ega,Musaev:2015ces,Berman:2015rcc,Bossard:2018utw,Bossard:2021jix} there are external coordinates (that here belong to $D=2$ space-time dimensions) as well as internal coordinates that transform in a representation of the hidden symmetry group, here in the infinite-dimensional basic representation of E$_9$.\footnote{The global symmetry group of `extended' E$_9$ exceptional field theory~\cite{Bossard:2021jix} also contains half of a Virasoro group related to reparametrisations of the spectral parameter occurring in the loop group description of the affine E$_9$ symmetry. This is discussed in more detail in~\cite{Bossard:2021jix,gss} and in Section~\ref{sec:review}.}
Importantly, the internal coordinates are constrained by the so-called section constraint that guarantees a consistent diffeomorphism algebra~\cite{Coimbra:2011nw,Berman:2012vc} and the correct counting of degrees of freedom. The dependence of the twist matrix on the internal coordinates determines which subgroup of E$_9$ is gauged along with the resulting dynamics and is constrained by the generalised Scherk--Schwarz consistency condition as discussed in the companion paper \cite{Hohm:2014qga}.

The SO($9$) gauge subgroup related to the $S^8$ sphere reduction sits inside an SL($9$) subgroup of E$_9$ as is usual for sphere reductions~\cite{Hohm:2014qga,Inverso:2017lrz}. This SL($9$) is different from the (geometric) SL($9$) arising in the $T^9$ torus reduction from $D=11$ to two dimensions. One determines the correct SL($9$)\,$\subset$\,E$_9$ through the identification of the fields supporting the one-half BPS pp-wave solution \cite{Nicolai:2000zt}. Remarkably, this reveals that the SL($9$) relevant for SO($9$) gauged supergravity can be obtained by spectral flow from the eleven-dimensional one. 
The relation between these two SL($9$) subgroups of E$_9$ will be central for deriving the explicit uplift formul\ae{} to $D=11$ dimensions in Section~\ref{sec:uplift}.

In order to give the reader an impression of the uplift formul\ae{}, we display here the reduction ansatz for the $D=11$ metric
\begin{align}
    \dd s^2_{\scalebox{0.6}{11D}} =  \rho^{-\frac{8}{9}}  e^{2\varsigma} \tilde{g}_{\mu\nu} {\rm d}x^\mu {\rm d}x^\nu +  \rho^{\frac{2}{9}}  G_{\Is\Js}  ( {\rm d}y^\Is +\mathcal{A}^\Is ) ( {\rm d}y^\Js +\mathcal{A}^\Js )\,.
\end{align}
Its components along the two external dimensions involve the conformal factor $e^{2\varsigma}$, the uni-modular metric of the two-dimensional space-time $\tilde g_{\mu\nu}$, and the internal volume density that reads
\begin{align}
    \rho(x,y) =  (\det\! \mathring g)^{\frac12} \sugrarho(x)\,,
\end{align}
in terms of the determinant of the round $S^8$ metric $\mathring g_{ij}$, as well as the two-dimensional dilaton $\varrho$. The unimodular internal $(9\times 9)$-part of the $D=11$ metric further decomposes into
\begin{align}
G_{\Is\Js}\,{\rm d}y^\Is {\rm d}y^\Js = G_{ij} {\rm d}y^i {\rm d}y^j + \left(\det  G_{ij}\right)^{-1} \, ( {\rm d}y^9 + K_i {\rm d}y^i )^2
\end{align}
with respect to the M-theory fibre. 
The inverse $G^{ij}$ is expressed, up to the conformal factor, as
\begin{align}
e^{2\varsigma} G^{ij} = \coup^2 \sugrarho^{\frac{2}{3}}e^{2\sigma}  (\det \mathring g)^{\frac{5}{9}}\,Y_{I}\, \mathring g^{ik} \partial_k Y_J \,Y_{K}\, \mathring g^{jl} \partial_l Y_L \bigl(2\M^{K[I} \M^{J]L} + \sugrarho^{-2/3} \M_{PQ} a^{IJP} a^{KLQ} \bigr)\,,
\end{align}
where the right-hand side contains the nine embedding coordinates $Y_J$ (in Euclidean $\mathds{R}^9$) of the reduction space that is homological to the eight-sphere, and that satisfy $\mathring g_{ij} = \partial_i Y_I \partial_j Y_J \delta^{IJ}$, as well as the propagating fields of the two-dimensional SO(9) gauged supergravity (see~\eqref{eq:gaugedLag}). The latter are the SL(9) metric $\M_{PQ}=\M_{QP}$ and the conjugate three-form $a^{IJK}=a^{[IJK]}$. Further equations for the remaining bosonic fields of $D=11$ supergravity expressed through those of the SO(9) gauged theory can be found in Section~\ref{sec:uplift}.
We stress that all uplift expressions are finite expression although they are constructed at intermediate steps from infinite-dimensional E$_9$ modules. Their structure is similar to that occurring in  lower rank cases, see for instance~\cite{deWit:1984nz,Godazgar:2013nma,Ciceri:2014wya,Lee:2014mla,Varela:2015ywx,Baguet:2015sma,Kruger:2016agp}.

Another important result of this paper is that we explain in detail how one can obtain a proper physical Lagrangian from the combined pseudo-Lagrangian and duality equation system. This hinges on choosing an appropriate parabolic gauge for the scalar fields, tantamount to a choice of duality frame, and then rewriting the pseudo-Lagrangian in a form of a finite set of terms plus an infinite set of terms that are all bilinear in components of the duality equation. The bilinearity implies that these terms can be ignored when varying the pseudo-Lagrangian as their contribution to the Euler--Lagrange equations will be set to zero by the duality equation that has to be imposed separately. However, it turns out the duality equations no longer constrain the finitely many fields occurring in the finitely many terms that were separated out, so that the latter constitute a proper physical Lagrangian for the propagating fields, potentially with a finite number of auxiliaries. This mechanism was already encountered in~\cite{Bossard:2021jix,Bossard:2021ebg} and will be described in detail in Section~\ref{sec:pseudotophys}.

The construction described in this paper can be extended to other gauge groups. In fact, our results include the case of CSO($p,q,r$) gaugings with $p{+}q{+}r{=}9$~\cite{Hull:1984qz} in a straight-forward manner by replacing the embedding tensor $\Theta_{IJ} \sim \delta_{IJ}$ by the appropriate invariant (degenerate) metric of CSO($p,q,r$). Here, $\Theta_{IJ}=\Theta_{JI}$, in the ${\bf 45}$ of SL(9), arises from the appropriate choice of twist matrix. Besides this minimal generalisation, one may also envisage the study of completely different gaugings in $D=2$ using different choices of twist matrix and following the steps of the present paper. 

We also address the question of which embedding tensors admit an uplift to eleven-dimen\-sional or type IIB supergravity. 
We show that any Lagrangian gauging admitting such an uplift is only parametrised by \emph{finitely} many components, which we identify explicitly.
This analysis, presented in Appendix~\ref{app:uplift}, relies on choosing the appropriate duality frame and decomposition of the embedding tensor.

The structure of this article is as follows. We begin with a review of the algebraic underpinnings of the construction, including E$_9$, its representations and spectral flow. We also explain the transition from the pseudo-Lagrangian to a proper Lagrangian in Section~\ref{sec:review}.
In Section~\ref{sec:11d} we present all the relevant steps for obtaining SO(9) gauged supergravity via a generalised Scherk--Schwarz reduction. Section~\ref{sec:uplift} is devoted to deriving explicit uplift expressions for any configuration in SO(9) gauged supergravity to $D=11$ supergravity. We describe in detail the SO(3)$\times$SO(6) invariant subsector and its relevance for the BMN matrix model. Several appendices contain additional, more technical, details on some aspects presented here.

\section{\texorpdfstring{Spectral flow and duality frames}{Spectral flow and duality frames}}
\label{sec:review}

In this section, we set up the algebraic preliminaries needed for describing the generalised Scherk--Schwarz reduction of E$_9$ ExFT that leads to SO(9) gauged supergravity as a consistent truncation of higher-dimensional supergravity. We begin by identifying various SL(9) subgroups of E$_9$ that have different physical interpretations. This will be illustrated by substituting them into the pseudo-Lagrangian (of $D=2$ supergravity) to generate proper physical Lagrangians in different duality frames.

\subsection{\texorpdfstring{SL(9) subgroups of E$_9$}{SL(9) subgroups of E9}}
\label{sec:a8e9}

At the level of the Lie algebra, we describe $\mf{e}_9$ as the loop extension of the split exceptional $\mf{e}_8$, together with a central element $\dK$ and a Virasoro operator $L_0$ that is part of a whole Virasoro algebra spanned by $L_m$ for $m\in\mathds{Z}$ and we follow the conventions of~\cite{gss} for the commutation relations. The Dynkin diagram of the affine Kac--Moody algebra $\mf{e}_9$ is shown in Figure~\ref{fig:e9dynk}, including a numbering of its nodes.

\begin{figure}[t!]
\centering
\begin{picture}(250,50)
\thicklines
\multiput(10,10)(30,0){8}{\circle*{10}}
\put(10,10){\line(1,0){210}}
\put(160,40){\circle*{10}}
\put(160,10){\line(0,1){30}}
\put(7,-5){$0$}
\put(37,-5){$1$}
\put(67,-5){$2$}
\put(97,-5){$3$}
\put(127,-5){$4$}
\put(157,-5){$6$}
\put(187,-5){$6$}
\put(217,-5){$7$}
\put(167,36){$8$}
\end{picture}
\caption{\label{fig:e9dynk}{\sl Dynkin diagram of $\mf{e}_{9}$ with labelling of nodes. Nodes $1,\ldots,8$ make up the diagram of $\mf{e}_8$.}}
\end{figure}

\subsubsection{\texorpdfstring{Branching of $\mf{e}_8$}{Branching of e8}}

In order to exhibit the various $\mf{sl}_9$ subalgebras we first need to  decompose $\mf{e}_8$.
The adjoint representation of $\mf{e}_8$ decomposes under the $\mf{gl}_8$ that is embedded along nodes $1,\ldots, 7$ as 
\begin{align}
\label{eq:e8a7}
{\bf 248} = \overline{\bf 8}^\ord{-1} \oplus {\bf 28}^\ord{-2/3} \oplus \overline{\bf 56}^\ord{-1/3} \oplus \left( \mf{gl}_8\right)^\dgr{0} \oplus {\bf 56}^\ord{1/3} \oplus \overline{\bf 28}^\ord{2/3} \oplus {\bf 8}^\ord{1}\,,
\end{align}
where the superscripts describe the eigenvalue of the $\mf{gl}_1$ of the reductive $\mf{gl}_8 \cong \mf{sl}_8 \oplus \mf{gl}_1$. Writing the generators of $\mf{gl}_8$ as $T^i{}_j$ with $i,j=1,\ldots,8$ and commutation relation
\begin{align}
\label{eq:gl8}
\lb  T^i{}_j , T^k{}_\ell \rb = \delta^k_j \, T^i{}_\ell - \delta^i_\ell \, T^k{}_\ell\,,
\end{align}
the various graded pieces in this decomposition can be given as tensor densities transforming under this $\mf{gl}_8$, explicitly
\begin{align}
\label{eq:e8gl8}
\overline{\bf 8}^\ord{-1}:\quad & T_k &  \lb T^i{}_j , T_k  \rb &= - \delta^i_k \, T_j \,,\nn\\
{\bf 28}^\ord{-2/3}:\quad & T^{k_1k_2}= T^{[k_1k_2]} &  \lb T^i{}_j , T^{k_1k_2} \rb &= - 2 \delta_j^{[k_1} T^{k_2] i}_{\phantom{i} } -\frac13 \delta^i_j T^{k_1k_2}\,,\nn\\
\overline{\bf 56}^\ord{-1/3}:\quad & T_{k_1k_2k_3}= T_{[k_1k_2k_3]} &  \lb T^i{}_j , T_{k_1k_2k_3} \rb &= -3 \delta^i_{[k_1} T_{k_2k_3] j}^{\phantom{i} } +\frac13 \delta^i_j T_{k_1k_2k_3}\,,\nn\\
{\bf 56}^\ord{1/3}:\quad & T^{k_1k_2k_3}= T^{[k_1k_2k_3]} &  \lb T^i{}_j , T^{k_1k_2k_3} \rb &= 3 \delta_j^{[k_1} T^{k_2k_3] i}_{\phantom{i} } -\frac13 \delta^i_j T^{k_1k_2k_3}\,,\nn\\
\overline{\bf 28}^\ord{2/3}:\quad & T_{k_1k_2}= T_{[k_1k_2]} &  \lb T^i{}_j , T_{k_1k_2} \rb &= 2 \delta^i_{[k_1} T_{k_2] j}^{\phantom{i} } +\frac13 \delta^i_j T_{k_1k_2}\,,\nn\\
{\bf 8}^\ord{1}:\quad & T^k &  \lb T^i{}_j , T^k  \rb &=  \delta^k_j \, T^i \,.
\end{align}
The absence of density terms in the transformation of $T^i$ and $T_i$ is the reason for our choice of normalisation of the $\mf{gl}_8$. Some relevant $\mf{e}_8$ commutation relations in this basis read
\begin{align}
\label{eq:e8c1}
\lb T^i , T_j \rb = T^i{}_j + \delta^i_j \,T^k{}_k\,,\quad
\lb T^{i_1i_2i_3}, T_{j_1j_2j_3} \rb = 18\, \delta^{[i_1i_2}_{[j_1j_2} T^{i_3]}_{\phantom{]}}{}_{j_3]}^{\phantom{]}} \,,
\end{align}
and further relations can be found in Appendix~\ref{app:e8}.

One consequence of this is that defining
\begin{align}
T^i{}_0 \coloneqq T^i  \quad\text{and}\quad
T^0{}_i \coloneqq  T_i\,,
\end{align}
leads to an $\mf{sl}_9$ subalgebra of $\mf{e}_8$ that we write as $T^I{}_J$, where now $I,J=0,1,\ldots,8$ are fundamental indices of $\mf{sl}_9$ with commutation relations
\begin{align}
\label{eq:sl9}
\lb T^I{}_J , T^K{}_L \rb = \delta^K_J T^I{}_L -\delta^I_L T^K{}_J \,.
\end{align}
This $\mf{sl}_9$ is a maximal subalgebra of $\mf{e}_8$.
The adjoint ${\bf 248}$ of $\mf{e}_8$ decomposes under this $\mf{sl}_9$ as\footnote{This is a $\mathds{Z}_3$-graded decomposition although this grading will not play a role in our analysis.}
\begin{align}
\label{eq:e8deca8}
{\bf 248} = \overline{\bf 84} \oplus {\bf 80} \oplus {\bf 84}\,,
\end{align}
where ${\bf 80}$ is the adjoint of $\mf{sl}_9$ and ${\bf 84}$ corresponds to a three-form $T^{IJK}$ of $\mf{sl}_9$ while $\overline{\bf 84}$ is a dual three-form $T_{IJK}$. The branching under $\mf{gl}_8\subset\mf{sl}_9$ gives the components shown in~\eqref{eq:e8gl8}, for example $T^{IJK} \to (T^{ijk}, T^{0ij}\equiv T^{ij})$.

The affine extension $\hat{\mf{e}}_8$ of $\mf{e}_8$ consists of infinitely many copies of the adjoint of $\mf{e}_8$, labelled by a mode number $m\in \mathds{Z}$, together with a central element $\dK$. The mode number means appending an index $m$ to all generators in~\eqref{eq:e8gl8}, leading for example to $T^i_m$ and $T_{m\, i}$. The mode number is additive in commutators. The central element $\dK$ occurs as an extension in commutators when the mode numbers add up to zero and we also make use of Virasoro generators $L_m$ for $m\in\mathds{Z}$ with the standard commutation relations.\footnote{The value of the Virasoro central charge will drop out of any final formula.}
The Virasoro generators act on the loop algebra elements by $[L_m, T_n^\bullet] = -n T_{m+n}^\bullet$, where $\bullet$ can be any of the $\mf{sl}_8$-representations in~\eqref{eq:e8gl8}. We will also make use of a non-degenerate bilinear form $\eta_{-k\,\alpha\beta}$ over $\hat{\mf{e}}_8\oplus \langle L_{-k}\rangle$ (for a fixed $k$) that pairs loop generators whose mode numbers add up to $-k$ as well as $\dK$ with $L_{-k}$. For more details on the algebraic structures we refer to~\cite{gss}.
Further details on this branching and commutation relations can be found in Appendix~\ref{app:e8}.

\subsubsection{\texorpdfstring{Spectrally flowed $\mf{sl}_9$ algebras}{Spectrally flowed sl9 algebras}}
\label{sec:flow}

The identification of the $\mf{sl}_9$ subalgebra can be generalised within $\mf{e}_9$ by using a version of spectral flow~\cite{Schwimmer:1986mf,Maldacena:2000hw}. We define for $p\in\mathds{Z}$ the generators
\begin{align}
\label{eq:sl9flow}
\T^i{}_j = T^i_{0\, j} + \frac{p}9 \delta^i_j\, \dK \,, \quad \T^i{}_0 = T^i_p\,,\quad \T^0{}_i = T_{-p\, i}\,,
\end{align}
where the generators in the ${\bf 8}$ and $\overline{\bf 8}$ of $\mf{gl}_8$ have been shifted by $p$ affine units in opposite directions. This means that the definition is different for every $p\in\mathds{Z}$, but we are not indicating by how many units $p$ we have flowed in the notation to avoid cluttering.\footnote{Only in Appendix~\ref{sec:inequiv}, where we make statements about inequivalent values of $p$, we will need the distinction. A specific convention for $p=1$ and $p=2$ which will play a special role will be introduced in Section~\ref{sec:p1flow}.}
The addition of the central term for the $\mf{gl}_8$ is necessary in order to maintain the $\mf{sl}_9$ commutation relations, viz.
\begin{align}
\label{eq:com11}
\lb \T^i{}_0 , \T^0{}_j \rb &= T_0^i{}_j + \delta^i_j \, T_0^k{}_k  + \delta^i_j \dK
  = \T^i{}_j + \delta^i_j \T^k{}_k
= \T^i{}_j - \delta^i_j \T^0{}_0 \,,
\end{align}
by the vanishing trace of $\T^I{}_J$. This last relation is still in agreement with the general $\mf{sl}_9$ structure~\eqref{eq:sl9}. 
The case $p=0$ leads to the maximal $\mf{sl}_9\subset \mf{e}_8$ and then $\T^I{}_J=T^I{}_J$ as defined in~\eqref{eq:sl9}.

The spectral flow of the $\mf{sl}_9$ subalgebra extends to all of $\mf{e}_9$. The generators of the affine extension of  $\mf{sl}_9$ are defined as 
\begin{align}
\label{eq:flowsl9}
\T^i_{n\, j} = T^i_{n\, j} + \frac{p}9 \delta^i_j\delta_{n,0}\, \dK \,, \quad \T^i_{n\, 0} = T^i_{n+p}\,,\quad \T^0_{n\, i} = T_{n-p\, i}\,,
\end{align}
for any mode number $n\in \mathds{Z}$. They satisfy the usual $\widehat{\mf{sl}_9}$  algebra relations
\begin{align}
\label{eq:sl9aff0}
    \lb \T^I_{m\, J} , \T^K_{n\, L} \rb = \delta^K_J \T^I_{m+n\, L} - \delta^I_L \T^K_{m+n\, J} + m \left( \delta^I_L \delta^K_J - \frac19 \delta^I_J \delta^K_L\right) \delta_{m,-n} \dK\,.
\end{align}
We have in particular 
\begin{align}
\label{eq:sl9aff}
\lb \T^i_{m\, 0}, \T^0_{n\, j} \rb = \T^{\,\,i}_{m+n\, j} + \delta^i_j \T^{\,\,k}_{m+n\, k} + m \,\delta_{m,-n} \delta^i_j \, \dK\,,
\end{align}
that extends~\eqref{eq:com11} for any $p$ and any mode numbers $m$ and $n$. 
The other flowed generators in~$\mf{e}_9$ are defined according to
\begin{align}
\label{eq:flowe8}
\T^{ijk}_{n-p/3} &= T^{ijk}_{n} \,, & \T^{ij0}_{n-p/3} &= T^{ij}_{n-p} \,,\nn\\
\T_{n+p/3\, ijk} &= T_{n\, ijk}\,, & \T_{n+p/3\, ij0} &= T_{n+p\, ij}\,.
\end{align}
Here, we have included in the definition of the level a shift that is related to the way the generators appear in the $\mf{gl}_8$ decomposition of $\mf{e}_8$, see~\eqref{eq:e8gl8}.

The commutation relations associated with these definitions are
\begin{align}
\lb \T^{I_1I_2I_3}_{m-p/3} , \T_{n+p/3\, J_1J_2J_3} \rb &= 18 \, \delta^{[I_1I_2}_{[J_1J_2} \T^{I_3]}_{m+n\, J_3]} + 6\, \delta^{I_1I_2I_3}_{J_1J_2J_3} \left(m-\frac{p}3\right) \delta_{m,-n} \dK\,,\nn\\
\lb \T^{I_1I_2I_3}_{m-p/3} , \T^{I_4I_5I_6}_{n-p/3} \rb &= - \frac16 \varepsilon^{I_1\ldots I_9} \T_{m+n-2/3\, I_7I_8I_9}\,.
\end{align}

In this flowed basis we also define the Virasoro generators $\LL_m$ 
\begin{align}
\label{eq:flowvir}
\LL_m = L_m + p \,T^k_{m\, k} + \frac{4p^2}{9} \delta_{m,0} \dK\,,
\end{align}
that satisfy the Virasoro algebra for the same central charge as the original $L_m$.
The action of these redefined Virasoro operators on the flowed $\mf{e}_9$ basis is
\begin{align}
\lb \LL_m , \T^I_{n\, J} \rb &= -n \T^{\,\,I}_{m+n\, J}\,,\nn\\
\lb \LL_m , \T^{IJK}_{n-p/3} \rb &= -\left(n-\frac{p}{3}\right) \T^{IJK}_{m+n-p/3}\,,\nn\\
\lb \LL_m , \T_{n+p/3\, IJK}\rb &= -\left(n+\frac{p}{3}\right) \T_{m+n+p/3\, IJK}\,.
\end{align}
The redefined Virasoro generators are thus tuned to the mode numbers of the generators given in~\eqref{eq:flowsl9} and~\eqref{eq:flowe8}.

The formulation of E$_9$ ExFT in~\cite{Bossard:2018utw,Bossard:2021jix} also makes use of shift operators $\cS_m$ that act on the original unflowed generators according to
\begin{align}\label{eq:shift op def}
\cS_m(T_n^A) = T_{m+n}^A \,,\quad \cS_m(L_n) = L_{m+n} \,,\quad \cS_m(\dK) = 0\,.
\end{align}
Here, $A$ is an adjoint E$_8$ index and this definition of shift operators is adapted to the $\mf{e}_8$ basis of~$\mf{e}_9$. 

One can similarly define shift operators that are adapted to the spectrally flowed $\mf{sl}_9$ basis of $\mf{e}_9$ and they appear in the generalised Scherk--Schwarz reduction in this paper. We will write these shift operators as $\sS_m$ and they act by
\begin{align}
\label{eq:shift2}
\sS_m \left( \T^I_{n\, J} \right) &= \T^{\,\, I}_{m+n\, J} \,, &
\sS_m \left(\T_{n-p/3}^{I_1I_2I_3}\right) &= \T_{m+n-p/3}^{I_1I_2I_3}\,,\\
\sS_m \left(\T_{n+p/3\, I_1I_2I_3}\right) &= \T_{m+n+p/3 \, I_1I_2I_3}\,,&
\sS_m (\LL_n) &= \LL_{m+n} \,,
\hspace{20mm} \sS_m(\dK)=0\,.\nn
\end{align}
It is important to note that the two shift operators are not identical but differ by central terms due to the explicit $\dK$ modifications appearing in~\eqref{eq:flowsl9} and~\eqref{eq:flowvir}. The relation between the two shift operators is given, in the flowed basis, by
\begin{align}
\label{eq:shifrelt}
\cS_m \left( \T^I_{n\, J} \right) &= \sS_m \left( \T^I_{n\, J} \right) -\frac{p}{9} \delta_{m,-n} \left(\delta^I_i \delta_J^i -8 \delta^I_0 \delta_J^0\right) \dK \,,\nn\\
\cS_m \left(\T_{n-p/3}^{I_1I_2I_3}\right) &= \sS_m \left(\T_{n-p/3}^{I_1I_2I_3}\right) \,,\nn\\
\cS_m \left(\T_{n+p/3\, I_1I_2I_3}\right) &= \sS_m \left(\T_{n+p/3\, I_1I_2I_3}\right)\,,\nn\\
\cS_m (\LL_n) &= \sS_m (\LL_n) - \delta_{m,-n} \frac{4p^2}{9} \dK \,.
\end{align}
The extra terms in the first and last relation show that the shift operators $\cS_m$ appearing in the construction of E$_9$ ExFT are not $\mf{sl}_9$ covariant. However, the additional central terms can be reabsorbed in the definition of the constrained fields $\langle \chi|$ and $\chi_\mu$ such that one may work with $\sS_m$ that are $\mf{sl}_9$ covariant throughout.\footnote{When relating two distinct flowed bases, one has to recall these extra terms.}

We finally note the following identity involving the shifted bilinear form $\eta_{-k\, \alpha \beta}$ for the flowed and unflowed generators
\begin{align}
\label{eq:shifteta}
\eta_{-k\, \alpha \beta} \T^\alpha \otimes \T^\beta = \eta_{-k\, \alpha \beta} T^\alpha \otimes T^\beta \,.
\end{align}
Following the conventions of~\cite{Bossard:2021jix}, the index $\alpha$ in this formula ranges over both $\hat{\mf{e}}_8$ and the Virasoro generators. We display the expansion of the bilinear form $\eta_{-k\,\alpha\beta}$ in the flowed $\mf{sl}_9$ basis in detail in~\eqref{eq:etak9}.

As we show in Appendix~\ref{sec:inequiv} there are only two different E$_9$ conjugacy classes of spectral flows, namely those with $p\equiv 0\mod 3$ and the remaining $p\equiv 1,2\mod 3$. 
The $p=0$ flowed basis corresponds physically to  the dimensional reduction of three-dimensional supergravity   on a circle. The E$_8$ subgroup commuting with $L_0$ is the three-dimensional Cremmer--Julia group of Ehlers type.
We will next explain the physical meaning of the other spectrally flowed bases.
In fact, even though $p=1$ and $p=2$ are conjugate, it is useful to consider them  separately. 
To distinguish and relate them explicitly, we will use the notation of the generators introduced in this section for the $p=2$ flowed basis only, while we shall write  $\widetilde{\T}_n^\Is{}_\Js, \widetilde{\T}_{n-1/3}^{\Is_1\Is_2\Is_3},\widetilde{\T}_{n+1/3\, \Is_1\Is_2\Is_3}$ for the generators in the $p=1$ flowed basis. Moreover, we also put a tilde on the fundamental index $\Is$ whose decomposition under $\mf{gl}_8$ we choose  as $\Is=(i,9)$ instead of $I = (0,i)$. So unless stated specifically, the generators ${\T}_n^I{}_J, {\T}_{n+1/3}^{I_1I_2I_3},\T_{n-1/3\, I_1I_2I_3}$ will  always be in the $p=2$ flowed basis.

Relating the supergravity field components then amounts to using the change of basis described in \eqref{eq:flowsl9} and \eqref{eq:flowe8} together with the corresponding choice of coset representative. As explained in more detail in the sequel, with this convention that eleven-dimensional supergravity is written in the $p=1$ flowed basis, while the consistent truncation on $S^8\times S^1$ leads to the gauging of the SO(9) $\subset$ SL(9) in the $p=2$ flowed basis. 
For short, we will therefore use the notation of the generators introduced in this section for the $p=2$ flowed basis only, while we shall write  $\widetilde{\T}_n^\Is{}_\Js, \widetilde{\T}_{n-1/3}^{\Is_1\Is_2\Is_3},\widetilde{\T}_{n+1/3\, \Is_1\Is_2\Is_3}$ for the generators in the $p=1$ flowed basis. Moreover, we also put a tilde on the fundamental index $\Is$ whose decomposition under $\mf{gl}_8$ we choose  as $\Is=(i,9)$ instead of $I = (0,i)$. So, unless stated specifically, the generators ${\T}_n^I{}_J, {\T}_{n+1/3}^{I_1I_2I_3},\T_{n-1/3\, I_1I_2I_3}$ will  always be in the $p=2$ flowed basis.

\subsubsection{\texorpdfstring{Spectral flow by $p=1$ unit}{Spectral flow by p=1 unit}}
\label{sec:p1flow}

If one carries out the dimensional reduction from eleven-dimensional supergravity on $T^9$, the SL(9) symmetry of Matzner--Misner type of the torus is the one commuting with the derivation $\LL_0$ in the  $p=1$ flowed basis. 
The degrees of freedom coming from the internal metric are associated to the generators $\LL_0$ and $\widetilde{\T}^{\Is}_{0}{}_{\Js}$, while those coming from the internal three-form are associated to  $\widetilde{\T}_{-1/3}^{\Is\Js\Ks}$. The relation to the Cremmer--Julia E$_8$ group in the $p=0$ flowed basis can be seen from the fact that the $\mf{gl}_8\subset \mf{e}_8$ that is common to all spectrally flowed $\mf{sl}_9$ corresponds to the eight-torus that is used in the reduction from $D=11$ to $D=3$ space-time dimensions. The generators $\widetilde{\T}_0^i{}_9$ in the ${\bf 8}$ that are used for $p=1$ correspond to the eight Kaluza--Klein vectors that appear additionally when reducing on $T^9$ instead of $T^8$.
The generators $\widetilde{\T}_0^i{}_9$ are equal to $ T^i_{-1}$ in the $p=0$ basis, which agrees with the fact that the eight Kaluza--Klein vectors translate into the first set of dual scalar fields in the E$_8$ formulation~\cite{Breitenlohner:1986um,Nicolai:1987kz,Samtleben:2007an}.

Another way of understanding this is by recalling general aspects of the so-called gravity line of hidden symmetries~\cite{Damour:2002cu,West:2002jj,Nicolai:2003fw,Kleinschmidt:2003mf}. The $D=11$ gravity line of a hidden exceptional symmetry corresponds to the horizontal line of Cartan type $A$ in Figure~\ref{fig:e9dynk}. This $A$-type algebra clearly  includes the $\mf{gl}_8$ that was used in the $\mf{gl}_8$ decomposition~\eqref{eq:e8a7}. In order to extend the algebra to also include node $0$ one must add an ${\bf 8}$ that uses the affine node generator exactly once. This means taking the ${\bf 8}$ for $m=1$ and this argument confirms that $p=1$ unit of spectral flow is related to $D=11$ dimensions.

\subsubsection{\texorpdfstring{Spectral flow by $p=2$ units}{Spectral flow by p=2 units}}
\label{sec:p=2flow}

With the convention that eleven-dimensional supergravity is written in the $p=1$ flowed basis, we will now see that the consistent truncation on $S^8\times S^1$ leads naturally to the gauging of the SO(9) $\subset$ SL(9) in the $p=2$ flowed basis. 

The SO(9) gauge group is associated to the isometries of $S^8$. The rotation group SO(8) $\subset$ SO(9)  appears as a subgroup of the GL(8) for the GL(8)$/$SO(8) coset entering the type IIA metric on $S^8$,\footnote{%
In the standard level decompositions of hidden symmetries~\cite{Damour:2002cu,West:2002jj,Nicolai:2003fw,Kleinschmidt:2003mf} the type IIA $\mf{gl}_8$ corresponds to nodes $0,1,\ldots,6$ of the E$_9$ Dynkin diagram in Figure~\ref{fig:e9dynk}. Upon conjugation under the M-theory SL(9) we can as well choose the type IIA $\mf{gl}_8$ as corresponding to the nodes  $1,\ldots,7$ that is common to all spectrally flowed algebras. This freedom can be interpreted as choosing a different M-theory circle.} 
but the full isometry group SO(9) is not a subgroup of the geometric GL(9) for the GL(9)$/$SO(9) coset representing the M-theory metric on $S^8 \times S^1$ and must instead lie in another SL(9) in a different spectrally flowed basis.

To identify the relevant spectrally flowed basis for the $S^8\times S^1$ compactification, it is useful to analyse the fields involved in the corresponding AdS$_2\times S^8\times S^1$ vacuum solution \cite{Nicolai:2000zt}. The $S^8$ metric is determined by the type IIA  GL(8)$/$SO(8) coset, but the M-theory circle is not fibered over $S^8$ and the solution does not involve the full GL(9)$/$SO(9) M-theory  coset. The circle is instead fibered over the AdS$_2$ space through the Kaluza--Klein vector field. The relevant duality frame is determined by the two-dimensional scalars, so one should instead interpret the Kaluza--Klein vector field as the dual of the  dual graviton field $h_{i_1\dots i_7 9,9}$ on $S^8\times S^1$, i.e. schematically 
\begin{align}
\varepsilon^{i_1\dots i_8} \partial_{i_1} h_{i_2\dots i_8 9,9} \sim \varepsilon^{\mu\nu} \partial_\mu A_{\nu 9} \; . 
\end{align}
In the $p=1$ flowed basis this is the component in $\widetilde{\T}_{\!-1\, i}^{ \; 9}$ that is ${\T}_{\! 0\;  i}^{  0}$ in the $p=2$ flowed basis. We find therefore that the scalar fields involved in the $S^8\times S^1$ compactification parametrise the SL(9)$/$SO(9) coset of zero $\LL_0$ level in the $p=2$ flowed basis. 

The general framework for reductions of maximal supergravities on $S^n$ to $D>3$ external dimensions was presented in \cite{Hohm:2014qga}. The resulting gauge group SO($n$+1) always lies inside an SL($n$+1) rigid symmetry group and the $n$ coordinates of the $S^n$ are components of the ExFT generalised coordinates sitting in an antisymmetric rank two tensor of this SL($n$+1). 
As we show in detail in the next section, the $S^8$ coordinates sit in the ${\bf 9}= {\bf 8}\oplus {\bf 1}$ of the $p=1$ flowed SL(9) and translate to the vector eight in the ${\bf 36}= {\bf 8}\oplus {\bf 28}$ of the $p=2$ flowed SL(9), thus confirming that the latter is the correct choice of basis. The argument relies on the study of the branching of the basic module in which the generalised coordinates are defined.

\subsection{Basic module and its decomposition}
\label{sec:basdec}

\noindent
In E$_9$ exceptional field theory, the derivatives $\partial_M$ take values in a lowest weight representation that was denoted $\overline{R(\Lambda_0)_{-1}}$ in~\cite{Bossard:2018utw,Bossard:2021jix}. This is the conjugate of the basic representation of $\mf{e}_9$ and the subscript denotes the conformal weight of the lowest weight vector. Following the notation of~\cite{Bossard:2018utw,Bossard:2021jix}, we shall write the derivatives as bra vectors $\langle \partial |$ that can be expanded over the lowest weight vector $\langle 0|$  in the $\mf{e}_8$-grading of $\mf{e}_9$ as
\begin{align}
\label{eq:derm}
\langle \partial | = \langle 0 |  \left( \partial_\psi  +T^A_1  \partial_A+ \ldots \right)
\end{align}
and where the lowest weight vector satisfies
\begin{align}
\langle 0 | L_m =  0 \quad \text{for $m \leq 1$} \quad\text{and} \quad \langle 0 | T_{m}^A =0  \quad\text{for all $m\leq 0$.}
\end{align}
Since the basic module is a level $\dK=1$ module, we have $\langle \partial | \dK = \langle \partial |$ for the action of the central element on the whole module. In fact, $\dK$ can be replaced by one in all the formulas. 

Under $\mf{e}_8$ the module decomposes as~\cite{Bossard:2017aae}
\begin{align}
\label{eq:dere8}
\overline{R(\Lambda_0)_{-1}} &= {\bf 1}_{0} \oplus {\bf 248}_{1} \oplus \left( {\bf 1} \oplus {\bf 248} \oplus {\bf 3875}\right)_{2} \oplus \left( {\bf 30380} \oplus {\bf 3875} \oplus 2{\times} {\bf 248} \oplus {\bf 1}\right)_3\nn\\
&\quad \oplus \left( {\bf 147250} \oplus {\bf 30380} \oplus {\bf 27000} \oplus 2{\times} {\bf 3875} \oplus 3{\times} {\bf 248} \oplus 2{\times} {\bf 1}\right)_4 \oplus \ldots\,.
\end{align}
The subscripts denote the eigenvalues under $L_0=L_0^\dagger$.

\medskip

For us it will be important how the states in $\overline{R(\Lambda_0)_{-1}}$ reorganise themselves under the spectrally flowed $\mf{sl}_9$ bases discussed in Section~\ref{sec:flow}.
We note that the groundstate of~\eqref{eq:dere8} satisfies $\langle 0| \LL_0 = \frac{4p^2}{9} \langle 0|$  under the $\LL_0$ generator that has been flowed by $p$ units, see~\eqref{eq:flowvir}, and it is not necessarily the state of lowest $\LL_0$ eigenvalue in the module.

As we analyse in Appendix~\ref{app:basbranch} in detail, the branching under $\mf{sl}_9$ is different for different units $p$ of spectral flow. Summarising the result from there, we have that for $p=0$
\begin{align}
\overline{R(\Lambda_0)_{-1}} = {\bf 1}_0 \oplus \left( \overline{\bf 84} \oplus {\bf 80} \oplus {\bf 84}\right)_1 \oplus
\left(\overline{\bf 240} \oplus \overline{\bf 1050} \oplus {\bf 1215} \oplus {\bf 80} \oplus {\bf 1050} \oplus {\bf 240}
\right)_2 
\oplus \ldots\,.
\end{align}
For $p=1$, we have by contrast
\begin{align}
\label{eq:flowdecGR}
\overline{R(\Lambda_0)_{-1}} = {\bf 9}_{\frac{4}{9}} \oplus \overline{\bf 36}_{\frac{7}{9}} \oplus {\bf 126}_{\frac{10}9} \oplus \left({\bf 9}\oplus \overline{\bf 315}\right)_{\frac{13}{9}} \oplus \left( \overline{\bf 36} \oplus \overline{\bf 45} \oplus \overline{\bf 720}\right)_\frac{16}{9} \oplus \ldots\,.
\end{align}
And finally for $p=2$
\begin{align}
\label{eq:flowdec}
\overline{R(\Lambda_0)_{-1}} = \overline{\bf 9}_{\frac{4}{9}} \oplus {\bf 36}_{\frac{7}{9}} \oplus \overline{\bf 126}_{\frac{10}9} \oplus \left(\overline{\bf 9}\oplus {\bf 315}\right)_{\frac{13}{9}} \oplus \left( {\bf 36} \oplus {\bf 45} \oplus {\bf 720}\right)_\frac{16}{9} \oplus \ldots\,.
\end{align}
The subscripts in each case correspond to the $\LL_0$ eigenvalues for the chosen value of $p$. We note that the decompositions for $p=1$ and $p=2$ differ by conjugating the $\mf{sl}_9$ representations.

The physical interpretation of this ${\bf 9}$ appearing for $p=1$ is that the corresponding nine derivatives are those with respect to the coordinates of the M-theory compact space that completes the two external coordinates to $D=11$ dimensions.

For $p=2$ and the decomposition~\eqref{eq:flowdec}, we introduce the notation
\begin{align}
\label{eq:bas2}
\langle 0|_I \,,\quad \langle 1/3|^{IJ}=- \langle 1/3|^{JI}\,,\quad
\langle 2/3|^{IJKL}= \langle 2/3|^{[IJKL]}\,,\quad
\langle 1|^I_{JK} = -\langle 1|^I_{KJ} 
\end{align}
for the first few levels, where the number in the bra vector denotes the difference of the eigenvalue with respect to $\LL_0-\tfrac49$. The precise definition of these states and their relations are given in Appendix~\ref{app:basbranch}. Of particular interest to us will also be the ${\bf 45}$ with $\LL_0=\tfrac{16}9$ appearing in~\eqref{eq:flowdec}. This symmetric tensor will be written as
\begin{align}
\langle 4/3 |^{IJ} = \langle 4/3 |^{JI}
\end{align}
and its components define a basis for the embedding tensor of CSO$(p,q,r)$ gaugings with $p+q+r=9$. These can be obtained by consistent truncations of type IIA and the type IIA coordinates are appearing inside the ${\bf 36}_{7/9}$ in~\eqref{eq:flowdec}.

\medskip

Similarly to~\eqref{eq:bas2} one can define a basis of $\overline{R(\Lambda_0)_{-1}}$ adapted to the $p=1$ decomposition~\eqref{eq:flowdecGR}. The corresponding generators are written with tildes for distinction and denoted by
\begin{align}
\label{eq:bas1}
\widetilde{\langle 0|}{}^\Is \,,\quad \widetilde{\langle 1/3|}{}_{\Is\Js}=- \widetilde{\langle 1/3|}{}_{\Js\Is}\,,\quad
\widetilde{\langle 2/3|}{}_{\Is\Js\Ks\Ls}= \widetilde{\langle 2/3|}{}_{[\Is\Js\Ks\Ls]}\,,\quad
\widetilde{\langle 1|}{}_\Is^{\Js\Ks} = -\widetilde{\langle 1|}{}_\Is^{\Ks\Js} \,.
\end{align}
In Appendix~\ref{sec:relmat}, we study the relation between the two bases~\eqref{eq:bas2} and~\eqref{eq:bas1}. By construction this relation cannot be $\mf{sl}_9$ covariant.
For example, the lowest ${\bf 9}$ in~\eqref{eq:flowdecGR} is expressed in terms of vectors of~\eqref{eq:bas2} as
\be\label{eq:sec in two bases}
 \widetilde{\langle 0|}{}^9 = \langle 4/3|^{00} \,, \qquad\qquad 
\widetilde{\langle 0|}{}^i = \langle 1/3|^{0i}\, . 
\ee
In the above equation, we have broken the two $\mf{sl}_9$ algebras to their common $\mf{gl}_8$ and denoted the extra (vector) index by $9$ for the $p=1$ flow and by $0$ for the $p=0$ to distinguish them.
Further relations are given in~\eqref{p1p2}. 

Throughout this paper, we will also encounter  ket vectors that belong to the representation $R(\Lambda_0)_{-1}$, such as the vector field $|A_\mu\rangle$.
All branchings and algebraic relations described above apply to $R(\Lambda_0)_{-1}$, with conjugated SL(9) representations and opposite signs for the $\LL_0$ grading.

We close this section with a comment on a subtle technical point.
On the representation $\overline{R(\Lambda_0)_{s}}$ (or another conformal weight) we can define a $K(\mf{e}_9)$ invariant pairing that we will write using a bra-ket notation, such as ${}_I |0\rangle = (\langle 0|_I)^\dagger$, and that will feature prominently for instance in the potential~\eqref{eq:Vpotg}. The `kets' in such expressions are still elements of $\overline{R(\Lambda_0)_{s}}$ and can be distinguished from `proper' kets from context or by the position of SL(9) indices.

\subsection{Interpretation of spectral flow as change of duality frame}
\label{sec:flow=duality}

Let us now give an interpretation of spectral flow in terms of the supergravity theory.
For simplicity, we first focus on the case of $D=2$ ungauged supergravity, in the language used in~\cite{gss,Bossard:2021jix}.

The theory can be formulated in terms of infinitely many scalar fields parametrising 
\begin{equation}\label{eq:E9coset}
\frac{\widehat{\mathrm{E}}_8\rtimes\Virm}{K(\mathrm{E}_9)}\,,
\end{equation}
with coset representative $V$, Hermitian currents $P_\mu$ 
and anti-Hermitian composite connection $Q_\mu$, defined from the Maurer--Cartan form $\Omega$:
\begin{equation}\label{eq:MCform}
\dd VV^{-1} = \Omega = P + Q\,.
\end{equation}
On shell, the currents obey the twisted self-duality constraint (in form notation, so that $P = P_\mu\dd x^\mu$)
\begin{equation}\label{eq:twsd}
\star P = \cS_{1}(P)+\breve\chi_1\dK\equiv P^{(1)}\,,
\end{equation}
where the shift operator is defined in \eqref{eq:shift op def} and the auxiliary one-form $\breve\chi_1$ is introduced to restore $K(\mathrm{E}_9)$ covariance.
For later convenience, we shall indeed define the $K(\mathrm{E}_9)$ covariant combinations
\begin{equation}\label{eq:shiftP def}
P^{(k)} = \cS_{k}(P) + \breve\chi_k \dK\,,
\end{equation}
with $\breve\chi_k = \breve\chi_{-k}$ are independent auxiliary one-forms, except for $\breve\chi_0=P_\dK$.
These new one-forms will be necessary in writing a pseudo-Lagrangian later on. 
They are related by iterating \eqref{eq:twsd} and writing
\begin{equation}\label{eq:twsd k times}
P^{(k)} = \star^{|k|}P\,,
\end{equation}
in which only the $\dK$ component is independent of the original twisted self-duality.
Equation \eqref{eq:twsd} determines the duality relations between physical and dual scalars, but the distinction between the two is only determined by fixing a parametrisation of the coset representative $V$. 
Different parametrisations provide dual description of the same physical system.
A first example is the E$_8$ covariant parametrisation based on the grading with respect to the $L_0$ generator
\begin{equation}\label{eq:E9cosetrep E8 basis}
V = \sugrarho^{-L_0}e^{-\varphi_1L_{-1}}e^{-\varphi_2L_{-2}}\cdots
\mathring{V} e^{Y_{1\,A}T^A_{-1}} e^{Y_{2\,A}T^A_{-2}}\cdots e^{-\sugrasigma\dK}\,,
\end{equation}
where $\mathring{V}$ is a coset representative for $\mathrm{E}_8/(\mathrm{Spin}(16)/\ints_2)$ and the central factor is identified with the determinant of the metric: $e^{2\sigma}=\sqrt{-g}$.
This parametrisation is naturally obtained when constructing $D=2$ maximal supergravity as Kaluza--Klein reduction of $D=3$ maximal supergravity.
The Lagrangian obtained from such a dimensional reduction only involves the E$_8$ scalars, the $D=2$ metric and the dilaton $\sugrarho$, corresponding to the size of the Kaluza--Klein circle:
\begin{equation}\label{eq:physLag E8}
\cL_{\text{sugra}} = \sqrt{-g}\left(\sugrarho R  - \varrho \,\eta^{AB} P_{\mu\,A}^{\hspace{2.3mm}0}P^\mu{}_{B}^{\,0}\right)\,,
\end{equation}
with $P_{\mu\,A}^{\hspace{2.2mm}0}$ the Hermitian projection of the Maurer--Cartan form of $\mathrm{E}_8/(\mathrm{Spin}(16)/\ints_2)$, constructed from $\mathring{V}$ and also corresponding to the degree 0 loop component of $P_\mu$. 
Twisted self-duality provides duality relations between these currents and the infinite series of dual potentials $Y_m^A$, which do not appear in the physical Lagrangian.%
\footnote{Relations bewteen $\sugrarho$ and all $\varphi_m$ are also obtained from \eqref{eq:twsd}, but the only nontrivial relation is $\dd\varphi_1 = 2\star\dd\sugrarho$, while all others are algebraically solved in terms of $\sugrarho$ and $\varphi_1$.}
In fact, once one establishes that the theory \eqref{eq:physLag E8} admits the duality relations \eqref{eq:twsd}, its dynamics are entirely encoded in the integrability conditions of the latter.
One may then investigate whether other Lagrangians lead to the same twisted self-duality relations and are hence (classicaly) equivalent to \eqref{eq:physLag E8}.

Theories equivalent to \eqref{eq:physLag E8}  must involve a different subset of the fields parametrising \eqref{eq:E9coset} and can be obtained by 
a procedure analogous to (non-)Abelian T-duality \cite{delaOssa:1992vci}. It amounts to 
gauging part of the symmetries of \eqref{eq:physLag E8} and introducing Lagrange multipliers (corresponding to some combination of the $Y_m^A$ fields) to impose flatness of the gauge connection.
Integrating out the latter produces a new Lagrangian based on a different non-linear sigma model.
We refer to this procedure as a change of \emph{duality frame}, in analogy with the choice of a symplectic frame for vector fields in $D=4$ theories (see e.g. \cite{Gaillard:1981rj,deWit:2005ub}).
We can reinterpret such changes of duality frame in terms of different parametrisations of $V$, associated with inequivalent choices of parabolic subalgebras of $\mathfrak{e}_9$.
Spectral flows such as the one described in the previous sections indeed determine a choice of parabolic subalgebra associated to the grading of the flowed derivation $\LL_0$.
We can for instance look at the duality frame associated with the $p=2$ spectrally flowed SL(9) and introduce the new parametrisation
\begin{equation}\label{eq:E9cosetrep SL9 basis}
V = \sugrarho^{-\LL_0} e^{-\varphi_1 \LL_{-1}}  e^{-\varphi_2 \LL_{-2}} \cdots
\ \V \  e^{-\frac16 a^{IJK} \T_{-1/3\,IJK}} e^{-\frac16 b_{IJK} \T_{-2/3}^{IJK}}
e^{-h^I{}_J \T_{-1}^{\ J}{}_I}\cdots\ e^{-\sugrasigma\dK}\,,
\end{equation}
where $\V$ is a coset representative for $\mathrm{SL(9)}/\mathrm{SO(9)}{}_{K}$, where from now on the subscript $K$ is used to distinguish the local reparametrisation invariance of the coset space from the SO(9) gauge group that will appear in later sections. 
The map between this expression and \eqref{eq:E9cosetrep E8 basis} involves a change of $K(\mathrm{E}_9)$ gauge and field redefinitions of the loop scalars, mixing in particular some of the original E$_8$ scalars with the dual potentials, thus reflecting the spectral flow relations \eqref{eq:flowsl9},~\eqref{eq:flowe8}.
We also stress that while the dilaton and other Virasoro scalars are not affected by the redefinitions, the conformal factor $e^{2\sugrasigma}=\sqrt{-g}$ is. We nevertheless keep the same symbol.
The physical field content consists of the $D=2$ metric $g_{\mu\nu}$, the dilaton $\sugrarho$, the scalar fields parametrising SL(9)/SO(9)$_K$ and the axions $a^{IJK}$ transforming as a three-form under SL(9).
All together, these scalars parametrise the coset space
\begin{equation}\label{eq:physicalcosetSL9}
\frac{\mathrm{GL\!}^+\!(9)\ltimes\reals^{84}}{\mathrm{SO}(9)_K}
\end{equation}
where the $\reals^{84}$ factor is parametrised by $a^{IJK}$.%
\footnote{Notice that $\mathrm{GL\!}^+(9)\ltimes\reals^{84}$ is not a subgroup of E$_9$, since the generators $\T_{-1/3\,IJK}$ do not commute but rather produce lower-degree generators. The correct way to interpret the numerator of \eqref{eq:physicalcosetSL9} is as a quotient of the parabolic subgroup of E$_9$ parametrised by \eqref{eq:E9cosetrep SL9 basis} by its further subgroup generated by algebra elements of degree smaller than $-1/3$.
In more physical terms, the non-commutativity of the $a^{IJK}$ axion shifts is hidden in the physical spectrum since it only affects dual potentials absent from the physical Lagrangian.
}
The physical Lagrangian reads
\begin{align}
\cL_{\text{sugra}} =\ & 
\sqrt{-g}\left( 
  \varrho R 
  +\frac14 \varrho \, g^{\mu\nu} \partial_\mu\M^{IJ}\partial_\nu\M_{IJ} 
-\frac{1}{12}\rho^{1/3}g^{\mu\nu}\partial_\mu a^{I_1I_2I_3} \partial_\nu a^{J_1J_2J_3} 
\M_{I_1J_1}\M_{I_2J_2}\M_{I_3J_3}
\right)  \nonumber
\\&
+\frac1{6^4} \varepsilon^{\mu\nu}
\varepsilon_{I_1\ldots I_9} a^{I_1I_2I_3} \partial_{\mu} a^{I_4I_5I_6} \partial_{\nu} a^{I_7I_8I_9}
\,.\label{eq:physLag SL9}
\end{align}
where we have introduced the  matrix  $\M_{IJ}=\M_{(IJ)}$ and its inverse $\M^{IJ}$ to parametrise the $\mathrm{SL}(9)/\mathrm{SO}(9)_K$ scalar fields.
It is mapped to the basic representation as the hermitian element 
\begin{equation}\label{eq:SL9 M def}
\M = \V^\dagger\V\, . 
\end{equation}
The relation between the operator $\M$ and matrix $\M_{IJ}$ is such that $\M^{-1}\dd\M = -\M^{IK}\dd\M_{JK} \T_0^J{}_I$.
In line with the comment at the end of Section~\ref{sec:basdec}, we can indeed write $\langle 0|_I \M^{-1} {}_J | 0\rangle = \M_{IJ}$.

It is instructive to look at the first few duality relations descending from \eqref{eq:twsd} in this duality frame
\bea
\varrho^{\frac{1}{3}} \M_{IL} \M_{JP} \M_{KQ}  \star {\rm d}  a^{LPQ} \hspace{-2mm} &=& \hspace{-2mm} {\rm d} b_{IJK} - \frac1{72} \varepsilon_{IJKP_1P_2P_3Q_1Q_2Q_3} a^{P_1P_2P_3}  {\rm d} a^{Q_1Q_2Q_3}  \; , \\
\varrho \star \M^{IK} {\rm d} \M_{KJ}\hspace{-2mm} &=& \hspace{-2mm} {\rm d} h^I{}_J + \frac12 a^{IKL} \Bigl({\rm d} b_{JKL} - \frac1{216} \varepsilon_{JKLP_1P_2P_3Q_1Q_2Q_3} a^{P_1P_2P_3}  {\rm d} a^{Q_1Q_2Q_3}  \Bigr)  \CR 
&& \hspace{2mm} - \frac1{18} \delta^I_J a^{KLP} {\rm d} b_{KLP}    \; . \nonumber
\eea
Eliminating $b_{IJK}$ in the second equation one finds that $h^I{}_J$ is dual to the SL(9) Noether current for the Lagrangian \eqref{eq:physLag SL9} 
\begin{align}
\label{eq:hdef}
{\rm d} h^{I}{}_J &= \varrho \star \M^{IK} {\rm d} \M_{KJ} - \frac12 \varrho^{\frac13} \M_{JP_1} \M_{KP_2} \M_{LP_3} a^{IKL} \star {\rm d} a^{P_1P_2P_3} \nn\\  &\quad  + \frac{1}{18}\varrho^{\frac13} \delta^I_J \M_{Q_1P_1} \M_{Q_2P_2} \M_{Q_3P_3} a^{Q_1Q_2Q_3} \star {\rm d} a^{P_1P_2P_3} \nn\\ 
&\quad - \frac1{216} \varepsilon_{JKLP_1P_2P_3Q_1Q_2Q_3} a^{IKL} a^{P_1P_2P_3} {\rm d} a^{Q_1Q_2Q_3} \; . 
\end{align}

Recall that the $p=2$ spectral flowed basis is conjugate under E$_9$ to the $p=1$ one, up to conjugation of all SL(9) representations. Therefore one obtains in the $p=1$ basis the same Lagrangian as~\eqref{eq:physLag SL9}, except that the position of the indices on the axions is interchanged, i.e. $a^{IJK} \to a_{IJK}$. This $p=1$ Lagrangian can be obtained by dimensional reduction of eleven-dimensional supergravity, after integrating out the non-dynamical fields. The axions $a_{IJK}$  (with lower indices)
are then the components of the eleven-dimensional three-form along the torus and $\varrho^{2/9}\, m_{IJ}$ is the internal components of the metric.\footnote{As noted in~\cite{Ortiz:2012ib}, the $p=1$ Lagrangian obtained from the reduction of $D=11$ supergravity contains a Chern--Simons-type term unlike its E$_8$ version~\eqref{eq:physLag E8}.}
The construction of the SO(9) gauged theory proceeds via the $p=2$ flowed basis and a central theme in Section~\ref{sec:uplift} will be how to relate this to the $p=1$ flow and $D=11$ supergravity.\footnote{%
In the rest of this paper we will deal with certain truncations on non-toroidal manifolds, such that the structure group of the internal space is indeed associated to a $p=1$ flowed SL(9), but the resulting $D=2$ gauged supergravity is naturally written in terms of the $p=2$ parametrisation and associated duality frame. For this reason we work with $p=2$ in all sections related to $D=2$ (gauged) supergravity.}

\subsection{From pseudo-Lagrangians to physical Lagrangians}
\label{sec:pseudotophys}

The relation between physical Lagrangians in specific duality frames on the one hand, and parametrisations of \eqref{eq:E9coset} in specific parabolic subgroups of E$_9$ on the other hand, is made systematic by rephrasing ungauged $D=2$ maximal supergravity in terms of a duality invariant pseudo-Lagrangian.
We will now describe this approach and demonstrate how physical Lagrangians can be extracted from the pseudo-Lagrangians. It should be noted that for ungauged supergravity this pseudo-Lagrangian is entirely redundant, since one must anyway impose after variation the twisted self-duality constraint, whose integrability already encodes the full dynamics of the theory.
The advantage of the pseudo-Lagrangian formulation is that it straightforwardly generalises to gauged supergravity (and in fact, to E$_9$ ExFT as well), as we will show in Section~\ref{sec:red}.
Throughout this section we will use for convenience the conformal gauge for the external metric
\begin{equation}
g_{\mu\nu} = e^{2\sigma} \eta_{\mu\nu}\,,
\end{equation}
such that $\star$ denotes Hodge duality with respect to the flat metric.

The pseudo-Lagrangian for ungauged supergravity is topological and can be written in terms of the currents $P_\mu$, their shifted versions \eqref{eq:shiftP def} as well as the associated composite connection $Q_\mu$.
The pseudo-Lagrangian  $\cL^{\text{pseudo}}_{\text{sugra}}$ is defined by~\cite{Bossard:2021jix}\footnote{Wedge products are understood on the right-hand side and the Lie algebra commutator is understood to be graded such that $[Q\,,\,P^{(1)}] = Q\wedge P^{(1)}+P^{(1)}\wedge Q$.}
\begin{align}
\frac1{2\varrho}\cL^{\text{pseudo}}_{\text{sugra}} \,\dd x^0\wedge\dd x^1\ \dK =\ &
\dd P^{(1)} + \big[Q\,,\,P^{(1)}\big] 
+ \sum_{k=1}^\infty P_{k}\big( P^{(k+1)}-P^{(k-1)} \big) \nonumber\\&
+\frac{c_\vir}{12}\,\sum_{k=2}^\infty (k^3-k) P_k\big( P_{k+1} +P_{k-1} \big)\ \dK\,,\label{eq:LtopK}
\end{align}
where the term in the second line is invariant by itself.\footnote{The second line is proportional to $c_\vir$, the Virasoro central charge associated to the representation in which $V$ and the currents have been defined. It is introduced to make the pseudo-Lagrangian independent of such choice by cancelling a similar term coming from the commutator in the first line. 
As a result, the topological term $\cL^{\text{pseudo}}_{\text{sugra}}$, as a whole, does not depend on $c_\vir$.
This will become apparent in the explicit expressions \eqref{eq:LtopE8} and \eqref{eq:LtopSL9}, in which $c_\vir$ indeed cancels out.}
The Maurer--Cartan equation for $P$ guarantees that the right-hand side of \eqref{eq:LtopK} is indeed entirely proportional to $\dK$. We shall explain below how to obtain well-defined equations of motion from this infinite sum of terms.

\subsubsection{E$_8$ duality frame}

As~\eqref{eq:LtopK} is written in terms of the $\hevir$-valued currents, the expression is 
independent of the choice of basis of $\hat{\mathfrak e}_8$ in which expand the currents (for instance, $T^\alpha$ rather than $\T^\alpha$ as defined in the previous sections).
We stress in particular that the forms $P_k$ are the Virasoro components of the current and are independent of whether or not we choose an expansion in terms of $L_k$ or the flowed $\LL_k$.
Writing $\hevir$ in the $\mf{e}_8$ decomposition
and using \eqref{eq:shiftP def} to identify the $\dK$ component of the shifted currents, we obtain
\begin{align}\label{eq:LtopE8}
\cL^{\text{pseudo}}_{\text{sugra}}\ \dd x^0\wedge\dd x^1  =\ &
{2\varrho}\,\dd\breve\chi_1 - {2\varrho}\,\eta^{AB}\sum_{n}n\,Q_{A}^n\wedge P_B^{-n-1}
+{2\varrho}\,\sum_{k=1}^\infty P_{k}\wedge(\breve\chi_{k+1}-\breve\chi_{k-1})\,.
\end{align}
The second term is the central component of the commutator in \eqref{eq:LtopK} in the $p=0$ flowed basis, in which the loop components of $P$ and $Q$ are expanded in terms of the generators $T^A_{n}$, for instance
\begin{align}\label{eq:P expansion E8}
P &= \sum_{m\in\ints}P_A^m T^A_m + \sum_{m\in\ints} P_m L_m + P_\dK\,.
\end{align}
The expansion~\eqref{eq:LtopE8} is the most convenient one when working in the E$_8$ duality frame, as we shall see shortly. Nonetheless, \eqref{eq:LtopE8} is valid regardless of the choice of parabolic gauge.
Expanding $\cL^{\text{pseudo}}_{\text{sugra}}$ in terms of other bases is more convenient (albeit not strictly necessary) in order to perform computations in other duality frames.
Different expansions amount to a field redefinition of the auxiliary one-forms, as we shall see below.

Let us first show how we can recover the physical Lagrangian \eqref{eq:physLag E8} from \eqref{eq:LtopE8}. 
The idea \cite{Bossard:2021jix,Bossard:2021ebg} is that, once a parabolic parametrisation of the coset representative $V$ is made, we can manipulate and reorganise the terms in \eqref{eq:LtopE8} to write it as the sum of a finite set of terms, involving only the fields of lowest degree in the parabolic expansion, and an infinite series of squares of the twisted-selfduality constraint \eqref{eq:twsd}.
Since \eqref{eq:twsd} must be imposed after variation of the pseudo-Lagrangian, 
one is then allowed to drop the squares of twisted self-duality and recovers a true Lagrangian for a finite set of `physical' fields.

To see this in practice, we begin with the sector involving Virasoro and central charge one-forms.
Integrating by parts the first term in \eqref{eq:LtopE8} and dropping for brevity the overall factor of $2\varrho$ we have that the first and last term in \eqref{eq:LtopE8} can be manipulated into the following expressions (recall that wedge products are understood)
\begin{align}
P_0\breve\chi_1&+\sum_{n\ge1} P_n (\breve\chi_{n+1}-\breve\chi_{n-1})
\\
&=\nonumber
-P_1 P_\dK +\sum_{n\ge1}(P_{n-1}-P_{n+1})\breve\chi_n\\
&=\nonumber
P_0\star P_\dK + (\star P_0-P_1)P_\dK
+\sum_{n\ge1}[(P_{n-1}-\star P_n)-(P_{n+1}-\star P_n)] \breve\chi_n\,,
\end{align}
where we isolated each $\breve\chi_n$, $n\ge0$ and in the second line we just added and subtracted $\star P_n$. 
Then, we separate the two pieces in the series and combine the last one with $(\star P_0-P_1)P_\dK$ to write
\begin{equation}
P_0\breve\chi_1+\sum_{n\ge1} P_n (\breve\chi_{n+1}-\breve\chi_{n-1})
=
P_0\star P_\dK
+\sum_{n=0}^{+\infty}(\star P_n - P_{n+1})(\breve\chi_n-\star \breve\chi_{n+1})\,.
\end{equation}
We see that the infinite series is a sum of bilinears of components of the duality relations \eqref{eq:twsd k times}.
Thus, only the first term contributes to the physical Lagrangian.
Indeed, using $P_0 = -\varrho^{-1}\dd\varrho$ and $P_\dK=-\dd\sigma$, it expands to (reinstating the overall $2\varrho$ factor)
\begin{equation}
2\varrho P_0\star P_\dK = 2\dd\varrho\star\dd\sigma = \sqrt{-g}\varrho\, R \, \dd x^0\wedge\dd x^1\,,
\end{equation}
where the last identity holds up to a total derivative in the conformal gauge.

Let us now focus on the cocycle term in \eqref{eq:LtopE8}.
Using the parametrisation \eqref{eq:E9cosetrep E8 basis} for $V$, we find the relation $Q_A^n = -\mathrm{sgn}(n)P^n_A$ for $n\neq0$, so that
\begin{equation}\label{eq:E8coccycle as PP}
- 2\varrho\,\eta^{AB} \sum_{n\in\ints} n \, Q^n_A P^{-n-1}_B
=
 2\varrho\,\eta^{AB} \sum_{n\in\ints} |n| \, P^n_A P^{-n-1}_B\,,
\end{equation}
We need to perform several manipulations analogous to the ones above in order to isolate a term depending only on the E$_8$ currents $P^0_A$. 
Details are given in Appendix~\ref{app:Ltop to physical}.
We arrive at the expression
\allowdisplaybreaks
\begin{align}\label{eq:E8kinterm squaring}
2\varrho\,\eta^{AB}\sum_{n\in\ints} |n| \, P^n_A P^{-n-1}_B
=\ &-\varrho\,\eta^{AB} P^0_A \star P^0_B
\\*
&\nonumber
+\varrho\,\eta^{AB} \sum_{n\ge0} ( P^{2n+1}_A - \star P^{-2n}_A ) (P^{-2n}_B-\star P^{2n+1}_B)
\\*
&\nonumber
-\varrho\,\eta^{AB} \sum_{n\ge0} ( P^{2n+1}_A - \star P^{-2n-2}_A ) (P^{-2n-2}_B-\star P^{2n+1}_B)\,,
\end{align}
so that the first term gives the physical kinetic term for the $\mathrm{E}_8/(\mathrm{Spin}(16)/\ints_2)$ non-linear sigma model, completing the physical Lagrangian \eqref{eq:physLag E8} as anticipated.
We have proved that, schematically,
\begin{equation}
\cL^{\text{pseudo}}_{\text{sugra}} = 
2\partial_\mu\varrho\,\partial^\mu\sigma 
- \varrho \, \eta^{AB} P_{\mu\,A}^{\hspace{2.3mm}0}  P^\mu{}^{\,0}_B
+\text{``self-duality square terms''} \,.
\end{equation}

Because we defined the topological term in the conformal gauge, we must also ensure that the Virasoro constraint, coming from the variation of the uni-modular component $\tilde{g}_{\mu\nu}$ of the metric in the physical Lagrangian, is correctly reproduced.
In the E$_8$ duality frame, this is written as
\begin{equation}\label{eq:Vircstr E8}
\delta \tilde{g}^{\mu\nu}\Big(2 P_{\mu\,\dK} P_{\nu\,0} 
- P_{\mu\,0} P_{\nu\,0}
+ \partial_\mu P_{\nu\,0} -\eta^{AB} P_{\mu}{}_{A}^0 P_{\nu}{}_{B}^0 \Big) = 0
\end{equation}
with $\delta \tilde{g}^{\mu\nu}$ symmetric traceless with respect to $\eta_{\mu\nu}$.
This equation can be obtained from the Einstein equations of \eqref{eq:physLag E8}. 
Alternatively, we can define $\cL^{\text{pseudo}}_{\text{sugra}}$ without imposing conformal gauge by following the same procedure as what was done for the minimal formulation of E$_9$ ExFT in~\cite{Bossard:2021ebg}.
The current $P_{\mu\,\dK}$ is shifted by a term $\tilde{g}^{\nu\sigma}\partial_\nu\tilde{g}_{\mu\sigma}$ and the topological term is complemented by the single extra term $\frac14 \varrho \varepsilon^{\mu\nu}\varepsilon^{\sigma\rho}\tilde{g}^{\kappa\lambda}\partial_\mu\tilde{g}_{\sigma\kappa}\partial_\nu\tilde{g}_{\rho\lambda} $.
Following the exact same steps as above to recover a physical action, these modifications combine to reproduce the term $\varrho R$ in~\eqref{eq:physLag E8}.

We therefore conclude that the dynamics captured by the pseudo-Lagrangian combined with twisted self-duality and the Virasoro constraint are the same as those of the physical Lagrangian \eqref{eq:physLag E8}.
The advantage of the pseudo-Lagrangian formulation is that it generalises to gauged supergravity (and ExFT) and guarantees that the resulting equations of motion are invariant under gauge transformations (generalised diffeomorphisms for ExFT) as well as local $K(\mathrm{E}_9)$ reparametrisations.\footnote{To see this, one uses that $\cL^{\text{pseudo}}_{\text{sugra}}$ is invariant by construction and that the gauge and $K(\mf{e}_9)$ variations of squares of twisted self-duality are again proportional to squares of twisted self-duality equations. It follows that the gauge and $K(\mf{e}_9)$ variations of the corresponding Euler--Lagrange equations are by construction proportional to the twisted self-duality equation and the Euler--Lagrange equations themselves.}

\subsubsection{SL$(9)$ duality frame}

We now perform a similar computation to recover the physical Lagrangian in the $p=2$ flowed SL(9) frame given in \eqref{eq:physLag SL9}. The computation for $p=1$ is completely analogous. Some intermediate steps are displayed in Appendix~\ref{app:Ltop to physical}.
The starting point is to notice that the definition \eqref{eq:shiftP def} of the shifted currents relied on the definition of the shift operators \eqref{eq:shift op def}.
We can equivalently use the shift operators $\sS_m$ defined in \eqref{eq:shift2}, absorbing the difference between the two into a redefinition of the auxiliary one-form:
\begin{equation}
P^{(m)} = \cS_m(P) + \breve\chi_m\,\dK = \sS_m(P) + \breve\upchi_m\,\dK\,.
\end{equation}
Furthermore, we will expand the currents in the spectrally flowed basis of generators $\T^\alpha$ defined in Section~\ref{sec:flow},
\begin{align}
P =\ &
\sum_{m\in\ints} \sP^m{}^{\mathsf{B}}{}_{\mathsf{A}} \T_m{}^{\mathsf{B}}{}_{\mathsf{A}}
+ \sum_{m\in\ints} P_m \LL_m 
+ \sP_\dK \dK\nonumber
\\&
+ \frac16\sum_{n\in\ints} \sP^{m-1/3\,\mathsf{A}\mathsf{B}\mathsf{C}} \T_{m-1/3\,\mathsf{A}\mathsf{B}\mathsf{C}}
+ \frac16\sum_{n\in\ints} \sP^{m-2/3}_{\,\mathsf{A}\mathsf{B}\mathsf{C} } \T_{m-2/3}^{\,\mathsf{A}\mathsf{B}\mathsf{C}}\,,\label{eq:P expansion SL9}
\end{align}
and analogously for $\sQ$.
Notice that we have changed symbol for the central charge component ($\sP_\dK$ instead of $P_\dK$) to reflect that it is redefined compared to \eqref{eq:P expansion E8}. We are using $\mathsf{A}$, $\mathsf{B}$, $\mathsf{C}$ for the indices transforming under local SO(9)$_K$ of $\mathrm{SL(9)}/\mathrm{SO}(9)_K$, whereas we use $I,J,K$ for the SL(9) indices. The SO(9)$_K$ vector indices $\mathsf{A}$, $\mathsf{B}$, $\mathsf{C}$ should hopefully not be confused with the E$_8$ adjoint indices $A$, $B$, $C$ used in the preceding section in the E$_8$ duality frame. 

Extracting $\cL_{\text{top}}$ from \eqref{eq:LtopK} in these variables, we find
\begin{align}\label{eq:LtopSL9}
\frac1{2\varrho}\,\cL^{\text{pseudo}}_{\text{sugra}}\,\dd x^0\wedge\dd x^1 =\ &
\dd\breve\upchi_1 
+\sum_{k=1}^\infty P_{k}\wedge(\breve\upchi_{k+1}-\breve\upchi_{k-1})
- \sum_{n\in\ints}n\, \sQ^n{}^{\mathsf{A}}{}_{\mathsf{B}}\wedge \sP^{-n-1}{}^{\mathsf{B}}{}_{\mathsf{A}}
\\&\nonumber
-\sum_{n\in\ints}\frac16\left[
\big(n-\tfrac23\big) \sQ^{n-2/3}_{\,\mathsf{A}\mathsf{B}\mathsf{C}} \sP^{-n-1/3}{}^{\mathsf{A}\mathsf{B}\mathsf{C}}
+\big(n+\tfrac23\big) \sQ^{n+2/3}{}^{\,\mathsf{A}\mathsf{B}\mathsf{C}} \sP^{-n-5/3}_{\,\mathsf{A}\mathsf{B}\mathsf{C}}
\right]\,.
\end{align}
The terms in the first line correspond to the dilaton/central sector plus the loop cocycle for the $\widehat{\mf{sl}}_9$ currents. The second line is the cocycle term for the axion sector.
Using the coset parametrisation \eqref{eq:E9cosetrep SL9 basis}, we can rewrite the twisted self-duality relations for the loop currents as follows:
\begin{subequations}
\label{eq:twsdsl9}
\begin{align}
\label{eq:sl9twsd}
\sP^m{}^{\mathsf{A}}{}_{\mathsf{B}} &= \star^{|m|} \sP^0{}^{\mathsf{A}}{}_{\mathsf{B}}\,,\\
\label{eq:1/3twsd}
\hspace{6em}P^{-1/3-m}{}^{\,\mathsf{A}\mathsf{B}\mathsf{C}} &= \star^m P^{-1/3}{}^{\,\mathsf{A}\mathsf{B}\mathsf{C}}\,,&&\hspace{-6em} m\ge0\,,\\
\label{eq:2/3twsd}
\hspace{6em}P^{-2/3-m}_{\,\mathsf{A}\mathsf{B}\mathsf{C}} &= \star^m P^{-2/3}_{\,\mathsf{A}\mathsf{B}\mathsf{C} }\,,&&\hspace{-6em} m\ge0\,,\\
\label{eq:1/3 2/3 twsd}
P^{-2/3}_{\,\mathsf{A}\mathsf{B}\mathsf{C}} &= \star P^{-1/3}_{\,\mathsf{A}\mathsf{B}\mathsf{C}}\,.
\end{align}
\end{subequations}
In the last line we have used the SO(9)$_K$-invariant metric $\delta_{\mathsf{A}\mathsf{B}}$ for lowering the indices on the right-hand side.

We now want to manipulate \eqref{eq:LtopSL9} into a physical Lagrangian plus bilinears of the relations \eqref{eq:twsdsl9}.
The manipulations of the first line are identical to the E$_8$ case above, so we focus on the axion sector.
After some steps displayed in Appendix~\ref{app:Ltop to physical}, we find the identity
\begin{align}
\label{eq:axion topterm squaring}
\cL^{\text{pseudo}}_{\text{sugra}} \,\dd x^0\wedge\dd x^1 \Big|_{\text{axions}}
=\ &
\frac2{9}\varrho\,P^{-1/3}{}^{\,\mathsf{A}\mathsf{B}\mathsf{C}}\,P^{-2/3}_{\,\mathsf{A}\mathsf{B}\mathsf{C}}
-\frac1{3}\varrho\,P^{-1/3}{}^{\,\mathsf{A}\mathsf{B}\mathsf{C}}\,\star P^{-1/3}_{\,\mathsf{A}\mathsf{B}\mathsf{C}}
+\ldots
\end{align}
where the dots correspond to squares of twisted self-duality equations.
Expanding the Maurer--Cartan form we then find
\begin{align}
\cL^{\text{pseudo}}_{\text{sugra}} \,\dd x^0\wedge\dd x^1 \Big|_{\text{axions}}
=\ &
-\frac{1}{12}\rho^{1/3}\dd a^{I_1I_2I_3} \star\dd a^{J_1J_2J_3} \M_{I_1J_1}\M_{I_2J_2}\M_{I_3J_3}\nonumber
\\&
+\frac1{6^4} \varepsilon_{I_1I_2I_3 J_1J_2J_3 K_1K_2K_3} \dd a^{I_1I_2I_3} \dd a^{J_1J_2J_3} a^{K_1K_2K_3}
\nonumber\\&
+\frac{1}{18}\dd a^{IJK}\dd b_{IJK}  +\ldots \ ,\label{eq:Ltop axions last step}
\end{align}
We see that $b_{IJK}$ only appears in a total derivative and can therefore be dropped. 
Adding back the dilaton/central sector as well as the SL(9) kinetic term, the physical Lagrangian \eqref{eq:physLag SL9} is reproduced.\footnote{It is useful to note the identity $-\varrho\,g^{\mu\nu}\sP_\mu^0{}^{\mathsf{A}}{}_{\mathsf{B}}\sP_\nu^0{}^{\mathsf{B}}{}_{\mathsf{A}} = \frac14 \varrho \, g^{\mu\nu} \partial_\mu\M^{IJ}\partial_\nu\M_{IJ}\,.$}

\medskip

One computes an equation analogous to~\eqref{eq:Vircstr E8} from the SL(9) frame Lagrangian \eqref{eq:physLag SL9}:
\begin{equation}\label{eq:Vircstr SL9}
\delta \tilde{g}^{\mu\nu}\Big(
2\sP_{\mu\,\dK} \sP_{\nu\,0} 
- \sP_{\mu\,0} \sP_{\nu\,0}
+ \partial_\mu \sP_{\nu\,0}
-\sP_\mu^0{}^{\mathsf{A}}{}_{\mathsf{B}} \sP_\nu^0{}^{\mathsf{B}}{}_{\mathsf{A}} 
-\frac13 \sP_\mu^{-1/3}{}^{\mathsf{A}\mathsf{B}\mathsf{C}}\sP_\nu^{-1/3}{}_{\!\! \mathsf{A}\mathsf{B}\mathsf{C}}
\Big)
= 0\,.
\end{equation}
One checks that the same equation is obtained from the Virasoro constraint~\eqref{eq:Vircstr E8} in the E$_8$ basis simply by relating the coefficients in the expansions \eqref{eq:P expansion E8} and \eqref{eq:P expansion SL9} of the current $P$ and using twisted self-duality to write the result exclusively in terms of the physical fields.

\section{\texorpdfstring{Consistent truncation on $S^8\times S^1$}{Consistent truncation on S8xS1}}
\label{sec:11d}

In this section, we apply the general procedure of gSS reduction of E$_9$ ExFT~\cite{gss} to obtain SO(9) gauged supergravity in $D=2$ space-time dimensions. SO(9) gauged supergravity has been constructed directly in $D=2$ using supersymmetry in~\cite{Ortiz:2012ib} as we shall review in Section~\ref{sec:so9rev}. Our gSS construction, presented in Section~\ref{sec:so9gss}, produces the same bosonic theory and moreover proves that the theory is obtained by consistent truncation from $D=11$. In Section~\ref{sec:uplift}, we shall use this to present general uplift formul\ae{} for $D=2$ solutions to $D=11$ dimensions, where the differently flowed SL(9) subgroups play an important role.

\subsection{Review of SO(9) gauged supergravity}
\label{sec:so9rev}

SO(9) maximal gauged supergravity was constructed in~\cite{Ortiz:2012ib} using supersymmetry starting from ungauged supergravity written in an SL(9) duality frame.  In this section, we briefly review some aspects of the construction of the reference translated into our conventions. 

The bosonic field content and Lagrangian of ungauged maximal supergravity in the $p=2$ flowed SL(9) duality frame was reviewed in Section~\ref{sec:flow=duality}.
The fermionic fields of the theory are given by a gravitino, transforming as a spinor under the local SO(9)$_K$, as well as matter fermions transforming as a vector-spinor under SO(9)$_K$. We will not use supersymmetry in this paper and therefore do not display fermions. Details on fermions can be found in \cite{Ortiz:2012ib}.

The construction of~\cite{Ortiz:2012ib} starts from an ungauged Lagrangian density whose bosonic part, in our conventions, is given by \eqref{eq:physLag SL9}.
The gauging of $\mathrm{SO}(9)\subset\mathrm{SL}(9)$ requires introducing vector fields $A_{\mu}^{IJ}= A_{\mu}^{[IJ]}$ in the adjoint representation.
These vector fields occur in the gauged covariant derivative
\begin{align}
\label{eq:covd1}
D_\mu = \partial_\mu -  A_\mu^{IJ} \Theta_{JK} \, \bbdelta^{K}{}_{\!I}
\end{align}
where $\bbdelta^{K}{}_{\!I}$ is the rigid $\mf{sl}(9)$ variation of the field on which the derivative is acting. 
{For instance, $\bbdelta^I{}_{\!\!J}(\M_{KL}) = 2\delta^I_{(K} \M^{\vphantom{I}}_{L)J}-\frac29\delta^I_J \M_{KL}$.} The constant symmetric tensor $\Theta_{IJ} = \Theta_{JI}$ is the embedding tensor describing the embedding of SO(9) in SL(9). We have written a more general symmetric tensor $\Theta_{IJ}$ in order to accommodate gaugings of the type CSO($p$,$q$,$r$), with $p+q+r=9$, in analogy with \cite{Hull:1984qz,Hull:1984vg}.
The SO(9) gauging corresponds to a positive-definite or negative-definite $\Theta_{IJ}$ which, up to a rigid SL$(9)$ transformation, can always be cast to the form $\Theta_{IJ} = \coup \delta_{IJ}$.
In~\eqref{eq:covd1} we have written the generators $\T_0^I{}_J$ that correspond to the $p=2$ flowed $\mf{sl}_9$.  The vector fields $A_\mu^{IJ} $ are not propagating in $D=2$. 

The gauged theory is then given by covariantising all derivatives in~\eqref{eq:physLag SL9} and introducing a topological term for the non-abelian field strength and a scalar potential in the form
\begin{align}
\label{eq:SO91}
\mathcal{L}_{\text{gsugra}} = \mathcal{L}_{{\rm sugra},{\rm cov}} +\frac12 \varepsilon^{\mu\nu} F_{\mu\nu}^{IJ} \Theta_{JK} h^K{}_I - V_{\text{gsugra}}\,,
\end{align}
where the vector field strength reads
\begin{equation}\label{eq:SO9 field strength}
F_{\mu\nu}^{IJ} = 2\,\partial_{[\mu} A_{\nu]}^{IJ} + 2\,\Theta_{KL} A_{[\mu}^{IK} A_{\nu]}^{JL}\,,
\end{equation}
and the scalar fields $h^I{}_J$ are identified with those introduced on-shell in the duality relation~\eqref{eq:hdef}.
The potential term $V_{\text{gsugra}}$ was obtained from supersymmetry in~\cite[Eq.~(5.5)]{Ortiz:2012ib}. We shall derive it from the gSS reduction of E$_9$ ExFT in Section~\ref{sec:red} and therefore do not display it here. 
It is however important to stress that only the anti-symmetric combination $h^K{}_{[I} \Theta_{J]K}$ appears in $V_{\text{gsugra}}$ as well as in \eqref{eq:SO91}.

Both fields $h^I{}_J$ and $A_\mu^{IJ}$ are auxiliary in SO(9) gauged supergravity. Their equations of motion are consistent with the gauge-covariantised version of the duality equation~\eqref{eq:hdef} when varying $A_\mu^{IJ}$, while varying $h^I{}_J$ fixes the curvature $F_{\mu\nu}^{IJ}$ of the non-propagating vector fields in terms of the remaining fields of the theory.
Integrating out the auxiliary field $h^I{}_J$ that occurs only algebraically in the Lagrangian leads to a Yang--Mills kinetic term for the vector fields. 

\medskip
In order to covariantise the duality relations \eqref{eq:twsd} properly, we need to identify which part of the infinite-dimensional rigid on-shell symmetry of ungauged $D=2$ supergravity is gauged. 
Both from the general structure of the gSS reduction of E$_9$ ExFT~\cite{gss} and the analysis of supergravity directly in $D=2$ dimensions~\cite{Samtleben:2007an}, one knows that Lagrangian gaugings utilise an embedding tensor that takes values in the basic representation $\overline{R(\Lambda_0)_{-1}}$ that was discussed in Section~\ref{sec:basdec}. 
The general coupling of vector fields $|A_\mu\rangle$ in $R(\Lambda_0)_{-1}$  to the embedding tensor is given through the pairing 
\begin{align}\label{eq:covd2}
D_\mu = \partial_\mu +  \eta_{-1\,\alpha\beta} \langle  \theta | \T^\alpha |A_\mu\rangle \bbdelta^\beta\,,
\end{align}
where the covariant derivative acts through the rigid $\mf{e}_9\oleft\langle\LL_{-1}\rangle$ variation $\bbdelta^\alpha$ on the various fields.\footnote{See~\cite{gss}. For instance, $\bbdelta^\alpha V = V\T^\alpha - k^\alpha V $ with $k^\alpha$ a compensating $K(\mf{e}_9)$ transformation. Notice that the choice of $\T^\alpha$ rather than, for example, $T^\alpha$, is linked to the choice of such basis in \eqref{eq:covd2}.}
The decomposition under the $p=2$ spectrally flowed $\mf{sl}_9$ was given in~\eqref{eq:flowdec}. 
The restriction to $\langle \theta|$ in the ${\bf 45}_{16/9}$
, with%
\begin{equation}\label{eq:bratheta SO9}
\bra\theta = -\Theta_{IJ} \bra{4/3}^{IJ}\,,
\end{equation}
reproduces the gauged covariant derivative~\eqref{eq:covd1} when acting on the physical fields.
The expression \eqref{eq:covd2} also determines the gauging of shift symmetries of the dual potentials.
The first few terms in its expansion are determined as follows
\begin{align}
\eta_{-1\,\alpha\beta}\bra\theta \T^\alpha \ket{A_\mu} \T^\beta =\ & 
-\Theta_{IK} A_\mu^{JK} \T_0{}^I{}_J
 -\frac{1}{2} \Theta_{KI_1} \bra{1}^K_{I_2I_3}\ket{A_\mu} \T_{-2/3}^{I_1I_2I_3} 
\nonumber\\&
+2 \, \Theta_{IJ} \bra{4/3}^{KI}\ket{A_\mu} \T_{-1}^{\,J}{}_K
+\Theta_{IJ} \bra{4/3}^{IJ} \ket{A_\mu} \LL_{-1}
+ \ldots\ ,
\end{align}
where the dots stand for generators of lower $\LL_0$ degree, and we identified {$A_\mu^{IJ} = \bra{1/3}^{IJ}\ket{A_\mu}$}.
Notice in particular that, based on the coset parametrisation \eqref{eq:E9cosetrep SL9 basis}, the shift symmetries of the dual axions $b_{IJK}$ are entirely gauged, while only the shifts of the symmetric combination $\Theta_{K(I}h^K{}_{J)}$, which does not enter the Lagrangian, are gauged.

We also provide the general expression for the field strengths, which is, in form notation
\begin{equation}
\ket{F} = \ket{\dd A} -\frac12 \eta_{-1\,\alpha\beta}  \bra\theta \T^\alpha\ket{A}\wedge \T^\beta\ket{A}
-  \eta_{-1\alpha\beta}  \langle \theta | \T^\alpha \otimes \T^\beta |\hspace{-0.5mm}| {C}\rangle\hspace{-1mm}\rangle\,
\label{eq:FSgs}
\end{equation}
where $|\hspace{-0.5mm}| {C}\rangle\hspace{-1mm}\rangle$ denotes the two-form, sitting in the symmetric tensor product of two $R(\Lambda_0)$ representations, with the $R(2\Lambda_0)$ representation subtracted.\footnote{In the companion paper~\cite{gss}, we denote $|\hspace{-0.5mm}| {C}\rangle\hspace{-.8mm}\rangle = \ket{C_{(1}}\otimes\ket{C_{2)}}$.}
Using \eqref{eq:bratheta SO9}, the first few entries in $\ket F$ are found to be
\bea \label{eq:gsugra Fs}  
\langle 1/3|^{IJ} |F\rangle  &=&  \langle 1/3|^{IJ} |\dd A\rangle + \Theta_{KL}  \langle 1/3|^{IK} |A\rangle \wedge \langle 1/3|^{JL} |A\rangle\,, \\*
 \langle 4/3|^{IJ} |F\rangle  &=&  \langle 4/3|^{IJ} |\dd A\rangle + 2 \Theta_{KL}  \langle 1/3|^{K(I} |A\rangle \wedge \langle 4/3|^{J)L} |A\rangle\,, \CR
  \langle 0|_I |F\rangle  &=&  \langle 0|_I |\dd A\rangle - \Theta_{IJ}  \langle 1/3|^{JK} |A\rangle \wedge \langle 0|_K |A\rangle +   \Theta_{IJ} {\langle 1/3|}{}^{JK}  \otimes   {\langle 0|}{}_{K}  |\hspace{-0.5mm}| {C}\rangle\hspace{-1mm}\rangle\,,\CR 
   \langle 1|_{IJ}^K  |F\rangle  &=&   \langle 1|_{IJ}^K  |\dd A\rangle - \Theta_{PQ}  \langle 1/3|^{KP} |A\rangle \wedge   \langle 1|_{IJ}^Q  |A\rangle+2  \Theta_{P[I}  \langle 1/3|^{PQ} |A\rangle \wedge   \langle 1|_{J]Q}^K |A\rangle \CR\nonumber
   && + 4  \Theta_{P[I}  \bigl( {\langle 1/3|}{}^{Q(K}  \otimes   {\langle 1|}{}^{P)}_{J]Q} + {\langle 0|}{}_{J]}  \otimes   {\langle 4/3|}{}^{KP}   \bigr) |\hspace{-0.5mm}| {C}\rangle\hspace{-1mm}\rangle + 2\delta^K_{[I}  X_{J]}\,,
\eea
where the definition of $ |\hspace{-0.5mm}| {C}\rangle\hspace{-1mm}\rangle $ has been modified {compared to~\eqref{eq:FSgs}} in order to reabsorb {some $|A\rangle\wedge|A\rangle$} terms. The gauge field $\langle 0|_I |A\rangle$ is therefore pure gauge, as well as all but the completely antisymmetric component $\Theta^{\vphantom{I}}_{L[I}  \langle 1|_{JK]}^L  |A\rangle$ of the weight $ \frac{13}{9}$ gauge field. 
The trace component of $\langle 1|_{IJ}^K  |F\rangle $ has a piece $\delta^K_{[I} X_{J]}$ which we do not display explicitly since it is projected out from all physically relevant quantities.
From \eqref{eq:gsugra Fs} we see that the restriction to the ${\bf36}_{-16/9}$ component {$F^{IJ}= \bra{1/3}^{IJ}\ket{F}$} reproduces \eqref{eq:SO9 field strength}.

\subsection{\texorpdfstring{The generalised Scherk--Schwarz ansatz on $S^8\times S^1$}{The generalised Scherk--Schwarz ansatz on S8 x S1}}
\label{sec:so9gss}

The generalised Scherk--Schwarz ansatz for the SO(9) gauging follows the general procedure for E$_9$ ExFT presented in~\cite{gss}. We recall from there that a complete gSS ansatz consists in identifying a twist matrix $\cU(y)\in \mathrm{E}_9$, along with an additional $y$-dependent  $\langle w^+|\in \overline{R(\Lambda_0)_{-2}}$, that together induce the constant embedding tensor of the SO(9) gauged theory and that factor out of the pseudo-Lagrangian. The twist matrix $\cU$ comprises a scalar component $\twistrho(y)$ along $\LL_0$ and the bra vector $\langle w^+|$ satisfies the `flat version' the section constraint, namely the bra vector $\bra{w^+}\cU$ must be on section.

The structure of the gSS ansatz for the exceptional field theory fields is~\cite{gss}
\begin{align}
\label{eq:gssans}
    \mathcal{V}(x,y) &= V(x)\, \mathcal{U}(y)\,,\nn\\
    \rho(x,y) &= \twistrho(y)\, \sugrarho(x)\,,\\
    \ket{\mathcal{A}(x,y)} &= \twistrho^{-1}(y)\, \mathcal{U}^{-1}(y) \ket{A(x)}\,.\nn
\end{align}
In these expression the ExFT fields are on the left-hand side and the twist matrix $\mathcal{U}(y)$ and the scalar $\twistrho(y)$ must be chosen such that the dependence on the internal ExFT coordinates $y$ factorises from all equations. The bra vector $\bra{w^+}$ appears in the expression for the embedding tensor below.
The ansatz~\eqref{eq:gssans} will be central for the derivation of the uplift formul\ae{} in Section~\ref{sec:uplift}.\footnote{E$_9$ ExFT contains additional constrained fields and two-forms that have to be considered in the general construction but will not play a role in this paper.}
Besides the ansatz for the fields we also record the ansatz for the parameter $\ket{\Lambda}$ of generalised diffeomorphisms in E$_9$ ExFT:
\begin{align}
    \ket{\Lambda(x,y)} = r^{-1}(y)\, \mathcal{U}^{-1}(y) \ket{\lambda(x)}\,.
\end{align}
The generalised diffeomorphism action on the ExFT fields reduces to the gauge symmetries of the gauged supergravity theory in two dimensions. Consistency of the resulting gauge algebra places constraints on $\mathcal{U}(y)$ and $\twistrho(y)$ that we review next.

The twist matrix $\cU$ gives rise to the Weitzenb\"ock connection $\langle W_\alpha|$ through its first internal derivative according to
\begin{align}
\label{eq:Weitz}
\langle \partial | \otimes \cU = \twistrho  \langle W_\alpha | \cU \otimes \T^\alpha \cU\,.
\end{align}
Notice that we expand the Weitzenb\"ock connection in terms of the $p=2$ flowed basis, as this will be the natural choice for the case of the SO(9) reduction ansatz.
Then, the embedding tensor resulting from a gSS reduction is made of two components $\bra\vartheta$ and $\bra\theta$ \cite{Bossard:2017aae,gss}\footnote{Notice that writing $\bra\theta$ in terms of $\sS_{+1}$ rather than $\cS_{+1}$ only amounts to a redefinition of $\bra{w^+}$.}
\begin{equation}\label{eq:theta from W}
\bra\vartheta = \bra{W_\alpha}\T^\alpha\,,\qquad
\bra\theta=-\bra{W_\alpha}\sS_{+1}(\T^\alpha)-\bra{w^+}\,.
\end{equation}
The component $\bra\vartheta$ corresponds to non-Lagrangian gaugings of the $\LL_0$ symmetry. 
We will only consider situations in which $\bra\vartheta=0$.
The components of the embedding tensor must be constant and integrability of its definition also enforces the quadratic constraint~\cite{Samtleben:2007an,gss}
\begin{align}
\label{QuadraCons}
  \eta_{-1\,\alpha\beta} \bra{\theta}T^\alpha \otimes \bra{\theta}T^\beta=0\,,    
\end{align}
where we have already set $\bra{\vartheta}=0$. 

We have argued for the relevance of the $p=2$ spectrally flowed $\mf{sl}_9\subset \mf{e}_9$ in the previous section. From the general analysis that we review in Appendix~\ref{app:uplift}, we know that the twist matrix must decompose as in equation~\eqref{eq:twistdeco}. This implies that $\cU$ belongs to the parabolic subgroup obtained from the generators of non-positive $\LL_0$ degree. Because we are looking for a consistent truncation contaning a purely gravitational pp-wave, it is also natural to restrict the ansatz to a parabolic element in the affine extension of SL(9). This is the relevant structure for eleven-dimensional gravity without three-form. One can moreover use gauge invariance to restrict the ansatz to the SL(9) subgroup. 
Using the notation $\T^I_{m\, J}$ etc. for the flowed generators as in Section~\ref{sec:flow}, we therefore make the following ansatz for the inverse of the twist matrix
\begin{align}
\label{eq:twistans}
\cU^{-1} =   \twistrho^{\LL_0}  e^{\twistsigma \dK} \,\mathsf{u}^{-1}\,,
\end{align}
where $\mathsf{u}$ belongs to the $p=2$ flowed SL(9).
Recall from~\eqref{eq:flowvir} that $\LL_0 = L_0 + 2\, T^k_{0\, k} + \frac{16}{9} \dK$. Except for the occurrence of  $\langle w^+|$ whose form will be determined below, the twist matrix and choice of section are analogous to what happens for other sphere reductions~\cite{Hohm:2014qga}.

We also have to specify a solution to the section constraint corresponding to the coordinates on $S^8\times S^1$. 
As argued in Section~\ref{sec:p=2flow}, the $S^8$ coordinates are expected to sit within the $\bf36$ of the $p=2$ flowed SL(9) and we indeed find such a representation at $\LL_0$-degree $7/9$ in the decomposition \eqref{eq:flowdec}. 
The solution to the section constraint for the derivatives $\langle \partial| \in \overline{R(\Lambda_0)_{-1}}$ is naturally written in the branching~\eqref{eq:bas1} with respect to the $p=1$ flowed SL(9).
It can be expressed in the branching~\eqref{eq:bas2} with respect to the $p=2$ flowed SL(9) relevant for the SO(9) gauging, by splitting  the index $\tilde{I}$ in~\eqref{eq:bas1} into $\tilde{I}=(i,9)$ and the index $I$ in~\eqref{eq:bas2} into $I=(0,i)$ and using~\eqref{eq:sec in two bases}:
\begin{equation}
\label{eq:secsol}
\langle \partial | = \widetilde{\langle 0 |}{}^{\tilde{I}} \partial_{\tilde{I}} = \langle 1/3|^{0i} \partial_i  +  \langle 4/3 |^{00} \partial_9  \,.
\end{equation}
In terms of the ${\bf 36}_{7/9}$ derivatives $\partial_{IJ}$  the solution~\eqref{eq:secsol} implies in particular that only $\partial_i\equiv \partial_{0i} \neq 0$ among the $\partial_{IJ}$.
That this corresponds to a solution of the section constraint can be verified in a straight-forward manner.
The gSS ansatz for the physical fields will only involve the eight coordinates along $\langle 1/3|^{0i}$, corresponding to $S^8$, while the component $\langle 4/3 |^{00}$ associated to the $S^1$ coordinate $y^9$ will only feature in the constrained fields.\footnote{%
In terms of the expansion~\eqref{eq:derm},
the nine non-vanishing components of the derivatives $\langle \partial| \in \overline{R(\Lambda_0)_{-1}}$ are rewritten as
$
\langle 0| T^i_{1} \partial_i \neq 0\,,\ 
\langle 0| \partial_\psi  \neq 0\,,
$
where the generators $T^i_1$ refer to the $\mf{gl}_8$ basis~\eqref{eq:e8gl8}. }
This kind of interplay of the $p=1$ flowed branching and $p=2$ flowed branching of the $\overline{R(\Lambda_0)_{-1}}$ representation will be central in Section~\ref{sec:uplift} when we determine the explicit uplift formul\ae.

Equipped with the choice of Section~\eqref{eq:secsol} and ansatz~\eqref{eq:twistans} for the twist matrix, we can evaluate the Weitzenb\"ock connection~\eqref{eq:Weitz}. Some details of this calculation are given in Appendix~\ref{app:Weitz}. From this we deduce the following expressions for the trombone embedding tensor $\langle \vartheta|\in \overline{R(\Lambda_0)_{-1}}$ and the standard embedding tensor $\langle \theta| \in \overline{R(\Lambda_0)_{0}}$ given by 
\begin{align}
\label{eq:vartheta9}
\langle \vartheta | &= \langle W_\alpha| \T^\alpha =- \twistrho^{-1} \partial_{i} \left( \twistrho^{\frac{7}{9}} e^\twistsigma \mathsf{u}^{-1\, 0}{}_I \mathsf{u}^{-1\, i}{}_J\right) \langle 1/3|^{IJ}\,, 
\end{align}
and
\begin{align}
\label{eq:theta9}
\langle \theta | &= -\bra{W_\alpha}\sS_{+1}(\T^\alpha)-\bra{w^+}\nn\\
&=\twistrho^{-\frac{2}{9}} e^\twistsigma \mathsf{u}^{-1\, 0}{}_K \mathsf{u}^{-1\, i}{}_L  \Big(\partial_{i}\mathsf{u}^S{}_{0} \mathsf{u}^{-1\, 0}{}_R  +\partial_{i}\mathsf{u}^S{}_{j} \mathsf{u}^{-1\, j}{}_R \Big)
\left( \langle 1/3|^{[KL}_{\phantom{[}} \T^{R]}_{1\, S} - \frac27  \langle 1/3|^{Q[K}_{\phantom{[}}  \T^{L\phantom{]}}_{1\, Q} \delta^{R]}_S \right)\nn\\
&\quad +\frac18 \twistrho^{-\frac{2}{9}}  e^\twistsigma  \left(   \mathsf{u}^{-1\, 0}{}_K \partial_{i} \mathsf{u}^{-1\, i}{}_L-  \mathsf{u}^{-1\, i}{}_K \partial_{i} \mathsf{u}^{-1\, 0}{}_L - W^+_{00} \, \mathsf{u}^{-1\,0}{}_K \mathsf{u}^{-1\, 0}{}_L \right)  \langle 1/3|^{P(K}_{\phantom{]}} \T^{L)}_{1\, P} \nn\\
&\quad +\frac{9}{14} \twistrho^{-16/9} e^\twistsigma   \partial_{i} \left( \twistrho^{14/9}  \mathsf{u}^{-1\, 0}{}_K \mathsf{u}^{-1\, i}{}_L \right)\langle 1/3|^{KL} \LL_1  
\,,
\end{align}
where we have simplified the constrained $\langle w^+|=W^+_{00} \langle 4/3|^{00}=W^+_{00} \widetilde{\langle 0|}{}^{9}$ as a specialisation of the solution~\eqref{eq:secsol} to the section constraint.

In order to obtain a Lagrangian gauging corresponding to a sphere reduction of type IIA supergravity we now consider an ansatz for the SL(9) twist matrix that is inspired by~\cite{Hohm:2014qga,Inverso:2017lrz}. We denote embedding coordinates or the sphere $S^8$ by $Y_I$ where $I=0,1,\ldots 8$ are Euclidean ambient space coordinates with $\sum_{I=0}^8 Y_I^2=1$ and we raise and lower these indices with the Euclidean ambient space metric $\delta_{IJ}$, invariant under SO(9). 
The SL(9) twist matrix components then are taken to be
\begin{align}
\mathsf{u}^{-1\, i}{}_I &= (\det \mathring g)^{1/9} \left( \mathring g^{ij} \partial_j Y_I + c^i Y_I\right)\,,\nn\\
\mathsf{u}^{-1\, 0}{}_I &= (\det \mathring g)^{-8/9+1/2} Y_I\,,
\end{align}
where $\mathring g_{ij} = \partial_i Y_I \partial_j Y_J \delta^{IJ}$ is the induced metric on the (round) sphere, $\mathring g^{ij}$ its inverse and $\det \mathring g$ its determinant. The field $c^i$ is akin to a Kaluza--Klein vector and related to the flux of the sphere compactification. The embedding coordinates satisfy the completeness relation and eigenvalue equation
\begin{align}
\label{eq:Yrels}
\mathring g^{ij} \partial_i Y_I \partial_j Y_J = \delta_{IJ} - Y_I Y_J
\quad\text{and}\quad
(\det \mathring g)^{-1/2} \partial_i \left( (\det  \mathring g)^{1/2} \mathring{g}^{ij} \partial_j Y_I \right) = -8 Y_I\,,
\end{align}
as can be checked easily by going to stereographic coordinates.
The components of the inverse SL(9) matrix are then
\begin{align}
\mathsf{u}^I{}_i &=  (\det \mathring g)^{-1/9} \partial_i Y^I\,, \label{gSSuY}
\nn\\
\mathsf{u}^I{}_0 &= (\det \mathring g)^{-1/9+1/2} \left( Y^I - c^i \partial_i Y^I\right)\,.
\end{align}

We next determine the conditions on the remaining components of the twist matrix~\eqref{eq:twistans}.
The requirement that the gauging be Lagrangian means that the trombone embedding tensor $\langle \vartheta|$ in~\eqref{eq:vartheta9} has to vanish.
Using~\eqref{eq:Yrels} and $[KL]$ anti-symmetry we find that this is tantamount to
\begin{align}
0 &\overset{!}{=} \partial_i \left( \twistrho^{\frac{7}{9}} e^\twistsigma (\det \mathring g)^{-7/9+1/2} \mathring g^{ij}\partial_j Y_K Y_L\right) \langle 1/3|^{KL}\nn\\
&= \partial_i \left( \twistrho^{\frac{7}{9}} e^\twistsigma (\det \mathring g)^{-7/9} \right) (\det \mathring g)^{1/2} \mathring g^{ij}\partial_j Y_K Y_L\langle 1/3|^{KL}\,,
\end{align}
so that we deduce
\begin{align}
\twistrho^{\frac{7}{9}} e^\twistsigma = \coup (\det \mathring g)^{7/9} \,,
\end{align}
for the vanishing of $\langle \vartheta|$. In the above relation we have introduced a convenient integration constant. 

In order to represent a consistent gSS reduction, the embedding tensor $\langle \theta|$ must be constant and we need it to be solely along the ${\bf 45}$ component according to the discussion in Section~\ref{sec:so9rev}. Substituting the sphere ansatz into~\eqref{eq:theta9}, we find that the component along the ${\bf 720}$ vanishes automatically and for the vanishing of the ${\bf 36}$ component along $\langle 1/3|^{KL} \LL_1$ we get
\begin{align}
0 &\overset{!}{=}  \partial_i \left( \twistrho^{14/9} (\det \mathring g)^{-5/18} Y_K \mathring g^{ij} \partial_j Y_L\right) \langle 1/3|^{KL} \LL_1\,,
\end{align}
whose vanishing according to~\eqref{eq:Yrels} requires
\begin{align}
\label{eq:rhosol}
\twistrho =  (\det \mathring g)^{1/2}
\quad \Rightarrow\quad
e^\twistsigma = \coup (\det \mathring g)^{7/18}\, .
\end{align} 
Here we fixed the integration constant for $\twistrho$ using the ExFT $\LL_0$ scaling symmetry.

Substituting this back into~\eqref{eq:theta9}, we are left with
\begin{align}
\langle \theta | &=  \coup \left[  - \delta_{IJ}  + \left( (\det \mathring g)^{-1/2} \partial_i \bigl( (\det \mathring g)^{1/2} c^i\bigr) - (\det \mathring g)^{1/2} W^+_{00} - 7 \right) Y_I Y_J \right] \langle 4/3|^{IJ}  \,.
\end{align}
For this to be a constant multiple of the SO(9) metric $\delta_{IJ}$ we need to make the anisotropic components vanish. Due to the presence of the component $W_{00}^+$ of $\langle w^+|$, the first condition $(\det \mathring g)^{-1/2} \partial_i \left( (\det \mathring g)^{1/2} c^i\right) -(\det \mathring g)^{1/2} W^+_{00} = 7$ is trivially satisfied. From the perspective of type IIA supergravity it is natural to set $W_{00}^+=0$, in which case this condition fixes the field $c^i$ corresponding to the seven-form type IIA potential in the ansatz. One then identifies the expected 8-form flux on $S^8$. From the perspective of eleven-dimensional supergravity, the field $c^i$ is a component of the dual graviton and it is natural to absorb the flux in the constrained field component $W_{00}^+$.

The summary of the analysis above is that we have achieved the form
\begin{align}
\label{eq:embfin}
\langle \vartheta | =0 \quad\text{and}\quad
\langle \theta | = - \coup \, \delta_{IJ} \langle 4/3|^{IJ} \,,
\end{align}
for the embedding tensor, which agrees with the identification $\Theta_{IJ}=\coup\,\delta_{IJ}$ in Section~\ref{sec:so9rev}.

\vskip 2mm

This construction can be easily adapted to accommodate also CSO($p,q,9{-}p{-}q$)-type gaugings~\cite{Hull:1984qz,Hull:1984vg} where the signature $(p,q,9{-}p{-}q)$ describes the number of positive, negative and vanishing eigenvalues of the symmetric tensor $\Theta_{IJ}$ in~\eqref{eq:bratheta SO9}. 
The internal space in these cases is $H^{p,q}\times T^{9{-}p{-}q}\times S^1$ with $H^{ p,q}$ the ($p{+}q{-}1$)-dimensional sphere or hyperboloid defined by the equation $Y^a \eta_{ab} Y^b = 1$ with $\eta_{ab}$ of signature $(p,q)$ in $p{+}q$ dimensions and $\Theta_{ab} = \coup\,  \eta_{ab}$. We then write accordingly $i=1,\ldots,p{+}q{-}1$ for the coordinate indices of  $H^{p,q}$ and $\hat{\imath}=p{+}q,\ldots,8$ for the indices of the coordinates on $T^{9{-}p{-}q}$. The twist matrix takes the form
\begin{align}
\mathsf{u}^a{}_i &=  |\det \mathring g|^{-1/9} \partial_i Y^a\,, \label{gSSuYpqr}
\nn\\
\mathsf{u}^a{}_0 &= |\det \mathring g|^{-1/9+1/2} \left( Y^a - c^i \partial_i Y^a\right) \nn\\
\mathsf{u}^{\hat{a}}{}_{\hat{\imath}} &= \delta^{\hat{a}}_{\hat{\imath}}   \, , 
\end{align}
where  $\mathring{g}_{ij}$ is the pseudo-Riemannian induced metric $\eta_{ab} \partial_i Y^a \partial_j Y^b$ and $c^i$ and $W^+_{00}$ satisfy 
\be  
|\det \mathring g|^{-1/2} \partial_i \bigl( |\det  \mathring g|^{1/2} c^i\bigr) - |\det \mathring g|^{1/2} W^+_{00} = p+q-2 \; . 
\ee

\subsection{From  pseudo-Lagrangian to  physical Lagrangian}
\label{sec:red}

As described in the companion paper \cite{gss}, for any embedding tensor $\bra\theta$ admitting a consistent uplift, the gSS reduction of the ExFT pseudo-Lagrangian leads to a pseudo-Lagrangian for gauged supergravity that decomposes into
\begin{align}
    \cL^{\text{pseudo}}_{\text{gsugra}}=\cL^{\rm top}_{\rm gsugra}-V_{\text{gsugra}}\,.\label{eq:pseudogL}
\end{align}
The general expression of the scalar potential $V_{\text{gsugra}}$ in terms of $\bra{\theta}$ will be recalled later on.
We stress that we include the measure factor in its definition.
The topological term $\cL^{\rm top}_{\rm gsugra}$ corresponds the gauged version of the (topological) pseudo-Lagrangian $\cL^{\rm pseudo}_{\rm sugra}$ for ungauged supergravity defined in \eqref{eq:LtopK}. It can be written schematically as the sum of two terms
\begin{align}\label{eq:Ltop gsugra schematic}
\cL^{\rm top}_{\rm gsugra}\ \dd x^0\wedge\dd x^1 
=
2\varrho\,\mathbf{D}\breve\upchi_1 + \varrho\,\langle\theta|\mathbf{O}(M)|F\rangle\,.
\end{align}
The first term corresponds to \eqref{eq:LtopSL9} with partial derivatives traded for gauge covariant ones.
The second term, linear in the field strengths, is new and descends from the proper covariantisation of the Maurer--Cartan equation used in \eqref{eq:LtopK} to define $\cL^{\rm pseudo}_{\rm sugra}$. 
See equation (B.4) of the companion paper \cite{gss} for the gauged supergravity version of the relation \eqref{eq:LtopK}.
Here we are already considering an expansion in terms of the $p=2$ spectrally flowed SL(9) basis.
Explicitly, we have
\begin{align}\label{eq:theta-OM-F}
\varrho\,\bra{\theta}\mathbf{O}(M)\ket{F} =\ &
\varrho\,\bra\theta \Big( \sS^{\sugraupgamma}_{+1}(\LL_{-1})+\sS^{\sugraupgamma}_{-1}(\LL_{-1})\Big) \ket{F}
\\*[1ex]\nonumber&
-\varrho\,\upomega^\alpha(V)\big[\sS^{\sugraupgamma}_{+1}(\T^\beta)+\sS^{\sugraupgamma}_{-1}(\T^\beta)\big]_\alpha \eta_{-1\beta\gamma}\bra{\theta}\T^\gamma\ket{F} \,,
\end{align}
where we have introduced the group cocycle
\begin{equation}\label{eq:groupcocycleSL9}
\upomega^\alpha(g)\dK = \sS_0(\T^\alpha)- g^{-1}\sS_0(g\T^\alpha g^{-1})g\,,\qquad\qquad g\in\widehat{\mathrm E}_8\rtimes\Virm\,,
\end{equation}
as well as a field-dependent version of the shift operators
\begin{align}\label{eq:SupgammaSL9}
\sS^{\sugraupgamma}_{m}(X) 
&\ =\ \sS_0\big(V^{-1}\sS_m(V\,X\,V^{-1})V\big)\\\nonumber
&\ =\ \varrho^{-m}\big(\sS_m(X) - m\varphi_1 \sS_{m-1}(X) + \ldots \big)\,,\qquad X\in\hevir\,.
\end{align}
The second line displays how these field-dependent shift operators are expanded in terms of the standard ones $\sS_k$, with $k\le m$, a fact that we will use shortly.
Details on this construction are found in \cite{Bossard:2021jix,gss}.

We will now present how to obtain a physical Lagrangian for gauged supergravity from  the pseudo-Lagrangian \eqref{eq:pseudogL}. Apart from the scalar potential, the other terms in the physical Lagrangian are obtained from \eqref{eq:Ltop gsugra schematic} by first repeating the same steps that we used in Section~\ref{sec:pseudotophys} to reproduce \eqref{eq:physLag SL9} from \eqref{eq:LtopSL9}, with covariant differentials instead of partial ones, and then adding $\langle\theta|\mathbf{O}(M)|F\rangle$ to the final result.
Let us now show this for the SO(9), or more generally for  $\mathrm{CSO}(p,q,r)$ gaugings of which the SO(9) gauging is a special case.

The first term in \eqref{eq:Ltop gsugra schematic} corresponds to \eqref{eq:LtopSL9} with covariant differentials. All steps carried out in Section~\ref{sec:pseudotophys} are still valid, until one arrives at the expression \eqref{eq:Ltop axions last step}.
Since the differentials are now covariantised, the last term there is no longer a total derivative but instead contributes with
\begin{align}\label{extra term to cancel}
-\frac{1}{18} a^{IJK} D^2 b_{IJK} &=
-\frac16 a^{L_1 L_2 L_3} \Theta_{L_1 K} F^{I K} b_{L_2L_3 I}
-\frac16 a^{L_1 L_2 L_3} \Theta_{L_1 K} F^{K}_{\,L_2 L_3}\,,
\end{align}
where we have defined the $\mathbf{36}_{-16/9}$ component $F^{IJ}=\bra{1/3}^{IJ}\ket{F}$
and the $(\overline{\bf9}\oplus {\bf315})_{-13/9}$ components $F^K_{IJ}=\bra{1}^K_{IJ}\ket{F}$
of the field strength $\ket{F}$ so that
\begin{equation}
\eta_{-1\,\alpha\beta}\bra\theta \T^\alpha \ket{F} \T^\beta = 
-\Theta_{IK} F^{JK} \T_0{}^I{}_J
- \frac12 \Theta^{\vphantom{I}}_{KI_1} F^K_{I_2I_3} \T_{-2/3}^{I_1I_2I_3} + \ldots\ .
\end{equation}

Looking now at the first line of \eqref{eq:theta-OM-F} and using the expansion in \eqref{eq:SupgammaSL9}, we see that acting on $\bra\theta$ is a series of Virasoro generators $\LL_{-k}$ with $k\ge0$.
One immediately finds that
\begin{equation}\label{eq:4/3 no vir}
\bra{4/3}^{(IJ)}\LL_{-k} = 0 \quad\forall k\ge1\,,
\end{equation}
hence for CSO$(p,q,r)$ gaugings, only the term proportional to $\bra\theta\LL_0\ket{F}$ survives.
But then we find
\begin{equation}\label{eq:theta OM F tot der}
\varrho\,\bra\theta \Big( \sS^{\sugraupgamma}_{+1}(\LL_{-1})+\sS^{\sugraupgamma}_{-1}(\LL_{-1})\Big) \ket{F} =
\Theta_{IJ}\bra{4/3}^{IJ} \mathsf{L_0}\ket{F}  
= \frac{16}{9} \Theta_{IJ}\bra{4/3}^{IJ}\ket{\dd A}\,,
\end{equation}
which is a total derivative.
We used the quadratic constraint to remove the $A\wedge A$ term coming from the field strength, since it is
proportional to
$
\eta_{-1\,\alpha\beta} \bra{\theta}T^\alpha\ket{A} \bra{\theta}T^\beta\ket{A} = 0
$.

We are left with computing the second line of \eqref{eq:theta-OM-F}.
Given the triangular gauge \eqref{eq:E9cosetrep SL9 basis} for $V$ and the fact that the CSO($p,q,r)$ embedding tensor does not gauge positive level generators in this decomposition, we conclude that within the square bracket, only the positive shift contributes, with terms proportional to  the gauging of $\mathsf T_0{}^I{}_J$ and $\mathsf T_{-2/3}^{IJK}$, on which acts the constant shift operator $\sS_{+1}$.
Then, computing the loop cocycle is just a matter of dressing such terms with $V$ and extracting the $\dK$ component.
We thus find
\begin{align}
\varrho\,\bra{\theta}\mathbf{O}(M)\ket{F}
&=
\Theta_{JK} F^{IK}\, V \,\T_1^{\,I}{}_J\, V^{-1} \big|_\dK
+ \frac12 \Theta^{\vphantom{I}}_{KI_1} F^K_{I_2I_3} V\,\T_{-2/3}^{I_1I_2I_3}\,V^{-1}\big|_\dK
\\*\nonumber
&=
\Theta_{JK}F^{IK}\, h^I{}_J
+\frac16 a^{L_1 L_2 L_3} \Theta_{L_1 K} F^{I K} b_{L_2L_3 I}
+\frac16 a^{L_1 L_2 L_3} \Theta_{L_1 K} F^{K}_{L_2 L_3}\,,
\end{align}
up to the total derivative in \eqref{eq:theta OM F tot der}. 
Here, $|_\dK$ denotes a projection on the central charge, defined as in \eqref{eq:groupcocycleSL9}.
We see that the last two terms cancel out the contribution obtained from \eqref{extra term to cancel}.
This means that the non-potential terms in the physical Lagrangian are given by the na\"ive covariantisation of \eqref{eq:physLag SL9}, plus the only extra term $\Theta_{JK}F^{IK}\, h^I{}_J$. The physical Lagrangian for CSO($p,q,r)$ gauged supergravity therefore reads,
\begin{align}\label{eq:gaugedLag}
\cL_{\rm gsugra} =\ & 
\sqrt{-g}\left( 
  \varrho R + \varrho \frac14 g^{\mu\nu}  D_\mu\M^{IJ} D_\nu\M_{IJ} 
-\frac{1}{12}\rho^{1/3}g^{\mu\nu} D_\mu a^{I_1I_2I_3}  D_\nu a^{J_1J_2J_3} 
\M_{I_1J_1}\M_{I_2J_2}\M_{I_3J_3} 
\right)  
\nonumber\\&
+\frac1{6^4} \varepsilon^{\mu\nu}
\varepsilon_{I_1\ldots I_9}
a^{I_1I_2I_3} D_\mu a^{I_4I_5I_6} D_\nu a^{I_7I_8I_9} 
+\frac12\varepsilon^{\mu\nu} F_{\mu\nu}^{IJ} h^{K}{}_I \Theta_{KJ} - V_{\text{gsugra}}\,.
\end{align}

For the determination of the scalar potential we start from the general formula~\cite{Bossard:2022wvi,gss}
\begin{align}
\label{eq:Vpotg}
\Vpot &=   \frac1{2\sugrarho^3} \langle \theta | M^{-1} |\theta \rangle  + \frac1{2\varrho} \eta_{-2\, \alpha\beta} \langle \theta| \T^\alpha M^{-1} \T^{\beta\dagger}  | \theta \rangle\nn\\*
 &=   \frac1{2\sugrarho^3} \langle \theta | V^{-1} V^{-1\dagger} |\theta \rangle  + \frac1{2\varrho^3} \eta_{-2\, \alpha\beta} \langle \theta| V^{-1} \T^\alpha  \T^{\beta\dagger} V^{-1\dagger} | \theta \rangle
 \,.
\end{align}
Notice that $\Vpot$ includes the measure factor in its definition because the $\dK$ component of $M=V^\dagger V$ contains $e^{2\sigma}$.
As discussed at the end of Section~\ref{sec:basdec}, we use the notation for Hermitian conjugation of bra-ket vectors  $\ket{\theta}=(\bra{\theta})^\dagger$  to simplify the contractions appearing in the potential.
The second line shows a rewriting where the coset representatives $V$ were moved through  the generators $\T^\alpha$, $\T^\beta$ in the second term.
This is allowed as all terms that might be generated by such a manipulation turn out to vanish, see equation~(3.71) of~\cite{gss}.
For both terms in the potential the basic ingredient to compute is then $\bra{\theta}V^{-1}$.

We plug the ansatz~\eqref{eq:E9cosetrep SL9 basis} for the supergravity scalar fields  as well as the embedding tensor~\eqref{eq:bratheta SO9} into this expression.  
As shown in more detail in Appendix~\ref{sec:relmat} we have
\begin{align}\label{eq:thetaV}
\bra\theta V^{-1} & =  \Theta_{IJ} e^\sugrasigma \sugrarho^{16/9}\bigg(  -  \V^{-1\, I}{}_{\mathsf{A}} \V^{-1\, J}{}_{\mathsf{B}} \bra{4/3}^{{\mathsf{A}}{\mathsf{B}}}
+ \sugrarho^{-1} h^I{}_K \V^{-1\, J}{}_{\mathsf{A}} \V^{-1\, K}{}_{\mathsf{B}} \bra{1/3}^{{\mathsf{A}}{\mathsf{B}}}\nn\\
&\hspace{13mm} -\frac12 \sugrarho^{-1/3} a^{IKL} \V^{-1 \, J}{}_{\mathsf{A}} \V^{\mathsf{B}}{}_K \V^{\mathsf{C}}{}_L \bra{1}^{\mathsf{A}}_{{\mathsf{B}}{\mathsf{C}}}- \sugrarho^{-4/3} a^{IKL} h^J{}_L \V^{\mathsf{A}}{}_K \langle 0|_{\mathsf{A}} \nn\\
&\hspace{13mm}+\frac18 \sugrarho^{-2/3} a^{IKL} a^{JPQ}\V^{\mathsf{A}}{}_K \V^{\mathsf{B}}{}_L \V^{\mathsf{C}}{}_P \V^{\mathsf{D}}{}_Q \bra{2/3}_{{\mathsf{A}}{\mathsf{B}}{\mathsf{C}}{\mathsf{D}}} \nn\\
&\hspace{13mm}+\frac1{288} \sugrarho^{-1} a^{IK_1K_2} a^{JK_3K_4} a^{K_5K_6K_7} \varepsilon_{K_1\ldots K_7 RS} \V^{-1\, R}{}_{\mathsf{A}} \V^{-1\, S}{}_{\mathsf{B}}
\bra{1/3}^{{\mathsf{A}}{\mathsf{B}}}\nn\\
&\hspace{13mm} +\frac{1}{1152} \sugrarho^{-4/3} a^{IK_1K_2} a^{JK_3K_4} a^{K_5K_6K_7} a^{K_8K_9L} \varepsilon_{K_1\ldots K_9} \V^{\mathsf{A}}{}_L \langle 0|_{\mathsf{A}}\bigg)\,,
\end{align}
where~\eqref{eq:null9} was used and the generators and states of the basic module were written with local SO(9)$_K$ indices using the action by $\V$. 
Notice that the Virasoro scalar fields $\varphi_n$, $n\ge1$ do not contribute. This is easy to check using  \eqref{eq:4/3 no vir} and that the contribution proportional to $\bra{1}^{\mathsf{A}}_{{\mathsf{B}}{\mathsf{C}}}\LL_{-1} = 2 \delta^{\mathsf{A}}_{[{\mathsf{B}}} \bra{0}^{\phantom{{\mathsf{A}}}}_{{\mathsf{C}}]}$ vanishes because  $\Theta_{IJ} a^{IJK}=0$.
We also note that the term in $b_{IJK} \T^{IJK}_{-2/3}$ in the ansatz for $V^{-1}$ disappears as would any term with a field multiplying $\T_{-4/3\, IJK}$ (by incompatible index symmetries) or more negative degrees (by grading).

From this we can determine the various terms in $\Vpot$. 
Collecting all the terms leads to the following potential
\begin{align}
\label{eq:pot9}
  \Vpot &= \frac{ e^{2\sugrasigma} \sugrarho^{5/9}}{2 }  \Theta_{IJ} \Theta_{KL}  \Biggl( \left( 2\M^{IK} \M^{JL} -\M^{IJ} \M^{KL} \right) \nn\\
&\quad+\frac12  \sugrarho^{-2/3}   \Big( a^{IPQ} a^{KRS} \M^{JL} \M_{PR} \M_{QS} - 2 a^{IKP} a^{JLQ} \M_{PQ} \Big)\nn\\
& \quad + 2 \sugrarho^{-2} h^I{}_P h^K{}_Q \M^{Q[P} \M^{J]L}
+  \sugrarho^{-8/3}  a^{IPR} h^J{}_P a^{KQS} h^L{}_Q \M_{RS} \nn\\
&\quad + \frac{\sugrarho^{-2} }{72}  h^J{}_P a^{KQ_1Q_2} a^{LQ_3Q_4} a^{Q_5Q_6Q_7} \varepsilon_{Q_1\ldots Q_9} \M^{IQ_8} \M^{PQ_9}\\
&\quad + \frac38  \sugrarho^{-4/3}  a^{I[M_1M_2} a^{M_3M_4]J} a^{K[N_1N_2} a^{N_3N_4]L}  \M_{M_1N_1} \M_{M_2N_2} \M_{M_3N_3} \M_{M_4N_4} \nn\\
&\quad +  \frac{ \sugrarho^{-2}}{2\cdot 144^2}  a^{IN_1N_2} a^{JN_3N_4} a^{N_5N_6N_7} \varepsilon_{N_1\ldots N_9}   a^{KP_1P_2} a^{LP_3P_4} a^{P_5P_6P_7} \varepsilon_{P_1\ldots P_9} \M^{N_8P_8} \M^{N_9P_9} \nn\\
&\quad + \frac{ \sugrarho^{-8/3}}{576}  a^{IRP} h^J{}_R a^{KN_1N_2}a^{LN_3N_4} a^{N_5N_6N_7}a^{N_8N_9Q} \varepsilon_{N_1\ldots N_9}  \M_{PQ}\nn\\
&\quad + \frac{\sugrarho^{-8/3}}{1152^2}  a^{IN_1N_2}a^{JN_3N_4} a^{N_5N_6N_7}a^{N_8N_9Q} \varepsilon_{N_1\ldots N_9} a^{KP_1P_2}a^{LP_3P_4} a^{P_5P_6P_7}a^{P_8P_9S} \varepsilon_{P_1\ldots P_9}  \M_{QS} \Biggr) \,.\nn
\end{align}
For the SO(9) gauging we choose $\Theta_{IJ} = \coup \delta_{IJ}$.

To conclude this section, we compare our Lagrangian for SO(9) gauged supergravity  to that presented in~\cite{Ortiz:2012ib}, whose scalar potential can be computed by expanding the Yukawa couplings in their eq.~(4.22).
We performed this computation and find perfect agreement with \eqref{eq:pot9} upon identifying their $Y_{IJ} =Y_{[IJ]} = \Theta_{K[I} h^K{}_{J]}$ and also taking into account the typo mentioned in footnote~1 of~\cite{Anabalon:2013zka}.\footnote{Note that the second term in the second line of \eqref{eq:pot9} above was overlooked in eq.~(5.5) of~\cite{Ortiz:2012ib}.}

\subsection{Duality equation for the gauge field strength}

In order to determine completely the uplift ansatz in eleven dimensions, it is also useful to derive the expressions of fields that do not appear in the physical two-dimensional Lagrangian. The field strength duality equation
\be   |{F}\rangle + \frac{1}{\varrho^{3}}\star \Big(M^{-1}  |\theta\rangle+\eta_{-2\,\alpha\beta} V^{-1} T^\alpha T^{\beta\dagger} V^{-\dagger}\ket{\theta} \Big)= 0\,, \label{FieldStrengthDuality} \ee 
determines all the gauge fields that appear in the uplift ansatz~\cite[Sec.~6]{gss}. The relevant field strength equations are \begin{subequations}
\begin{align} 
\langle 1/3|^{IJ} |F\rangle  &=  -  \sugrarho^{-3} \star \langle 1/3|^{IJ} M^{-1} |\theta\rangle \; , \label{FSO9duality} \\*
 \langle 4/3|^{IJ} |F\rangle  &= - \star \frac{ \partial V_{\text{gsugra}}}{\partial \Theta^{IJ}}\; , \label{FKKduality} \\*
  \langle 0|_I |F\rangle  &= -  \sugrarho^{-3} \star \langle 0|_I M^{-1} |\theta\rangle \; , \label{F0dual} \\* 
   \langle 1|_{IJ}^K  |F\rangle  &=  -\sugrarho^{-3} \star \langle 1|^{K}_{IJ}  M^{-1} |\theta\rangle- \sugrarho^{-1}\eta_{-2\alpha\beta}\star   \langle 1|^{K}_{IJ} T^\alpha  M^{-1}  T^{\beta\dagger} |\theta\rangle \; . \label{F1dual}
\end{align} 
\end{subequations}
The right-hand side of \eqref{FSO9duality} gives 
\bea  \sugrarho^{-3} \langle 1/3|^{IJ} M^{-1} |\theta\rangle \hspace{-2mm} &=&\hspace{-2mm}  e^{2\sigma} \sugrarho^{-\frac{13}{9}} \Theta_{KL} \Bigl( \bigl( 2 \M^{K[I} \M^{J]P} + \sugrarho^{-\frac23} \M_{RS} a^{IJR} a^{KPS}\bigr) h^L{}_P \\
&& + \frac1{144} \varepsilon_{P_1\dots P_9} \bigl(  \M^{P_1[I} \M^{J]P_2} + \tfrac18 \sugrarho^{-\frac23} \M_{RS} a^{IJR} a^{P_1P_2S}\bigr) a^{P_3P_4K} a^{LP_5P_6} a^{P_7P_8P_9} \Bigr)\,, \nonumber \eea
and can be checked to be compatible with the equation of motion of $h^I{}_J$ using  
\be  \frac{ \partial V_{\text{gsugra}}}{\partial h^I{}_J }  = - \Theta_{IK}  \sugrarho^{-3} \langle 1/3|^{JK} M^{-1} |\theta\rangle  \; . \ee
This duality equation is identical to the equation of motion of $h^I{}_J$ for the SO(9) gauging $\Theta_{IJ} = \coup \delta_{IJ}$, but includes more components if $\Theta_{IJ}$ is degenerate. 

Similarly one computes the right-hand side of \eqref{F0dual} 
\be \sugrarho^{-3} \langle 0|_I M^{-1} |\theta\rangle =-e^{2\sigma} \sugrarho^{-\frac{19}{9}} \Theta_{KL} \M_{IJ} \Bigl( a^{JPK} h^L{}_P - \frac1{1152} \varepsilon_{P_1\dots P_9} a^{P_1P_2K} a^{LP_3P_4} a^{P_5P_6P_7} a^{P_8P_9J}\Bigr)\,,\ee
and of  \eqref{F1dual} 
\bea && \sugrarho^{-3} \langle 1|^{K}_{IJ}  M^{-1} |\theta\rangle+ \sugrarho^{-1}\eta_{-2\alpha\beta}  \langle 1|^{K}_{IJ} T^\alpha  M^{-1}  T^{\beta\dagger} |\theta\rangle \CR
&=&  -e^{2\sigma} \sugrarho^{-\frac{1}{9}} \Theta_{PQ} \biggl(  \M_{IR} \M_{JS} a^{RSP} \M^{QK} + 2 \delta^P_{[I} \M_{J]L} a^{QKL} \CR
&& + \frac{1}{2} \sugrarho^{-\frac23} \M_{IL_1} \M_{JL_2} \M_{RL_3} \M_{SL_4} a^{KRS} a^{L_1L_2P}a^{QL_3L_4} \CR
&& +\frac{1}{48\times 144} \sugrarho^{-\frac43} \varepsilon_{IJL_1\dots L_7} \varepsilon_{R_1\dots R_9} a^{KL_1L_2} a^{L_3L_4L_5} \M^{L_6R_1} \M^{L_7R_2} a^{R_3R_4P} a^{Qr_5R_6} a^{R_7R_8R_9}   \CR
&& +\frac{1}{24} \sugrarho^{-\frac43} \varepsilon_{IJL_1\dots L_7}  h^P{}_R \M^{QL_1} \M^{RL_2} a^{KL_3L_4} a^{L_5L_6L_7}   \CR
&& - \sugrarho^{-2} \Bigl( 2 h^K{}_{[I} \M_{J]L} + \frac1{144} \M_{LT} \varepsilon_{IJR_1\dots R_7} a^{KR_1R_2} a^{R_3R_4R_5} a^{R_6R_7T} \Bigr) \CR
&& \hspace{10mm} \times \Bigl( a^{LSP} h^Q{}_S - \frac1{1152} \varepsilon_{S_1\dots S_9} a^{S_1S_2P} a^{QS_3S_4} a^{S_5S_6S_7} a^{S_8S_9L}  \Bigr)  + \delta^{K}_{[I} Z^{PQ}_{J]}  \biggr) \CR
&& + b_{IJL} \sugrarho^{-3} \langle 1/3|^{KL} M^{-1} |\theta\rangle \; ,  \label{F1DualityRight}  \eea
where the tensor $Z^{PQ}_{J}$ is  not spelt out because it does not contribute to the eleven-dimensional fields.

\section{Uplift formul\ae}
\label{sec:uplift}
In this section we will present the uplift formul\ae{} for the eleven-dimensional metric and the three-form potential. We use the standard Kaluza--Klein ansatz for the metric 
\be \dd s^2_{\scalebox{0.6}{11D}} =  \rho^{-\frac{8}{9}}  e^{2\varsigma} \tilde{g}_{\mu\nu} {\rm d}x^\mu {\rm d}x^\nu +  \rho^{\frac{2}{9}}  G_{\Is\Js}  ( {\rm d}y^\Is +\mathcal{A}^\Is ) ( {\rm d}y^\Js +\mathcal{A}^\Js ) \; ,  \ee
and the three-form
 \be  A^{\scalebox{0.6}{11D}} \hspace{-1mm}=  \tfrac16 \alpha_{\Is\Js\Ks}  ( {\rm d}y^\Is +\mathcal{A}^\Is)  \wedge ( {\rm d}y^\Js +\mathcal{A}^\Js) \wedge ( {\rm d}y^\Ks +\mathcal{A}^\Ks) + \tfrac12  \mathcal{A}_{\Is\Js}  \wedge  ( {\rm d}y^\Is +\mathcal{A}^\Is )  \wedge ( {\rm d}y^\Js +\mathcal{A}^\Js )  + \mathcal{B}_\Is  \wedge  ( {\rm d}y^\Is +\mathcal{A}^\Is )  \; ,   \label{ThreeForm}
 \ee
such that $e^{2\varsigma} \tilde{g}_{\mu\nu} $ is the two-dimensional metric, $\mathcal{A}^\Is, \mathcal{A}_{\Is\Js} $ are one-forms and $\mathcal{B}_\Is$ is a two-form in two dimensions. 
The coordinates $y^\Is$ decompose into the eight coordinates $y^i$ on the space homological to $S^8$ and the circle coordinate $y^9$, with the internal metric splitting accordingly as
 \be
 G_{\Is\Js}{\rm d}y^\Is {\rm d}y^\Js = G_{ij} {\rm d}y^i {\rm d}y^j + \det \! G^{-1} \, ( {\rm d}y^9 + K_i {\rm d}y^i )^2\; .
 \ee
The external coordinates  $x^\mu$ can be fixed in the conformal gauge $\tilde{g}_{\mu\nu} = \eta_{\mu\nu}$ to $x^0 = t$ and $x^1 = z$. The components can be identified with the E$_9$ ExFT fields using the basis of generators in the $p=1$ spectral flowed basis \eqref{eq:bas1}. As explained in Section~\ref{sec:basdec}, we  write  $\widetilde{\T}^\alpha$ the generators in the $p=1$ basis and  ${\T}^\alpha$ the generators in the $p=2$ spectral flowed basis \eqref{eq:bas2} associated to SO(9) gauged supergravity. 
It is convenient to gauge-fix the additional Virasoro fields to zero to write the uplift ansatz. The exceptional field theory scalar fields then parametrise the $\mathrm{E}_9/ K(\mathrm{E}_9)$ coset representative
\be 
\mathcal{V}^{-1} = \dots  e^{-\frac16 \beta^{\Is\Js\Ks} \widetilde{\T}_{-2/3 \Is\Js\Ks}}   e^{-\frac16 \alpha_{\Is\Js\Ks} \widetilde{\T}_{-1/3}^{\Is\Js\Ks}} e^{K_i  \widetilde{\T}_{0}^{\; i}{}_0} \widetilde{\upsilon}^{-1} \rho^{\widetilde{\LL}_0} e^{\varsigma}\,, \label{Vtilde11D} 
\ee
where the components of $\widetilde{\upsilon}^{-1}\in $ GL(8) in the basis $\widetilde{\T}_0^{\; j}{}_i$ are the vielbeins for the metric $G_{ij}$.\footnote{For example, in the symmetric gauge $ \widetilde{\V}^{-1 \dagger}  = \exp( \tilde{h}_i{}^j \widetilde{\T}_0^{\; i}{}_j)  $, one would have $G_{ij} = \exp(\tilde{h})_i{}^k \delta_{kl}  \exp(\tilde{h})_j{}^l  $.}
The normalisations of the fields in this ansatz are determined in Appendix~\ref{SignConvention}. Note that $\rho$ is independent of the parabolic decomposition of the coset and is therefore the same in all bases.   
One identifies similarly the one-forms in the ansatz \eqref{ThreeForm} with the following components of the exceptional field theory gauge field $|\mathcal{A}\rangle$
 \be \mathcal{A}^\Is = \widetilde{\langle 0|}{}^{\Is} |\mathcal{A}\rangle \; , \quad   \mathcal{A}_{\Is\Js}  = \widetilde{\langle 1/3|}{}_{\Is\Js} |\mathcal{A}\rangle\; ,   \label{ExFTVinKK}
 \ee
while the two-form in~\eqref{ThreeForm} is the first component of the exceptional field theory two-form. Recall that the unconstrained ExFT two-form 
 belongs to the symmetric tensor product of two copies of $R(\Lambda_0)$ with the representation $R(2\Lambda_0)$ removed \cite{Bossard:2021jix}. We therefore use the notation $ |\hspace{-0.6mm}| \mathcal{C} \rangle\hspace{-1mm}\rangle$ to express that it belongs to the tensor product. This representation decomposes as
 \be 
 R(\Lambda_0) \textrm{\large{$\vee$}}  R(\Lambda_0) \ominus R(2\Lambda_0) \supset R(\Lambda_7) = {\bf 9}_{\frac{11}{9}} \oplus {\bf 126}_{\frac{14}{9}} \oplus \dots 
 \ee
where the first component ${\bf 9}_{11/9}$ comes from the eleven-dimensional supergravity 3-form, the second ${\bf 126}_{14/9}$ from the supergravity 6-form, etc. In components we have 
\be  \mathcal{B}_\Is  =2 \widetilde{\langle 0|}{}^\Js  \otimes   \widetilde{\langle 1/3|}{}_{\Is\Js}  |\hspace{-0.6mm}| \mathcal{C}^\prime \rangle\hspace{-1mm}\rangle   =2 \widetilde{\langle 0|}{}^\Js  \otimes   \widetilde{\langle 1/3|}{}_{\Is\Js}  |\hspace{-0.6mm}| \mathcal{C} \rangle\hspace{-1mm}\rangle  + \tfrac12  \widetilde{\langle 0|}{}^\Js  | \mathcal{A} \rangle \widetilde{\langle 1/3|}{}_{\Is\Js}  |\mathcal{A}\rangle \; , \label{ExFTTinKK} \ee  
where we have redefined for convenience the two-form $|\hspace{-0.6mm}| \mathcal{C}^\prime \rangle\hspace{-1mm}\rangle$ from 
\be |\hspace{-0.6mm}| \mathcal{C} \rangle\hspace{-1mm}\rangle = |\mathcal{C}_{(1}\rangle \otimes |\mathcal{C}_{2)} \rangle\,, \ee
in \cite{Bossard:2021jix}, see Appendix~\ref{SignConvention} for details. 
Following \cite{Hohm:2014qga}, one identifies the uplift ansatz  in terms of the relevant exceptional field theory matrix elements
\be \rho^{-1} \widetilde{\langle 0|}{}^\Is \mathcal{M}^{-1} {}^\Js\widetilde{|0\rangle} = e^{2\varsigma} \rho^{-\frac19} G^{\Is\Js}\; , \qquad \rho^{-1} \widetilde{\langle 1/3|}{}_{\Is\Js}  \mathcal{M}^{-1} {}^\Ks\widetilde{|0\rangle} = -e^{2\varsigma} \rho^{-\frac19} \alpha_{\Is\Js\Ls} G^{\Ls\Ks} \; .  \label{MatrixSugra} 
\ee  
In this equation and the ones below, we use the notation 
\be \langle h_1 |^{A_{h_1}} \mathcal{M}^{-1} {}^{B_{h_2}}| h_2\rangle  = \langle e_{h_1}^{A_{h_1}} | \mathcal{M}^{-1} | e_{h_2}^{B_{h_2}}\rangle  \,, \ee
for the matrix elements of the $E_9$ group element $\mathcal{M}^{-1} $ between the basis elements $\langle h_1 |^{A_{h_1}}$ and $\langle h_2 |^{B_{h_2}}$, as defined in \eqref{eq:bas1}.

Before exposing the computations, we shall display the result in terms of the gauged supergravity fields through the matrix components of the supergravity E$_9$ group element $M$,
\be \rho(x,y) =  (\det\! \mathring g)^{\frac12} \sugrarho(x) \; , \ee
while the other metric components are determined by the conditions 
\bea \rho^{-\frac{8}{9}}   e^{2\varsigma} G^{ij} \hspace{-2mm}&=&\hspace{-2mm} \coup^2 \sugrarho^{-\frac{16}{9}} (\det \mathring g)^{\frac{1}{9}}Y_{I} \mathring g^{ik} \partial_k Y_J Y_{K} \mathring g^{jl} \partial_l Y_L \langle 1/3|^{IJ} M^{-1} \, {}^{KL}|1/3\rangle  \; , \CR
\rho^{-\frac{8}{9}}   e^{2\varsigma} G^{ij} K_j \hspace{-2mm}&=&\hspace{-2mm} -\coup^2  \sugrarho^{-\frac{16}{9}} (\det \mathring g)^{\frac{1}{9}} Y_{I}  Y_J Y_{K} \mathring g^{ij} \partial_j Y_L \langle 4/3|^{IJ} M^{-1} \, {}^{KL}|1/3\rangle \; ,  \CR
\rho^{-\frac{8}{9}}   e^{2\varsigma} \bigl( \det\! G\,  + G^{ij}K_i  K_j  \bigr) \hspace{-2mm}&=&\hspace{-2mm}  \coup^2 \sugrarho^{-\frac{16}{9}} (\det \mathring g)^{\frac{1}{9}} Y_{I}  Y_J Y_{K} Y_L  \langle 4/3|^{IJ} M^{-1} \, {}^{KL}|4/3\rangle \; . \label{Ginverse} 
\eea 
These matrix elements without tilde are evaluated in the $p=2$ flowed basis of the module \eqref{eq:bas2} and the conversion between the two bases is given in  Appendix~\ref{sec:relmat}.

One can in particular obtain the expression of the external metric's conformal factor from the determinant\footnote{Where we write the 9 by 9 matrix  as $\begin{pmatrix} A^i{}_j\!\! &\!\!   B^i{}_9 \\ C^9{}_j \!\! &\! \!\!  D^9{}_9 \end{pmatrix} $.} 
\bea 
\hspace{-2mm} &&  \hspace{-2mm} (\rho^{-\frac{8}{9}}   e^{2\varsigma}/\coup^2 )^9 \\
\hspace{-2mm} &=&\hspace{-2mm} \sugrarho^{-16} \det\! \left(\begin{array}{cc}  Y_{I} \mathring g^{ik} \partial_k Y_J Y_{K} \partial_j Y_L \langle 1/3|^{IJ} M^{-1} \, {}^{KL}|1/3\rangle   & Y_{I} \mathring g^{ik} \partial_k Y_J Y_{K} Y_L \langle 1/3|^{IJ} M^{-1} \, {}^{KL}|4/3\rangle  \\
Y_{I}  Y_J Y_{K} \partial_j Y_L \langle 4/3|^{IJ} M^{-1} \, {}^{KL}|1/3\rangle  & Y_{I} Y_J Y_{K}  Y_L \langle 4/3|^{IJ} M^{-1} \, {}^{KL}|4/3\rangle \end{array}\right)\, .  \nonumber 
\eea
We will show below that one can rewrite the components of $\alpha_{\Is\Js\Ks}$ in terms of SL(9) tensors  $\upalpha_{IJ}(Y) $ and $\upalpha_{IJK}(Y)$ as follows
\be 
\alpha_{9ij} =  \partial_i Y^I \partial_j Y^J   \upalpha_{IJ}(Y) \; , \qquad \alpha_{ijk} = \partial_i Y^I \partial_j Y^J  \partial_k Y^K  \upalpha_{IJK}(Y)\; . \label{AnsatzAlpha} 
\ee
These only depend on the sphere coordinates through the harmonic variables $Y_I$
and are determined by   
\bea \upalpha_{IJ}(Y) Y_P  \partial_i Y^I Y_K  \langle 1/3|^{PJ} M^{-1}\, {}^{KL}|1/3\rangle &=&  \partial_i Y^I  Y_K   \langle 0|_I M^{-1}\, {}^{KL}|1/3\rangle \; , \label{alphaIJ} \CR
\upalpha_{IJ}(Y) Y_P  \partial_i Y^I Y_K Y_L \langle 1/3|^{PJ} M^{-1}\, {}^{KL}|4/3\rangle &=&  \partial_i Y^I  Y_K Y_L  \langle 0|_I M^{-1}\, {}^{KL}|4/3\rangle \; ,\eea
and
\bea \hspace{-2mm}&& \hspace{-2mm}   \upalpha_{IJQ}(Y)  \partial_i Y^I  \partial_j Y^J Y_P Y_K   \langle 1/3|^{PQ} M^{-1}\, {}^{KL}|1/3\rangle  \label{alphaIJK} \\
\hspace{-2mm} &=&\hspace{-2mm}   -\partial_i Y^I  \partial_j Y^J Y_P Y_K   \langle 1|^{P}_{IJ} M^{-1}\, {}^{KL}|1/3\rangle-\upalpha_{IJ}(Y) Y_Q  \partial_i Y^I  \partial_j Y^J Y_P Y_K   \langle 4/3|^{PQ} M^{-1} \,{}^{KL}|1/3\rangle    \; ,  \CR[2mm]
\hspace{-2mm} &&\hspace{-2mm}    \upalpha_{IJQ}(Y)  \partial_i Y^I  \partial_j Y^J Y_P Y_K  Y_L  \langle 1/3|^{PQ} M^{-1}\, {}^{KL}|4/3\rangle \CR
\hspace{-2mm}&=&\hspace{-2mm}   -\partial_i Y^I  \partial_j Y^J Y_P Y_K Y_L  \langle 1|^{P}_{IJ} M^{-1}\, {}^{KL}|4/3\rangle-\upalpha_{IJ}(Y) Y_Q  \partial_i Y^I  \partial_j Y^J Y_P Y_K Y_L  \langle 4/3|^{PQ} M^{-1} \,{}^{KL}|4/3\rangle    \; .  \nonumber \eea
Relevant matrix elements for the two-dimensional scalar fields can be determined from the expressions given in Appendix~\ref{sec:relmat} and take the form
\begin{subequations}
\begin{align}
\langle 0|_I M^{-1} {}^{KL}|1/3\rangle &=  e^{2\sigma}  \sugrarho^{\frac{8}{9}}  a^{KLJ} \M_{IJ} \; , \\
  \langle 1/3|^{IJ} M^{-1} {}^{KL}|1/3\rangle  &= e^{2\sigma} \sugrarho^{\frac{14}{9}} \bigl(2\M^{K[I} \M^{J]L} + \sugrarho^{-2/3} \M_{PQ} a^{IJP} a^{KLQ} \bigr)\; , \\
 \langle 1|_{IJ}^P M^{-1} {}^{KL}|1/3\rangle  &=  e^{2\sigma} \sugrarho^{\frac{14}{9}} \Bigl( \delta^P_R   b_{SIJ} +  \delta^P{}_{\hspace{-2.6mm}[I} b_{J]RS}\Bigr) \bigl( 2 \M^{R[K} \M^{L]S} +\sugrarho^{-\frac23} \M_{TU} a^{RST} a^{KLU}\bigr)\CR
&\quad +2 e^{2\sigma}  \sugrarho^{\frac{8}{9}} \Bigl( \delta^P{}_{\hspace{-2.6mm}[I} h^R{}_{J]} a^{KLQ} \M_{RQ} - h^P{}_{[I} \M_{J]Q} a^{KLQ}  \Bigr)  \\*
&\quad - \frac{e^{2\sigma}}{48}  \sugrarho^{\frac{14}{9}}  \varepsilon_{IJQRSTUVW} a^{PQR} a^{STU}  \bigl( \M^{V[K} \M^{L]W} {+}\tfrac13 \sugrarho^{-\frac23} \M_{XY} a^{VWX} a^{KLY}\bigr)\; ,  \nonumber 
\end{align}
and
\begin{align} 
\langle 4/3|^{IJ} M^{-1}\,  {}^{KL}|1/3\rangle &=  e^{2\sigma} \sugrarho^{\frac{14}{9}} \bigl(2\M^{P[K} \M^{L]Q} + \sugrarho^{-2/3} \M_{RS} a^{PQR} a^{KLS} \bigr) \CR
& \hspace{20mm} \times \left( \delta_Q^{(I} h^{J)}_{\phantom{Q}}{}_P^{\phantom{I)}} - \frac1{288} \varepsilon_{PQT_1\dots T_7} a^{T_1T_2(I} a^{J)T_3T_4} a^{T_5T_6T_7} \right) \CR
&\quad +\frac{1}{384} e^{2\sigma} \sugrarho^{\frac{8}{9}} \varepsilon_{P_1\dots P_9} a^{P_1P_2(I} a^{J)P_3P_4} a^{P_5P_6P_7} a^{P_8P_9Q} a^{KLR} \M_{QR}\; .  
\end{align}
\end{subequations}
One also needs other components such as $ \langle 4/3|^{IJ} M^{-1} {}^{KL}|4/3\rangle$, which can straightforwardly be computed from \eqref{Vfourthird}.

The one-forms \eqref{ExFTVinKK} are determined similarly in terms of the gauged supergravity one-form~$|A\rangle$
\begin{subequations}
\begin{alignat}{3}
  \mathcal{A}^i &=
  \widetilde{\langle 0|}{}^i|\mathcal{A}\rangle &&=\,\, \coup\,  Y_{I} \mathring g^{ij} \partial_jY_{J} \langle 1/3|^{IJ} |A\rangle \; , \label{eq:KKveci}\\
 \mathcal{A}^9 &= \widetilde{\langle 0|}{}^9 |\mathcal{A}\rangle &&=\,\,\coup\,  Y_{I} Y_J \langle 4/3|^{IJ} |A\rangle  \; , \label{eq:KKvec9}\\
\mathcal{A}_{9i} &=  \widetilde{\langle 1/3|}{}_{9i} |\mathcal{A}\rangle   &&=\,\, -  \coup\,   \partial_i Y^I  \langle 0|_{I} |A\rangle \; , \\
\mathcal{A}_{ij} &=   \widetilde{\langle 1/3|}{}_{ij}  |\mathcal{A}\rangle &&=\,\,  \coup\,  Y_K \partial_i Y^I \partial_j Y^J     \langle 1|^K_{IJ} |A\rangle \; . 
\end{alignat}
\end{subequations}
As usual the SO(9) Yang--Mills fields $A^{IJ} = \langle 1/3|^{IJ} |A\rangle$ appear in the Kaluza--Klein one-forms contracted with the sphere Killing vectors. The other components of $|A\rangle$ do not appear in the gauged supergravity Lagrangian, they are determined by the first order equation \eqref{FieldStrengthDuality}. 
The two-forms \eqref{ExFTTinKK} are given in terms of the gauged supergravity two-form $  |\hspace{-0.5mm}| {C}\rangle\hspace{-1mm}\rangle$ as
\bea    
\mathcal{B}_9=2 \widetilde{\langle 0|}{}^\Js  \otimes   \widetilde{\langle 1/3|}{}_{9\Js}  |\hspace{-0.5mm}| \mathcal{C}^\prime \rangle\hspace{-1mm}\rangle \hspace{-2mm} &=&\hspace{-2mm}  -    \coup^2 Y_I  \, {\langle 1/3|}{}^{IJ}  \otimes   {\langle 0|}{}_{J}  \bigl(  |\hspace{-0.5mm}| {C}\rangle\hspace{-1mm}\rangle + 2 |A\rangle \wedge |A\rangle \bigr)  \label{TwoFormKKansatz} \\*[2mm]
 \mathcal{B}_i= 2 \widetilde{\langle 0|}{}^\Js  \otimes   \widetilde{\langle 1/3|}{}_{i\Js}  |\hspace{-0.5mm}| \mathcal{C}^\prime \rangle\hspace{-1mm}\rangle \hspace{-2mm} &=& \hspace{-2mm}  -  \coup^2 Y_I Y_J \partial_i Y^{K}    \bigl(  {\langle 4/3|}{}^{IJ}  \otimes   {\langle 0|}{}_{K}  +{\langle 1/3|}{}^{LI}  \otimes   {\langle 1|}{}^J_{KL} \bigr)   |\hspace{-0.5mm}| {C}\rangle\hspace{-1mm}\rangle\; .   \nonumber  
 \eea    
 
Note that the Kaluza--Klein ans\"atze for the eleven-dimensional three-form and metric depend on the fields $b_{IJK}$ and $\delta_{K(I} h^K{}_{J)}$ that do not appear in the  gauged supergravity Lagrangian and moreover turn out to be pure gauge. Consistently, we show in Appendix~\ref{sec:gaugeinv} that $b_{IJK}$ can be eliminated in eleven-dimensional supergravity using a three-form gauge transformation of two-form parameter 
\be \Lambda^{\scalebox{0.6}{11D}}(x,y) =   \frac \coup 2 \lambda(x)^K_{IJ} Y_K \partial_i Y^I \partial_j Y^J ( {\rm d}y^i + \mathcal{A}^i  ) \wedge  ( {\rm d}y^j + \mathcal{A}^j   ) \; , \ee
while $\delta_{K(I} h^K{}_{J)}$ can be gauged away using the diffeomorphism 
\be y^9 =  y^{9 \prime} - \coup Y_I Y_J \xi^{IJ}(x) \; . \ee 
One may therefore set them to zero.

\subsection{Derivation of the uplift formulas}
To derive the metric ansatz one uses the relations between the two spectral flowed basis \eqref{p1p2} and substitutes the generalised Scherk--Schwarz ansatz \eqref{eq:twistans} to get 
\begin{subequations}
\begin{align} 
\langle 1/3|^{0i} \mathcal{M}^{-1} \, {}^{0j}|1/3\rangle &= e^{2\twistsigma} \twistrho^{\frac{14}{9}} \mathsf{u}^{-1 0}{}_{I} \mathsf{u}^{-1 i}{}_{J} \mathsf{u}^{-1 0}{}_{K} \mathsf{u}^{-1 j}{}_{L} \langle 1/3|^{IJ} M^{-1} {}^{KL}|1/3\rangle \; , \\
 \langle 4/3|^{00} \mathcal{M}^{-1} \, {}^{0i}|1/3\rangle &= e^{2\twistsigma} \twistrho^{\frac{23}{9}} \mathsf{u}^{-1 0}{}_{I} \mathsf{u}^{-1 0}{}_{J} \mathsf{u}^{-1 0}{}_{K} \mathsf{u}^{-1 i}{}_{L} \langle 4/3|^{IJ} M^{-1} {}^{KL}|1/3\rangle \; , \\
 \langle 4/3|^{00} \mathcal{M}^{-1}\,  {}^{00}|4/3\rangle &=  e^{2\twistsigma} \twistrho^{\frac{32}{9}} \mathsf{u}^{-1 0}{}_{I} \mathsf{u}^{-1 0}{}_{J} \mathsf{u}^{-1 0}{}_{K} \mathsf{u}^{-1 0}{}_{L} \langle 4/3|^{IJ} M^{-1} {}^{KL}|4/3\rangle\; . 
\end{align}
\end{subequations}
 Comparison with \eqref{MatrixSugra} and using the explicit form  \eqref{gSSuY} of the twist matrix $(\mathsf{u}^I{}_0,\mathsf{u}^I{}_j)$ gives immediately \eqref{Ginverse}. To understand the dependence in the embedding coordinates $Y^I$ it is useful to combine these matrix elements into the nine by nine matrix 
 \begin{multline} 
\widetilde{G}^{IJ} =  Y_K Y_L  \langle 1/3|^{KI} M^{-1} \, {}^{LJ}|1/3\rangle - Y^I Y_P Y_Q Y_L \langle 4/3|^{PQ} M^{-1} \, {}^{LJ}|1/3\rangle \\
- Y^J Y_P Y_Q Y_K \langle 1/3|^{KI} M^{-1} \, {}^{PQ}|4/3\rangle  + Y^I Y^J Y_K Y_L Y_P Y_Q  \langle 4/3|^{PQ} M^{-1} \, {}^{KL}|4/3\rangle \,, \end{multline}
that satisfies 
\bea && \left(\begin{array}{cc}  Y_{I} \mathring g^{ik} \partial_k Y_J Y_{K} \partial_j Y_L \langle 1/3|^{IJ} M^{-1} \, {}^{KL}|1/3\rangle   & -Y_{I} \mathring g^{ik} \partial_k Y_J Y_{K} Y_L \langle 1/3|^{IJ} M^{-1} \, {}^{KL}|4/3\rangle  \\
-Y_{I}  Y_J Y_{K} \partial_j Y_L \langle 4/3|^{IJ} M^{-1} \, {}^{KL}|1/3\rangle  & Y_{I} Y_J Y_{K}  Y_L \langle 4/3|^{IJ} M^{-1} \, {}^{KL}|4/3\rangle \end{array}\right)\CR
&=& \left(\begin{array}{c}  \mathring g^{ik} \partial_k Y_I  \\ Y_I  \end{array}\right) \widetilde{G}^{IJ}  \left(\begin{array}{c}   \partial_j Y_J  , Y_J  \end{array}\right) \; .  \eea
In this form it is manifest that $\widetilde{G}^{IJ}$ only depends on the sphere coordinates through the embedding coordinates $Y^I$, and therefore admits an expansion in spherical harmonics. The expression of  $  e^{2\varsigma}$ takes the form 
\be   e^{2\varsigma} = \coup^2 \det\!\mathring g^\frac{4}{9}  \sugrarho^{-\frac{8}{9}} \det \!\widetilde{G}^{-\frac19} \; ,\ee
and the metric  $G_{\tilde I\tilde J}$
\be  \left(\begin{array}{cc}  G_{ij} & G_{i9} \\ G_{9j} & G_{99} \end{array}\right) = \det \! \mathring g^{-\frac19} \det\!\widetilde{G}^{- \frac19} \left(\begin{array}{c}   \partial_i Y^I  \\ Y^I  \end{array}\right) \widetilde{G}_{IJ}  \left(\begin{array}{c}   \partial_j Y^J  , Y^J  \end{array}\right)\; . \ee
It is useful to introduce this inverse matrix  $\widetilde{G}_{IJ}$  to exhibit the dependence of the uplift ansatz in the embedding coordinates and their derivatives, but it may not be the best way to obtain the explicit uplift for a given solution. We will also use it to prove that the three-form scalar components  satisfy   \eqref{alphaIJ} and \eqref{alphaIJK}.

For the three-form, \eqref{eq:twistans}, \eqref{MatrixSugra} and \eqref{p1p2}  give us
\begin{subequations}
\begin{align}
\langle 0 |_i \mathcal{M}^{-1} {}^{0j} |1/3\rangle &=   e^{2\twistsigma} \twistrho^{\frac{11}{9}} \mathsf{u}^K{}_i \mathsf{u}^{-1 0}{}_I \mathsf{u}^{-1 j}{}_J \langle 0|_K M^{-1} {}^{IJ}|1/3\rangle \; , \\
 \langle 0 |_i \mathcal{M}^{-1} {}^{00} |4/3\rangle &=   e^{2\twistsigma} \twistrho^{\frac{20}{9}} \mathsf{u}^K{}_i \mathsf{u}^{-1 0}{}_I \mathsf{u}^{-1 0}{}_J \langle 0|_K M^{-1} {}^{IJ}|4/3\rangle \; , \\
  \langle 1 |^0_{ij} \mathcal{M}^{-1} {}^{0k} |1/3\rangle  &= e^{2\twistsigma} \twistrho^{\frac{20}{9}} \mathsf{u}^I{}_i \mathsf{u}^J{}_j \mathsf{u}^{-1 0}{}_P \mathsf{u}^{-1 0}{}_K \mathsf{u}^{-1 k}{}_L \langle 1|_{IJ}^P M^{-1} {}^{KL}|1/3\rangle\; , \\
  \langle 1 |^0_{ij} \mathcal{M}^{-1} {}^{00} |4/3\rangle  &= e^{2\twistsigma} \twistrho^{\frac{29}{9}} \mathsf{u}^I{}_i \mathsf{u}^J{}_j \mathsf{u}^{-1 0}{}_P \mathsf{u}^{-1 0}{}_K \mathsf{u}^{-1 0}{}_L \langle 1|_{IJ}^P M^{-1} {}^{KL}|4/3\rangle  \; . 
\end{align}
\end{subequations}
One combines these equations into the matrix equation 
\bea \hspace{-2mm}&&\hspace{-2mm} -e^{2\varsigma} \rho^{\frac89} \left(\begin{array}{cc} \alpha_{ij\Ls} G^{\Ls k} & \alpha_{ij\Ls} G^{\Ls 9} \\ \alpha_{9 j\Ls} G^{\Ls k} & \alpha_{9 j\Ls} G^{\Ls 9} \end{array}\right)\label{Galpha=M} \\
\hspace{-2mm}&=&\hspace{-2mm}\coup^2 \det\! \mathring{g}   \left(\begin{array}{c} \partial_i Y^I \partial_j Y^J Y_P Y_Q \bigl( \langle 1|_{IJ}^P M^{-1} {}^{QL}|1/3\rangle + Y^L Y_R \langle 1|_{IJ}^P M^{-1} {}^{QR}|4/3\rangle\bigr) \\
- \partial_j Y^J Y_P  \bigl( \langle 0|_{J} M^{-1} {}^{PL}|1/3\rangle + Y^L Y_Q \langle 0|_{J} M^{-1} {}^{PQ}|4/3\rangle\bigr)  \end{array}\right) \left( \begin{array}{cc} \mathring g^{kl}\partial_l Y_L, Y_L \end{array}\right) \,,\nonumber  \eea
such that one can give the solution for $\alpha_{ijk}$ and $\alpha_{9ij}$ in terms of the inverse matrix $\widetilde{G}_{IJ}$ as
\bea \hspace{-2mm}&&\hspace{-2mm} \left(\begin{array}{cc} \alpha_{ijk}  & \alpha_{9ij}  \\ \alpha_{9jk}  & 0 \end{array}\right) \label{alphaGMM}  \\
\hspace{-2mm}&=&\hspace{-2mm}  -\left(\begin{array}{c} \partial_i Y^I \partial_j Y^J Y_P Y_Q \bigl( \langle 1|_{IJ}^P M^{-1} {}^{QL}|1/3\rangle + Y^L Y_R \langle 1|_{IJ}^P M^{-1} {}^{QR}|4/3\rangle\bigr) \\
- \partial_j Y^J Y_P  \bigl( \langle 0|_{J} M^{-1} {}^{PL}|1/3\rangle + Y^L Y_Q \langle 0|_{J} M^{-1} {}^{PQ}|4/3\rangle\bigr)  \end{array}\right) \widetilde{G}_{LK} \left( \begin{array}{cc} \partial_k Y^K, Y^K \end{array}\right) \nonumber\; .  \eea
Computing the inverse matrix $\widetilde{G}_{IJ}$ is straightforward, but would be rather cumbersome, and might not be the easiest way to get the explicit uplift ansatz. We will rather use this formula to prove equation \eqref{alphaIJ} and \eqref{alphaIJK} that are a priori easier to use in practice. Note first that \eqref{alphaGMM} is compatible with  the ansatz \eqref{AnsatzAlpha}. Putting back this ansatz in \eqref{MatrixSugra} one obtains 
\be  \alpha_{\Is\Js\Ls}  \widetilde{\langle 0|}{}^\Ls \mathcal{M}^{-1} {}^\Ks\widetilde{|0\rangle}= -\widetilde{\langle 1/3|}{}_{\Is\Js}  \mathcal{M}^{-1} {}^\Ks\widetilde{|0\rangle} \; . \label{RewriteAlpha} \ee
We note that the right ket on both sides of this equation reads 
\be \mathcal{U}^{-1}  {}^\Ks\widetilde{|0\rangle} =\bigl(  \mathcal{U}^{-1}  {}^k\widetilde{|0\rangle} \; , \;    \mathcal{U}^{-1}  {}^9\widetilde{|0\rangle}\bigr) =\coup \det\! \mathring g^{\frac12}  \bigl(  \mathring g^{kl} \partial_l Y_L  Y_Q\;  {}^{QL}{}|1/3\rangle  \; , \;   Y_Q  Y_S\;  ^{QS}{}|4/3\rangle  \bigr)\,.  \ee
Using this expression one rewrites \eqref{RewriteAlpha} as\footnote{This equation should be read such that$\biggl( \begin{array}{c} \! \! \langle \psi_1 |  \!\!  \\ \!\!  \langle \psi_2 |  \! \! \end{array} \biggr) \otimes \bigl( |\psi_3\rangle , |\psi_4\rangle \bigr) = \biggl( \begin{array}{cc} \langle \psi_1 | \psi_3\rangle & \langle \psi_1 | \psi_4\rangle \\ \langle \psi_2 | \psi_3\rangle & \langle \psi_2 | \psi_4\rangle \end{array}\biggr)$.Note that one cannot eliminate the multiplication by the right vector on both sides of the equation because of the Hilbert space scalar product.}
\bea 
&& \hspace{-4mm}  \left( \begin{array}{cc} \hspace{-2mm}  \partial_i Y^I \partial_j Y^J \partial_l Y^L  \upalpha_{IJL} \! &\!  \partial_i Y^I \partial_j Y^J \upalpha_{IJ}    \hspace{-2mm} \\[1mm]\!  \partial_j Y^J \partial_l Y^L \upalpha_{JL} & 0  \! \end{array} \right)
\left( \begin{array}{c} \hspace{-2mm} \mathring g^{lp} \partial_p Y_R  Y_P \langle \tfrac13|^{PR} M^{-1} \hspace{-2mm} \\[1mm] \!  Y_P Y_R \langle \tfrac43|^{PR} M^{-1} \!  \end{array} \right)  
\otimes
\Bigl( {}^{QK}{}|\tfrac13\rangle Y_Q ,\,\, {}^{QS}|\tfrac43\rangle Y_Q  Y_S \Bigr)  \nn\\[2mm]
&=& \hspace{-4mm}  \left( \begin{array}{c} \partial_i Y^I \partial_j Y^J  \Bigl( \upalpha_{IJL}  Y_P \langle \tfrac13|^{PL} M^{-1}  + \upalpha_{IJ} Y_L Y_P \langle \tfrac43|^{PL} M^{-1} \Bigr)   \\[1mm] \partial_j Y^J \upalpha_{JL} Y_P \langle \tfrac13|^{PL} M^{-1}  \end{array} \right) 
\otimes
\Bigl( {}^{QK}{}|\tfrac13\rangle Y_Q ,\,\, {}^{QS}|\tfrac43\rangle Y_Q  Y_S \Bigr)  \nn\\[2mm]
&=&\hspace{-4mm} \left( \begin{array}{c}  - \partial_i Y^I \partial_j Y^J Y_L \langle 1|^{L}_{IJ}  M^{-1}   \\[1mm] \partial_j Y^J  \langle 0|_J M^{-1}  \end{array} \right) 
\otimes
\Bigl( {}^{QK}{}|\tfrac13\rangle Y_Q , \,\,{}^{QS}|\tfrac43\rangle Y_Q  Y_S \Bigr) \label{alpha=Mfinal} \; ,
\eea
where one used the identity
 \be 
 \partial_j Y^{[J} \partial_l Y^{L]} \mathring g^{lk} \partial_k Y_K =  \partial_j Y^{[J}_{\phantom{K}} \delta_K^{L]} \; , 
 \ee
to factor out the tensor product with the vector 
\be\Bigl( {}^{QK}{}|1/3\rangle Y_Q , ^{QS}{}|4/3\rangle Y_Q  Y_S \Bigr)\,. \label{vectfactor}  
\ee 
Recombining these equations into the four components of a matrix gives precisely \eqref{alphaIJ} and \eqref{alphaIJK}. These equations are equivalent to \eqref{alphaGMM}, and therefore completely determine $\alpha_{ijk}$ and $\alpha_{9ij}$.

For the one-forms one computes from \eqref{p1p2} and \eqref{eq:gssans} that 
\bea  \widetilde{\langle 0|}{}^i |\mathcal{A}\rangle &=& \twistrho^{-\frac2 9} e^{\twistsigma} \mathsf{u}^{-1 0}{}_I\mathsf{u}^{-1 i}{}_J \langle 1/3|^{IJ} |A\rangle= \coup\,  Y_{I} \mathring g^{ij} \partial_jY_{J} \langle 1/3|^{IJ} |A\rangle \,,\CR
 \widetilde{\langle 0|}{}^9 |\mathcal{A}\rangle &=& \twistrho^{\frac7 9} e^{\twistsigma} \mathsf{u}^{-1 0}{}_I\mathsf{u}^{-1 0}{}_J \langle 4/3|^{IJ} |A\rangle = \coup\,  Y_{I} Y_J \langle 4/3|^{IJ} |A\rangle  \,,\CR
  \widetilde{\langle 1/3|}{}_{9i} |\mathcal{A}\rangle &=&- \twistrho^{-\frac5 9} e^{\twistsigma} \mathsf{u}^{I}{}_i \langle 0|_{I} |A\rangle  = -  \coup\,   \partial_i Y^I  \langle 0|_{I} |A\rangle \,,\CR
   \widetilde{\langle 1/3|}{}_{ij}  |\mathcal{A}\rangle &=& \twistrho^{\frac4 9} e^{\twistsigma} \mathsf{u}^{-1 0}{}_K  \mathsf{u}^I{}_i  \mathsf{u}^J{}_j  \langle 1|^K_{IJ} |A\rangle = \coup\,  Y_K \partial_i Y^I \partial_j Y^J     \langle 1|^K_{IJ} |A\rangle \; . 
   \eea
Note that this does not involve the degree $2/3$ component of the gauge supergravity one-form that contribute instead to the six-form potential in eleven-dimensional supergravity, with 
\bea \frac1{24} \varepsilon_{ijklpqrs} \widetilde{\langle 2/3|}{}^{pqrs} |\mathcal{A}\rangle  &=& -\twistrho^{\frac1 9} e^{\twistsigma}   \mathsf{u}^I{}_i  \mathsf{u}^J{}_j \mathsf{u}^K{}_k \mathsf{u}^L{}_l  {\langle 2/3|}{}_{IJKL} |A\rangle \CR
&=& - \coup\,  \partial_i Y^I \partial_j Y^J \partial_k Y^K \partial_l Y^L  {\langle 2/3|}{}_{IJKL} |A\rangle\; .    \eea
For the two-form one computes that 
\bea   \widetilde{\langle 0|}{}^\Js  \otimes   \widetilde{\langle 1/3|}{}_{9\Js}  |\hspace{-0.5mm}| \mathcal{C}^\prime \rangle\hspace{-1mm}\rangle \hspace{-2mm} &=&\hspace{-2mm}  -   \frac12  \twistrho^{-\frac7 9} e^{2\twistsigma} \mathsf{u}^{-10}{}_I \ {\langle 1/3|}{}^{IJ}  \otimes   {\langle 0|}{}_{J}  \bigl(  |\hspace{-0.5mm}| {C}\rangle\hspace{-1mm}\rangle + 2 |A\rangle \wedge |A\rangle \bigr)\,,  \\
 \widetilde{\langle 0|}{}^\Js  \otimes   \widetilde{\langle 1/3|}{}_{i\Js}  |\hspace{-0.5mm}| \mathcal{C}^\prime \rangle\hspace{-1mm}\rangle \hspace{-2mm} &=& \hspace{-2mm}  - \frac12 \twistrho^{\frac2 9} e^{2\twistsigma} \mathsf{u}^{-10}{}_I  \mathsf{u}^{-10}{}_J \mathsf{u}^{K}{}_i     \bigl(  {\langle 4/3|}{}^{IJ}  \otimes   {\langle 0|}{}_{K}  +{\langle 1/3|}{}^{LI}  \otimes   {\langle 1|}{}^J_{KL} \bigr)   |\hspace{-0.5mm}| {C}\rangle\hspace{-1mm}\rangle\,,  \nonumber  \eea
which gives \eqref{TwoFormKKansatz}.

\subsection{\texorpdfstring{Truncation to the $\mathrm{SO}(3)\times \mathrm{SO}(6)$ invariant sector}{Truncation to the SO(3)xSO(6) invariant sector}}

The solutions of SO(9) gauged supergravity  are expected to be relevant to the study of the holographic dual of the D0-brane matrix quantum mechanics  \cite{Banks:1996vh}. It is natural to wonder if its massive supersymmetric  deformation known as the BMN matrix model \cite{Berenstein:2002jq} can also be analysed in gauged supergravity. The latter  deformation breaks SO(9) to  SO(3)$\times$SO(6). There is a large set of vacua in the BMN matrix model that are holographically dual to one-half BPS solutions in eleven-dimensional supergravity with a non-vanishing four-form field strength \cite{Lin:2004nb,Lin:2005nh}. These solutions are generally too complicated to be uplifts of solutions of SO(9) gauged supergravity, because they involve arbitrary combinations of the SO(3)$\times$SO(6) invariant harmonics on $S^8$. Moreover, one can check from the supersymmetry transformations given in~\cite{Ortiz:2012ib} that the one-half BPS  solutions within the SO(3)$\times$SO(6) invariant truncation of SO(9) gauged supergravity necessarily have a vanishing axion, and therefore uplift in eleven dimensions to solutions with vanishing four-form field strength.\footnote{In principle the eleven-dimensional solution could involve sixteen Killing spinors that are not all contained within the truncation to $D=2$ supergravity.} 
It is therefore very unlikely that SO(9) gauged supergravity reproduces any BMN vacuum solutions. Nevertheless, we will argue that this  truncation is relevant for describing the BMN matrix models at finite temperature.

The general SO(3)$\times$SO(6) invariant ansatz for the fields of SO(9) gauged supergravity can be written as follows 
\be \M = e^{-2\phi} \mathds{1}_3  \oplus  e^{\phi} \mathds{1}_6 \; \Rightarrow \; \M^{ab} = e^{2\phi} \delta^{ab} \; , \quad \M^{\hat{a}\hat{b}} = e^{-\phi} \delta^{\hat{a}\hat{b}}  \; , \qquad a^{abc} =  \varepsilon^{abc} a \,, \ee
where $a=1,2,3$ and $\hat{a}=4$ to $9$. We set to zero the pure gauge fields $b_{abc}$, $h^a{}_b$ and $h^{\hat{a}}{}_{\hat{b}}$. One parametrises  the $S^8$ embedding coordinates $Y^I$ as 
\be Y^a = \y\,  Y_2^a\; , \quad Y^{\hat{a}} = \sqrt{1-\y^2} Y_5^{\hat{a}} \; , \ee
in terms of the $S^2$ and $S^5$ embedding coordinates $Y_2^a$ and $Y_5^{\hat{a}}$ and  $\zeta\in[0,1]$. One defines accordingly the round metrics 
\be \mathring g_{2\, \alpha\beta} = \delta_{ab} \partial_\alpha Y_2^a \partial_\beta  Y_2^b\; , \qquad \mathring g_{5\, \hat{\alpha}\hat{\beta}} = \delta_{\hat{a}\hat{b}} \partial_{\hat{\alpha}} Y_5^{\hat{a}} \partial_{\hat{\beta}}  Y_5^{\hat{b}}\; . \ee

Within this truncation, the internal metric is determined by the matrix elements
\bea \langle 1/3|^{ab} M^{-1} {}^{cd}|1/3 \rangle &=& e^{2\sigma} \varrho^\frac{14}{9} \bigl( e^{4\phi} + \varrho^{-\frac23} e^{-2\phi} a^2) 2 \delta^{c[a} \delta^{b]d} \; , \CR
\langle  1/3|^{a\hat{b}} M^{-1} {}^{c\hat{d}}|1/3 \rangle &=& e^{2\sigma} \varrho^\frac{14}{9}  e^{\phi} \delta^{ac} \delta^{\hat{b}\hat{d}} \; , \CR
\langle  1/3|^{\hat{a}\hat{b}} M^{-1} {}^{\hat{c}\hat{d}}|1/3 \rangle &=& e^{2\sigma} \varrho^\frac{14}{9}  e^{-2\phi}  2 \delta^{\hat{c}[\hat{a}} \delta^{\hat{b}]\hat{d}} \; , \CR
 \langle 4/3|^{ab} M^{-1} {}^{cd}|4/3 \rangle &=& e^{2\sigma} \varrho^\frac{32}{9} \bigl( e^{4\phi} + \varrho^{-\frac23} e^{-2\phi} a^2)  \delta^{c(a} \delta^{b)d} \; , \CR
\langle  4/3|^{a\hat{b}} M^{-1} {}^{c\hat{d}}|4/3 \rangle &=& \frac12 e^{2\sigma} \varrho^\frac{32}{9}  \bigl( e^{\phi}  + \tfrac12  \varrho^{-\frac23} e^{-5\phi} a^2 \bigr) \delta^{ac} \delta^{\hat{b}\hat{d}} \; , \CR
\langle  4/3|^{\hat{a}\hat{b}} M^{-1} {}^{\hat{c}\hat{d}}|4/3 \rangle &=& e^{2\sigma} \varrho^\frac{32}{9}  e^{-2\phi}   \delta^{\hat{c}(\hat{a}} \delta^{\hat{b})\hat{d}} \; , 
\eea
while the other components appearing in \eqref{Ginverse} vanish. One obtains the inverse matrix $G^{-1}$ as a block diagonal matrix 
\begin{multline} e^{2\varsigma} \rho^{\frac89} G^{\tilde{I}\tilde{J}}\partial_{\tilde{I}} \partial_{\tilde{J}} = \coup^2 \y^4 (1-\y^2)^4 \det\! \mathring g_2  \det\! \mathring g_5 e^{2\sigma} \sugrarho^{\frac{14}{9}} \Bigl( \frac{e^{2\phi}\Delta (1+f) }{\y^2} \mathring g_2^{\alpha\beta} \partial_\alpha \partial_\beta +  \frac{e^{-\phi}\Delta }{1-\y^2}  \mathring g_5^{\hat{\alpha}\hat{\beta}} \partial_{\hat{\alpha}} \partial_{\hat{\beta}} \\* +  e^\phi (1-\y^2)\partial_\zeta^{\; 2}  +  \sugrarho^2 \Delta^2 (1+f) \partial_{\psi}^{\; 2} \Bigr)\,, \end{multline}
where 
\be \Delta = \y^2 e^{2\phi} + (1-\y^2) e^{-\phi} >0 \; , \qquad f = \frac{\y^2 e^{-4\phi} \sugrarho^{-\frac23} a^2}{\Delta}\ge 0 \; .  \ee
We write $\psi = y^9$ the coordinate of the M-theory fibre, and $t,z$ the two-dimensional coordinates. 
Altogether, we obtain the eleven-dimensional metric is \footnote{Here $\dd\mathring{s}^2_{2} = \mathring{g}_{\alpha\beta} \dd y^\alpha \dd y^\beta $ denotes the sphere round metric and $\dd\mathring{s}^2_{5}$ the round metric on $S^5$.}
\begin{multline} \dd s^2_{\scalebox{0.6}{11D}}  = \coup^2 e^{2\sigma} \Delta ( 1+ f)^{\frac13}  \bigl( -\dd t^2+\dd z^2\bigr) + \sugrarho^{\frac49} (1+f)^{\frac13} \Bigl( e^{-\phi} \Delta \frac{\dd\y^2}{1-\y^2} + \frac{e^{-2\phi}}{(1+f)} \y^2 \dd\mathring{s}^2_{2} + e^{\phi}(1- \y^2) \dd\mathring{s}^2_{5} \Bigr)  \\
 + \sugrarho^{-\frac{14}{9}} \frac{1}{ \Delta (1+f)^{\frac23}} \left( {\rm d} \psi   + \y^2 {\omega}_3 +  (1-\y^2) {\omega}_6 \right)^2 \,, \end{multline}
where the  Kaluza--Klein one-form \eqref{eq:KKvec9} is given in terms of the two gauge supergravity one-forms
\be \omega_3\, \delta^{ab} =  \coup \langle 4/3|^{ab} |A\rangle \; , \qquad \omega_6 \,\delta^{\hat{a}\hat{b}} =  \coup \langle 4/3|^{\hat{a}\hat{b}} |A\rangle \; ,
\ee
that satisfy 
\bea \dd {\omega}_3 &=& - \coup^2 e^{2\sigma} \sugrarho^{\frac{5}{9}} \bigl( e^{4\phi} + 6 e^{\phi} + e^{-2\phi} \sugrarho^{-\frac23} a^2 \bigr) \dd t \wedge \dd z\; , \CR
\dd {\omega}_6 &=& - \coup^2 e^{2\sigma} \sugrarho^{\frac{5}{9}} \bigl( 4 e^{-2\phi} +3  e^{\phi}  \bigr)  \dd t \wedge \dd z\; . 
\eea
This uplift ansatz was already written in \cite{Anabalon:2013zka,Ortiz:2014aja} for a vanishing axion $a=0$. 

For $a\ne 0$, one also gets a non-zero three-form in eleven dimensions. It is determined by the matrix elements 
\bea \langle 0|_c M^{-1} {}^{ab}|1/3\rangle &=& \varepsilon^{ab}{}_c e^{2\sigma} \sugrarho^{\frac89} e^{-2\phi} a \; , \CR
\langle 1|^e_{ab} M^{-1} {}^{cd}|4/3\rangle &=& \varepsilon_{ab}{}^{(c} \delta^{d)e} e^{2\sigma} \sugrarho^{\frac{26}9} e^{-2\phi} a + \delta^e_{[a} Z_{b]}^{cd}\; ,  \CR
\langle 1|^{\hat{e}}_{ab} M^{-1} {}^{c\hat{d}}|4/3\rangle &=& \frac12 \varepsilon_{ab}{}^{c} \delta^{\hat{d}\hat{e}} e^{2\sigma} \sugrarho^{\frac{26}9} e^{-5\phi} a \; ,  \eea
where the unspecified tensor $ Z_{b}^{cd}$ does not contribute to the three-form ansatz, while the other components of \eqref{Galpha=M} vanish. One obtains 
\be \alpha_{9ij} = \varepsilon_{abc} Y^a_2 \partial_i Y^b_2\partial_j Y^c_2  \sugrarho^{-\frac23} \y^3 \frac{e^{-4\phi}}{\Delta(1+f)} a \;  , \qquad \alpha_{ijk} = 0 \; , \ee
and 
\be 
\dd \langle 1|^c_{ab} |A\rangle =  -  \coup e^{2\sigma} \sugrarho^{-\frac19} e^{-2\phi} a\,  \varepsilon_{ab}{}^c \dd t\wedge \dd z\; . \ee
The three-form expression~\eqref{ThreeForm} then reduces to
\be A^{\scalebox{0.6}{11D}} =  \Bigl( \sugrarho^{-\frac23}  \frac{e^{-4\phi}}{\Delta(1+f)} a  \bigl( {\rm d}\psi  +  \y^2 {\omega}_3 + (1-\y^2) {\omega}_6 \bigr) + {A}_3  \Bigr) \wedge  \y^3 \dd \Omega_{S^2} \,,
\ee
with the one-form $A_3$ defined by $A_3 \varepsilon_{ab}{}^c = \coup \langle 1|^c_{ab} | A\rangle$ and satisfying 
\be 
\dd  {A}_3 = - \coup^2 e^{2\sigma} \sugrarho^{-\frac19} e^{-2\phi} a \, \dd t\wedge \dd z  \; . 
\ee

Let us now describe a few properties of the corresponding solutions. 
The topology of the sphere $S^8$ is not modified by the deformation. The coordinate  singularities at $\y\rightarrow 0$
\be 
e^{-\phi} \Delta \frac{\dd\y^2}{1-\y^2} + \frac{e^{-2\phi}}{(1+f)} \y^2 \dd\mathring{s}^2_{2} + e^{\phi}(1- \y^2) \dd\mathring{s}^2_{5}  \sim e^{-2\phi} \bigl( \dd \y^2 + \y^2 \dd\mathring{s}^2_{2}\bigr) + e^{\phi}  \dd\mathring s^2_{5} \; ,  
\ee
and  $\y\rightarrow 1$ 
\bea 
e^{-\phi} \Delta \frac{\dd\y^2}{1-\y^2} + \frac{e^{-2\phi}}{(1+f)} \y^2 \dd\mathring{s}^2_{2} + e^{\phi}(1- \y^2) \dd\mathring{s}^2_{5} \hspace{-2mm}&\sim&\hspace{-2mm} e^{\phi} \bigl( (\dd\sqrt{1-\y^2})^2 + (1-\y^2) \dd\mathring{s}^2_{5}\bigr) \CR
&&  +  \frac{e^{-2\phi}}{(1+e^{-6\phi} \sugrarho^{-\frac23} a^2)} \dd\mathring{s}^2_{2} \; ,  
\eea
can indeed be removed by a change of variables for all finite values of the fields $\phi$ and $a$. In particular the internal space parametrised by the segment $\zeta \in [0,1]$ and the two spheres is a squashed $S^8$.
 
The Killing vector field  $\partial_t$ is light-like in the one-half BPS purely gravitational pp-waves in eleven dimensions  \cite{Ortiz:2014aja}, however, it is never light-like for a non-trivial axion profile. The norm squared of $\partial_t$ is indeed proportional to  
\be  
\coup^2 e^{2\sigma} \Delta^2 ( 1+ f)  -\sugrarho^{-\frac{14}{9}}    ( \y^2 {\omega}_{3\,  t} +  (1-\y^2) {\omega}_{6\,  t} )^2 \; ,  
\ee 
which never vanishes for $a\ne 0$. 

Let us now compare the SO(3)$\times$SO(6) truncation ansatz derived in this section to the SO(3)$\times$SO(6) invariant ansatz in eleven dimensions considered in \cite{Costa:2014wya}. To do so we introduce the inverse radius holographic coordinate $x(z)$, that is related to the conformal gauge coordinate $z$ as
\be \frac{\partial x(z)}{\partial z } = \frac{1-x(z)^7}{x(z)^{\frac32}} \; .\ee
The coordinates we use here are related to the one used by CGPS in  \cite{Costa:2014wya} as follows 
\be t = \eta_{\scalebox{0.6}{CGPS}}\; , \quad \zeta = x_{\scalebox{0.6}{CGPS}}\sqrt{2-(x_{\scalebox{0.6}{CGPS}})^{2}}\; ,\quad  \psi =- \zeta_{\scalebox{0.6}{CGPS}} - \eta_{\scalebox{0.6}{CGPS}} \; , \quad x = y_{\scalebox{0.6}{CGPS}} \; . \ee
For simplicity we set $\coup=1$. Following the ansatz considered in  \cite{Costa:2014wya}, one writes 
\be e^{2\sigma} = \frac{1-x^7}{x^7} h_1(x) \; , \quad \varrho = x^{ -\frac{9}{2}}  h_2(x)   \; , \ee
for the functions $h_1(x)$, $h_2(x)$, $a(x)$ and $\phi(x)$ that are analytic at $x=0$. For $a=\phi=0$ and $h_1=h_2=1$, one gets back the black hole solution \cite{Cvetic:1999xp,Wiseman:2013cda}, which was shown to be a solution of SO(9) gauged supergravity in \cite{Anabalon:2013zka}. This solution interpolates between the one-half BPS pp-wave solution \cite{Nicolai:2000zt} at $x=0$ and the near horizon of a ten-dimensional Schwarzschild black hole solution times a circle at $x=1$. It is therefore interpreted as the holographic dual of the BFSS matrix quantum mechanics at finite temperature \cite{Itzhaki:1998dd,Wiseman:2013cda,Costa:2014wya}. Here we use dimensionless coordinates as in \cite{Costa:2014wya}, such that the radius of the M-theory circle and the mass of the black hole are reabsorbed in the rescalings  $e^{2\sigma}(x)\rightarrow \ell^2 e^{2\sigma}(r_0 x)$, $\varrho(x)\rightarrow \ell^{\frac{9}{2}}\varrho(r_0 x)$.

According to  \cite{Costa:2014wya}, one can consider the high temperature limit of the BMN matrix model by including a perturbation associated to the non-normalisable mode of the three-form potential. Within gauged supergravity, one can consider the linearised solutions for the axion and the dilaton expressed in terms of hypergeometric
functions~${}_2\hspace{-0.2mm}F_1$
\bea a(x) \hspace{-2mm} &=&\hspace{-2mm}  - \frac{3}{4} \Bigl(  \hat{\mu} x \, {}_2\hspace{-0.2mm}F_1(\tfrac1{7},\tfrac{4}{7};\tfrac{5}{7};x^7)   + \alpha x^3 \, {}_2\hspace{-0.2mm}F_1(\tfrac3{7},\tfrac{6}{7};\tfrac{9}{7};x^7) \Bigr) + \mathcal{O}(x^5) \;  , \CR
\phi(x)\hspace{-2mm}&=&\hspace{-2mm}  \beta x^2 \, {}_2\hspace{-0.2mm}F_1(\tfrac2{7},\tfrac{2}{7};\tfrac{4}{7};x^7)   + \gamma  x^5 \, {}_2\hspace{-0.2mm}F_1(\tfrac{5}{7},\tfrac{5}{7};\tfrac{10}{7};x^7) +  \mathcal{O}(x^4) \;  ,   
 \eea
where the neglected orders are needed to solve the  non-linear  equations and are indicated in the limit $x\rightarrow 0$. They can be solved perturbatively in $x$ near the asymptotic boundary at $x=0$, and perturbatively in $(1-x)\log (1-x)$ and $1-x$ near the black hole horizon at $x=1$. Here the parameters $\hat{\mu}$ and $\beta$ are associated to the non-normalisable modes, while $\alpha$ and $\gamma$ are associated to normalisable modes (in $D=2$) that must be determined by the regularity of the solution at the horizon. It is the parameter  $\hat{\mu} = \frac{7}{12\pi} \frac{\mu}{T}$  that triggers the BMN deformation, where $\mu$ is the BMN mass parameter and $T$ the temperature \cite{Costa:2014wya}.  The corresponding system we obtain from SO(9) gauged supergravity is a truncation of the ansatz considered in \cite{Costa:2014wya} to the lowest harmonics on the sphere $S^8$, and on which one imposes the gauge
$g_{\scalebox{0.6}{11D}}(\partial_\zeta,\partial_x) = 0$ using a reparametrisation $x = x(x^\prime, \zeta)$. 

Solving this system perturbatively in small $x$ one obtains the expansion 
\bea e^{2\sigma} \hspace{-3mm}&=&\hspace{-3mm} \frac{1-x^7}{x^7}\Bigl( 1- \frac{14 \beta^2}{13} x^4 -\frac{11\hat \mu^2}{1200} x^5   + \mathcal{O}(x^6) \Bigr) \; , \qquad \varrho = x^{ -\frac{9}{2}}  \Bigl(1- \frac{9 \beta^2}{13} x^4- \frac{3\hat \mu^2 }{100 }x^5  +  \mathcal{O}(x^6) \Bigr)  \; ,  \CR
a  \hspace{-3mm}&=&\hspace{-3mm} - \frac{3}{4} \Bigl( \hat{\mu}   x + \alpha x^3 + \Bigl( 3\alpha\beta - \frac{24 \beta^2 \hat{\mu}}{13} \Bigr) x^5    + \mathcal{O}(x^6) \Bigr)  \; , \quad \phi =  \beta x^2 +\frac{ \beta^2}{2} x^4  + \gamma x^5   + \mathcal{O}(x^6)\; ,  \eea
such that 
\be A^{\scalebox{0.6}{11D}} =  \frac{1}{x^3}  \Bigl( \hat{\mu}   + \frac{3( \alpha-3 \beta \hat{\mu} ) }{2} x^2 + \mathcal{O}(x^4) \Bigr) \dd t \wedge  \y^3 
\dd \Omega_{S^2} - \frac{3}{4 }\hat{\mu} x^4 ( 1 + \mathcal{O}(x^2)  ) \dd \psi \wedge  \y^3 
\dd \Omega_{S^2}   \,.
\ee
The leading term of  $A^{\scalebox{0.6}{11D}} $ proportional to $\hat{\mu}$ reproduces the asymptotic expansion of the non-normalisable mode that triggers the BMN deformation. 
However, one can check that the uplifted solution does not include all the harmonics in $\zeta$ that appear in the numerical solution  \cite{Costa:2014wya} and therefore, we cannot reproduce the latter from SO(9) gauged supergravity. 
It would be very interesting to investigate whether a regular solution exists within the consistent truncation.

\medskip

SO(9) gauged maximal supergravity captures important features of the BFSS matrix model both at zero and finite temperature~\cite{Anabalon:2013zka,Ortiz:2014aja} and we have argued that it can also be relevant in describing the BMN model at finite temperature. 
An interesting application in this direction would be to study axionic perturbations of the so-called rotating D0 brane solutions~\cite{Cvetic:1999xp,Anabalon:2013zka}, along the lines of the discussion above. More generally, having access to the full SO(9) theory and its uplift opens up the possibility of studying many other deformations of the BFSS model with different symmetry breaking patterns.

\section*{Acknowledgements}

We would like to thank Thomas Fischbacher, Benedikt K\"onig, Emanuel Malek, Hermann Nicolai and Henning Samtleben for discussions.
AK is grateful to \'Ecole Polytechnique for its hospitality during the early stages of this paper.
Part of this work was carried out at the workshop on Higher Structures, Gravity and Fields at the Mainz Institute for Theoretical Physics of the DFG Cluster of Excellence PRISMA+ (Project ID 39083149). We would like to thank the institute for its hospitality. 
This work has received funding from the European Research Council (ERC) under the European Union’s Horizon 2020 research and innovation programme (grant agreement No 740209).


\appendix

\section{Algebras and decompositions}
\label{app:e8e9}

In this appendix, we collect more details on several decompositions of $\mf{e}_8$, $\mf{e}_9$ and their representations that are used in the main body of the paper.

\subsection{\texorpdfstring{The $\mf{gl}_8$ branching of $\mf{e}_8$ and $\mf{e}_9$}{The gl8 branching of e8 and e9}}
\label{app:e8}

The graded decomposition of $\mf{e}_8$ under its $\mf{gl}_8$ subalgebra associated with nodes $1,\ldots,7$ of Figure~\ref{fig:e9dynk} was given in~\eqref{eq:e8a7} along with a convention of the generators and their transformation under $\mf{gl}_8$ in~\eqref{eq:e8gl8}.

In order to complete this description to $\mf{e}_8$, we begin by giving the normalisations of all the generators
\begin{align}
 \left\langle  T^i{}_j \middle| T^k{}_\ell \right\rangle &= \delta^i_\ell \delta^k_j -\frac{1}{9} \delta^i_j \delta^k_\ell\,,\nn\\
\left\langle  T^{i_1i_2i_3} \middle| T_{j_1j_2j_3} \right\rangle  &= 3! \,\delta^{i_1i_2i_3}_{j_1j_2j_3}   \,,\nn\\
 \left\langle T^{ij} \middle| T_{k\ell} \right\rangle &= 2 \,\delta^{ij}_{k\ell} \,,\nn\\
\left\langle T^{i} \middle| T_{j} \right\rangle  &= \delta^i_j\,.
\end{align}

The induced bilinear form is the Cartan--Killing form on $\mf{e}_8$ and invariant under  the commutation relations~\eqref{eq:e8c1} and\footnote{The numerical $\varepsilon^{i_1\ldots i_8} \in \{-1,0,1\}$ commutes with $T^i{}_j$.}
\begin{align}
\label{eq:e8com}
\lb T^{i_1i_2i_3}, T^{i_4i_5i_6} \rb &= \frac12 \varepsilon^{i_1\ldots i_6 k \ell} T_{k\ell}\,,&
\lb T^{i_1i_2i_3} , T^{j_1j_2} \rb &= -\frac16 \varepsilon^{i_1i_2i_3j_1j_2k_1k_2k_3} T_{k_1k_2k_3}\,,\nn\\
\lb T^{i_1i_2i_3}, T_{j_1j_2} \rb&= 6 \, \delta^{[i_1i_2}_{j_1j_2} T^{i_3]}_{\phantom{[}}\,,&
\lb T_{i_1i_2i_3}, T^j \rb &= 3 \,\delta^j_{[i_1} T_{i_2i_3]}^{\phantom{[}}\,,\nn\\
\lb T^i , T_j \rb &= T^i{}_j + \delta^i_j \,T^k{}_k\,,&
\lb T^{ij} , T_{k\ell} \rb &=  4 \, \delta^{[i}_{[k}\, T^{j]}_{\phantom{ii} \ell]} -2 \, \delta^{ij}_{k\ell} T^p{}_p\,,\nn\\
\lb T_{ij}, T_k \rb &= T_{ijk}
\end{align}
In particular, we note $T^i= \frac1{42} \lb T^{ij_1j_2} , T_{j_1j_2}\rb$.

As explained in Section~\ref{sec:a8e9}, the generators $T^i{}_j$, $T^i$ and $T_{-i}$ form an $\mf{sl}_9$ algebra whose generators are denoted by $T^I{}_J$ and commutation relation given in~\eqref{eq:sl9}. The $\mf{e}_8$ Killing form restricted to this $\mf{sl}_9$ is
\begin{align}
\label{eq:kfsl9}
\langle T^I{}_J | T^K{}_L \rangle = \delta^I_L \delta^K_J - \frac19 \delta^I_J \delta^K_L\,.
\end{align}
The branching of $\mf{e}_8$ under $\mf{sl}_9$ is (see~\eqref{eq:e8deca8})
\begin{align}
{\bf 248} = \overline{\bf 84} \oplus {\bf 80} \oplus {\bf 84}
\end{align}
where the ${\bf 84}$ generators $T^{IJK}=T^{[IJK]}$ are made out of $T^{ijk}$ and $T^{ij}$ and similarly for the downstairs indices and the induced normalisation is
\begin{align}
\langle  T^{IJK} | T_{LMN} \rangle = 3! \, \delta^{IJK}_{LMN}\,,
\end{align}
while the $\mf{e}_8$ commutation relations become
\begin{align}
\label{eq:e8a8com}
\lb T^I{}_J , T^{K_1K_2K_3} \rb &= 3 \delta_J^{[K_1} T^{K_2K_3]I}_{\phantom{K]}} -\frac13 \delta^I_J T^{K_1K_2K_3}\,,\nn\\
\lb T^I{}_J , T_{K_1K_2K_3} \rb &= -3 \delta^I_{[K_1} T_{K_2K_3]J}^{\phantom{K]}} +\frac13 \delta^I_J T_{K_1K_2K_3}\,,\nn\\
\lb T^{I_1I_2I_3}, T_{J_1J_2J_3} \rb &= 18\, \delta^{[I_1I_2}_{[J_1J_2} T^{I_3]}_{\phantom{II} J_3]}\,,\\
\lb T^{I_1I_2I_3} , T^{I_4I_5I_6} \rb &= - \frac16 \varepsilon^{I_1\ldots I_9} T_{I_7I_8I_9}\,,\nn\\
\lb T_{I_1I_2I_3} , T_{I_4I_5I_6} \rb &=  \frac16 \varepsilon_{I_1\ldots I_9} T^{I_7I_8I_9}\,.\nn
\end{align}
Note that~\eqref{eq:e8deca8} is \textit{not} a graded decomposition of $\mf{e}_8$ as exemplified by the last two commutators above.

\medskip

The $\mf{gl}_8$ basis of $\mf{e}_8$ can be extended to the loop algebra $\hat{\mf{e}}_8$ of $\mf{e}_8$ as
\begin{align}
\label{eq:e8loop}
T_{n\, i} \,,\quad T^{ij}_n\,,\quad T_{n\, ijk} \,,\quad T^i_{n\, j} \,,\quad T^{ijk}_n \,,\quad T_{n\, ij} \,,\quad T^i_n
\end{align}
with $n\in \mathds{Z}$ corresponding to the loop number.

The central element $\dK$ occurs in the central extension of the loop commutators:
\begin{align}
\label{eq:e9com}
\lb T_m^A , T_n^B \rb = f^{AB}{}_C \,T^C_{m+n} + m \eta^{AB} \delta_{m,-n} \dK\,,
\end{align}
where $f^{AB}{}_C$ are the $\mf{e}_8$ structure constants and $\eta^{AB}$ the Killing form. For example, we have
\begin{align}
\lb T^i_{1} , T_{-1\, j} \rb = T^i{}_j + \delta^i_j T^k{}_k + \delta^i_j \dK\,.
\end{align}

\subsection{Basic representation}
\label{app:bas}

The dual of the basic representation $\overline{R(\Lambda_0)_{-1}}$, in which derivatives take their values, is written in terms of bra vectors as in~\eqref{eq:derm} with $\mf{e}_8$ decomposition given in~\eqref{eq:dere8}. 
Using the $\mf{gl}_8\subset\mf{e}_8$ subalgebra defined in~\eqref{eq:e8gl8}, this can be further decomposed under $\mf{gl}_8$ according to the following doubly graded decomposition
\begin{align}
\label{eq:basa7}
\overline{R(\Lambda_0)_{-1}} = {\bf 1}_{0}^\ord{0} \oplus \left(\overline{\bf 8}^\ord{-1} \oplus {\bf 28}^\ord{-2/3} \oplus \overline{\bf 56}^\ord{-1/3} \oplus \left( \mf{gl}_8\right)^\dgr{0} \oplus {\bf 56}^\ord{1/3} \oplus \overline{\bf 28}^\ord{2/3} \oplus {\bf 8}^\ord{1}\right)_1 \oplus \ldots 
\end{align}
The superscripts denote the $\mf{gl}_1\subset \mf{gl}_8$ weights whereas the subscripts are the affine levels (in $\mf{e}_9$) with respect to $L_0$.
For example, the ${\bf 8}_1^\ord{1}$ corresponds to the state $\langle 0| T_{1\, i}$. 
We also note the decomposition
\begin{align}
\label{eq:3875a7}
{\bf 3875} &= {\bf 8}^\ord{-5/3} \oplus {\bf 70}^\ord{-4/3} \oplus \left( \overline{\bf 28} \otimes {\bf 8}\right)^\ord{-1} \oplus \left( {\bf 56}\otimes \overline{\bf 8} \oplus  {\bf 36} \right)^\ord{-2/3} \oplus \left( {\bf 70}\otimes {\bf 8} \oplus \overline{\bf 168}\right)^\ord{-1/3}\nn\\
&\quad  \oplus \left( {\bf 720} \oplus 2{\times} {\bf 63} \oplus {\bf 1}\right)^\ord{0} \oplus 
\left({\bf 70}\otimes \overline{\bf 8} \oplus {\bf 168}\right)^\ord{1/3}\ldots
\end{align}
of the next $\mf{e}_8$ representation under $\mf{gl}_8$ which enters at affine level two in~\eqref{eq:dere8}. Some of the representations were written reducibly as tensor products for conciseness. The following mixed tensors appear: $\overline{\bf 168}=R(\lambda_1{+}\lambda_2)$ and ${\bf 720} = R(\lambda_2{+}\lambda_6)$, where the weights refer to $\mf{sl}_8$ and ${\bf 8}=R(\lambda_7)$ in these conventions.

The module $\overline{R(\Lambda_0)_{-1}}$ is irreducible and therefore there are null vectors that are generated when acting with the loop generators on the groundstate $\langle 0|$, i.e., not all states
\begin{align}
\langle 0| \prod_i T_{n_i}^{A_i}
\end{align}
are non-vanishing, where $n_i>0$ since the groundstate is $\mf{e}_8$ invariant, and $A_i$ denotes an adjoint $\mf{e}_8$ index.
As we shall make use of some them, we work out a few examples of such null states in the further $\mf{gl}_8$ decomposition.

At $L_0$-level two, we could write the vector
\begin{align}
\langle 0 | T_1^{(i} T_1^{j)} \,.
\end{align}
By construction this would have to be part of the ${\bf 27000}$ in the symmetric tensor product of two $\mf{e}_8$ adjoints. However, by inspecting~\eqref{eq:dere8}, we know the generic symmetric ${\bf 27000}$ of $\mf{e}_8$ is absent at level two. Therefore, the above vector has to be a null vector in the Verma module, i.e., it vanishes in the irreducible  module $\overline{R(\Lambda_0)_{-1}}$.  

This can be checked by direct computation by using~\eqref{eq:sl9aff} and the $\mf{gl}_8$ invariance of the ground state:
\begin{align}
\langle 0 | T_1^{(i} T_1^{j)} T_{-1\, k} &=  \langle 0| \left( \delta^{(i}_{k} T_1^{j)}  + T_1^{(i} T^{j)}_{0\, k}  +  \delta_k^{(i} T_1^{j)} T^{\ell}_{0\,\ell} + \delta_k^{(i} T_1^{j)}  \right)\nn\\
&=   \langle 0| \left( \delta^{(i}_{k} T_1^{j)} - \delta_k^{(i} T_1^{j)}  - \delta_k^{(i} T_1^{j)}  + \delta_k^{(i} T_1^{j)}\right)\nn\\
&=0\,.
\end{align}
Since the generic anti-symmetric ${\bf 30380}$ is also absent (and since $[T_m^i,T_n^j]=0$ by $\mf{gl}_8$ grading), one actually has the null vector
\begin{align}
\langle 0 | T_1^{i} T_1^{j}  =0
\end{align}
without any specific symmetry assumptions.

This null state has as a descendant
\begin{align}
\langle 0 | T_1^{(i} T_2^{j)}= \frac12 \langle 0 | T_1^{(i} T_1^{j)} L_1
\end{align}
since again $[T^i_1,T_2^j]=0$. Therefore in the module $\overline{R(\Lambda_0)_{-1}}$ the following relation holds
\begin{align}
\label{eq:28}
\langle 0 | T_1^{i} T_2^{j} = \langle 0 | T_1^{[i} T_2^{j]} \,,
\end{align}
which thus automatically projects to the anti-symmetric rank-two representation of $\mf{gl}_8$ at $L_0$ eigenvalue three.

\subsection{\texorpdfstring{Branching of the basic module under spectrally flowed $\mf{sl}_9$}{Branching of the basic module under spectrally flowed sl9}}
\label{app:basbranch}

In Section~\ref{sec:flow} we have introduced spectrally flowed $\mf{sl}_9$ subalgebras of $\mf{e}_9$ for any $p\in\mathds{Z}$. The case $p=0$ corresponds to the $\mf{sl}_9\subset \mf{e}_8$ with generators $T^I{}_J$ that appear in~\eqref{eq:sl9}. For any $p\in\mathds{Z}$, we have defined the flowed $\mf{sl}_9$ in~\eqref{eq:sl9flow}.
We also record here that the shifted bilinear form $\eta_{-k\,\alpha\beta}$ introduced in~\eqref{eq:shifteta} takes the following form in the basis where the $\mf{sl}_9$ was flowed by $p\in \mathbb{Z}$ units
\begin{align}
\label{eq:etak9}
 \eta_{-k\,\alpha\beta} \T^\alpha \otimes \T^\beta 
 &= \sum_{m\in\mathbb{Z}}\bigg( \T^{\hspace{2mm}I}_{m-k\, J}\otimes  \T^J_{-m\, I} + \frac16 \T^{IJK}_{m-k-p/3} \otimes  \T_{-m+p/3\, IJK} \nn\\*
 &\hspace{15mm}
 + \frac16\T_{-m+p/3\, IJK}   \otimes  \T^{IJK}_{m-k-p/3}
 \bigg)
 - \LL_{-k} \otimes \dK - \dK \otimes \LL_{-k}\,.
\end{align}

In the following we work out some details of the decomposition of the basic module $\overline{R(\Lambda_0)_{-1}}$ under the various flowed $\mf{sl}_9$ subalgebras of $\mf{e}_9$. 
A summary of the results was given in Section~\ref{sec:basdec}.

\subsubsection{\texorpdfstring{Spectral flow by $p=1$ unit}{Spectral flow by p=1 unit}}

The case $p=1$ corresponds to the $D=11$ gravity line and we recall from Section~\ref{sec:p1flow} that we use the convention to denote generators in the $p=1$ with a tilde. The corresponding fundamental indices are written as $\Is=(i,9)$, where, contrary to the $p=2$ flow, we denote the index extending the $\mf{gl}_8$ by $9$ rather than $0$.

The lowest eigenvalue of $\widetilde{\LL}_0= L_0+ T^\ell_{0\,\ell} + \frac49 \dK$ is $\widetilde{\LL}_0=\tfrac49$ and it is realised by the states
\begin{align}
\langle 0 | =  \frac18 \langle 0 |  T^k _1 T_{-1\, k} = \frac18 \langle 0 | \widetilde{\T}^k_{0\, 9} \widetilde{\T}^9_{0\, k} 
\quad\text{and}\quad
\langle 0| T_{1}^i = \langle 0| \widetilde{\T}^i_{0\, 9} \,.
\end{align}
in the module $\overline{R(\Lambda_0)_{-1}}$.
We have written the states in several forms to emphasise that we can identity among these lowest $\widetilde{\LL}_0$ states an $\mf{sl}_9$ representation ${\bf 9}$ of the $p=1$ flowed $\mf{sl}_9$ that we write as
\begin{align}
\label{eq:co9p1}
\widetilde{\langle 0 |}{}^{\Is}  \quad\text{with}\quad \widetilde{\langle 0|}{}^9 = \langle 0| \,,\quad \widetilde{\langle 0|}{}^i = - \langle 0| T_1^i\,.
\end{align}
Under the $p=1$ flowed $\mf{sl}_9$ this transforms as
\begin{align}
\widetilde{\langle 0 |}{}^\Is \widetilde{\T}^\Js_{0\, \Ks} = - \delta^\Is_\Ks \widetilde{\langle 0|}{}^\Js + \frac19 \delta^\Js_\Ks \widetilde{\langle 0|}{}^\Is\,,
\end{align}
where the extra term is required by the tracelessness of $\mf{sl}_9$ and the minus sign in~\eqref{eq:co9p1} is related to the minus sign in the $\mf{sl}_9$ action. 

The physical interpretation of this ${\bf 9}$ is that the corresponding nine derivatives are  those of the coordinates of the M-theory solution of the section constraint that completes the two external coordinates to $D=11$ dimensions.

The next possible $\widetilde{\LL}_0$ eigenvalue is $\tfrac79$ and is obtained by the action with $\widetilde{\T}_{1/3\,\Is\Js\Ks}$ on the ${\bf 9}$:
\begin{align}
\widetilde{\langle 1/3|}{}_{\Is\Js} = \frac17 \widetilde{\langle 0|}{}^\Ks \widetilde{\T}_{1/3\, \Is\Js\Ks}\,.
\end{align}
Plugging in the definition of $\widetilde{\T}_{1/3\,\Is\Js\Ks}$ from~\eqref{eq:flowe8} we find for example explicitly
\begin{align}
\widetilde{\langle 1/3|}{}_{ij} &= \frac17 \langle 0| T_{1 ij} - \frac17 \langle 0| T^k_1  T_{0 ijk} \,,\nn\\
\widetilde{\langle 1/3|}{}_{i9} &=  \frac17 \langle 0| T^k_1  T_{1 ik} \,.
\end{align}

One can continue the construction of the module $\overline{R(\Lambda_0)_{-1}}$  along these lines and ends up with the following decomposition
\begin{align}
\label{eq:flowdecGRapp}
\overline{R(\Lambda_0)_{-1}} = {\bf 9}_{\frac{4}{9}} \oplus \overline{\bf 36}_{\frac{7}{9}} \oplus {\bf 126}_{\frac{10}9} \oplus \left({\bf 9}\oplus \overline{\bf 315}\right)_{\frac{13}{9}} \oplus \left( \overline{\bf 36} \oplus \overline{\bf 45} \oplus \overline{\bf 720}\right)_\frac{16}{9} \oplus \ldots\,.
\end{align}
Some specific basis elements of this decomposition are defined as 
\begin{align} 
\label{eq:st1}
\widetilde{\langle 1/3|}{}_{\Is\Js} &= \frac17 \widetilde{\langle 0|}{}^\Ks \widetilde{\T}_{1/3\,\Is\Js\Ks} \; , \quad
\widetilde{\langle 2/3|}{}^{\Is\Js\Ks\Ls} =  \widetilde{\langle 0|}{}^\Is \widetilde{\T}_{2/3}^{\Js\Ks\Ls}\; , \CR
\widetilde{ \langle 1|}{}^{\Is\Js}_\Ks &= \widetilde{\langle 0 |}{}^{\Is} \widetilde{\T}_1^{\Js}{}_{\Ks} - \tfrac1{10} \delta_\Ks^\Js \widetilde{\langle 0 |}{}^\Ls \widetilde{\T}_1^\Is{}_\Ls\; , \quad 
\widetilde{\langle 4/3|}{}_{\Is\Js} = \frac18 \widetilde{\langle 1/3|}{}_{\Ks(\Is} \widetilde{\T}_1^{\Ks}{}_{\Js)} \; , 
 \end{align}
where we have labelled the state by the $\widetilde{\LL}_0$ weight relative to that of the lowest ${\bf 9}$. Note that, due to the irreducibility of the module, some symmetries are implied for the left-hand sides that are not manifest on the corresponding right-hand sides. For instance, the state $\widetilde{\langle 2/3|}{}^{\Is\Js\Ks\Ls}$ is completely anti-symmetric in its four indices and belongs to the ${\bf 126}$ representation. The na\"ive $(3,1)$ mixed symmetry term on the right-hand side of its definition is a null state. Similarly, the state $\widetilde{ \langle 1|}{}^{\Is\Js}_\Ks$ is anti-symmetric in $\Is\Js$ and contains a trace and thus represents reducibly the two components ${\bf 9}\oplus\overline{\bf 315}$ at $\widetilde{\LL}_0$ level $\tfrac{13}9$. For $\widetilde{\LL}_0$ level $\tfrac{16}{9}$ we have only written out the definition of the component in the $\overline{\bf 45}$ since this is the only one that appears in our analysis.

\subsubsection{\texorpdfstring{Spectral flow by $p=2$ units}{Spectral flow by p=2 units}}

The generators and indices for the spectral flow by $p=2$ are the prevalent ones in the paper and therefore written without tilde. The lowest $\LL_0=L_0 + 2 \,T^\ell_{0\, \ell} + \frac{16}{9} \dK $ eigenvalue that can be obtained is again $\LL_0=\frac49$ and arises for the states
\begin{align}
\langle 0| T_1^j T_{1\, ij} = \langle 0| \T^j_{-1\, 0} \T_{-1/3\, ij 0}
\quad\text{and}\quad
\langle 0| T_1^j T_{1\, ij} T_2^j = \langle 0| \T^j_{-1\, 0} \T_{-1/3\, ij 0} \T^i_{0\, 0}
\end{align}
that together form a $\overline{\bf 9}$ under the $p=2$ flowed $\mf{sl}_9$. We have written the states both in the standard $\mf{gl}_8$ basis of $\mf{e}_9$ and in terms of the flowed affine generators from Section~\ref{sec:flow}. We will write the corresponding ground state as
\begin{align}
\label{eq:co9}
\langle 0 |_I
\quad\text{with}\quad
\langle 0 |_i = \langle 0| T_1^j T_{1\, ij}\,,\quad 
\langle 0 |_0 = \frac18\langle 0| T_1^j T_{1\, ij} T_2^i 
\end{align}
that transforms under the flowed $\mf{sl}_9$ as
\begin{align}
\langle 0 |_I \T^J_{0\, K} = \delta^J_I \langle 0|_K - \frac19 \delta^J_K \langle 0 |_I\,,
\end{align}
where the extra term is due to the tracelessness of $\T^J_{0\, K}$.

The next $\LL_0$ eigenvalue that arises is $\LL_0=\tfrac79$ which occurs for the states
\begin{align}
\langle 0 | T_1^i = \langle 0 | \T^i_{-1\, 0} 
\quad\text{and}\quad
\langle 0 | T_1^i T_2^j = \langle 0 | \T^i_{-1\, 0}  \T^j_{0\, 0} \,.
\end{align}
Due to the structure of the module $\overline{R(\Lambda_0)_{-1}}$ we know (see~\eqref{eq:28}) that the second state is automatically anti-symmetric in $[ij]$ and therefore these two states together form a ${\bf 36}$ of the flowed $\mf{sl}_9$. 
We write it and subsequent states as
\begin{align} 
\label{eq:st2}
\langle 1/3|^{IJ} &= \frac17 \langle 0|_K \T_{1/3}^{IJK} \; ,& \quad\langle 2/3|_{IJKL} &=  \langle 0|_I \T_{2/3\, JKL}\; , \CR
 \langle 1|_{IJ}^K &= \langle 0 |_{I} \T_1^{K}{}_{J} - \tfrac1{10} \delta^K_J \langle 0 |_L \T_1^L{}_I\; ,& \quad \langle 4/3|^{IJ} &= \frac18 \langle 1/3|^{K(I} \T_1^{J)}{}_K \; .
\end{align}
Note that $\bra{1}^K_{IJ}$ is anti-symmetric in $IJ$ even though this is not manifest on the right-hand side of its definition.
This formula is similar to~\eqref{eq:st1} and corresponds to the branching
\begin{align}
\label{eq:flowdecapp}
\overline{R(\Lambda_0)_{-1}} = \overline{\bf 9}_{\frac{4}{9}} \oplus {\bf 36}_{\frac{7}{9}} \oplus \overline{\bf 126}_{\frac{10}9} \oplus \left(\overline{\bf 9}\oplus {\bf 315}\right)_{\frac{13}{9}} \oplus \left( {\bf 36} \oplus {\bf 45} \oplus {\bf 720}\right)_\frac{16}{9} \oplus \ldots
\end{align}

It is also useful to write out some consequences of the irreducibility of the module in this basis, i.e., the structure of the null vectors. We have
\begin{align}
\label{eq:null9}
\langle 0 |_L \T_{1/3}^{IJK} &= 3\, \delta_L^{[I} \langle 1/3|^{JK]}\,,\nn\\
\langle 0|_I \T_{2/3\, JKL} &= \langle 0|_{[I} \T_{2/3\, JKL]}\,,\\
\langle 0 |_K \T^{IKL}_{1/3} \T_{2/3\, L[J_1J_2} \, \Theta_{J_3] I} &= -42 \langle 0|_{[J_1} \T^I_{1\, J_2} \, \Theta_{J_3] I}\,,\nn
\end{align}
where in the last relation $\Theta_{IJ}=\Theta_{(IJ)}$ is any symmetric $\mf{sl}_9$ tensor (in the $\overline{\bf 45}$).

Moreover, since the ${\bf 36}_{16/9}$ in~\eqref{eq:flowdecapp} is multiplicity-free, one can show that
\begin{align}
\label{eq:36rel}
\langle 1/3|^{P[I}_{\phantom{[}} \T^{J]}_{1\, P} = 5 \langle 1/3|^{IJ} \LL_1
\end{align}
by relating the two ways of reaching this representation.

\subsubsection{Relation between the two bases and matrix elements}
\label{sec:relmat}

The basis elements in the two decompositions~\eqref{eq:st1} and~\eqref{eq:st2} are related by
\begin{align}
 \label{p1p2}
 \widetilde{\langle 0|}{}^9 &= \langle 4/3|^{00} \,,& 
\widetilde{\langle 0|}{}^i &= \langle 1/3|^{0i}\,,\nn\\
\widetilde{\langle 1/3|}{}_{i9} &= \langle 0|_i \,,&
\widetilde{\langle 1/3|}{}_{ij} &= \langle 1|^{0}_{ij} \,,\nn\\
 \widetilde{\langle 2/3|}{}^{ijkl} &=  -\frac1{24} \varepsilon^{ijklpqrs} \langle 2/3|{}_{pqrs} \,,&
  \widetilde{\langle 1|}{}^{ij}_9 &=  \langle 1/3|^{ij}  \,,\nn\\
  \widetilde{\langle 1|}{}^{i9}_9 &=  2 \langle 4/3|^{i0} \, . 
\end{align}
This can be verified by following through the definitions of all objects.

\medskip

For the uplift formul\ae{} we also require the dressing by $V$ of the basis states~\eqref{eq:st2} in the $p=2$ flowed basis of the basic representation. Here, $V$ is the E$_9$ element given in~\eqref{eq:E9cosetrep SL9 basis} and the dressing results in 
\begin{align}
\langle 0|_I V^{-1} &=  e^{\sugrasigma} \sugrarho^{\frac{4}{9}} \V^{\mathsf{A}} {}_I \langle 0|_{\mathsf{A}} \CR
\langle 1/3|^{IJ} V^{-1}  &=  e^{\sugrasigma} \sugrarho^{\frac{7}{9}} \bigl( \V^{-1 I}{}_{\mathsf{A}}  \V^{-1 J}{}_{\mathsf{B}}  \langle 1/3|^{{\mathsf{A}} {\mathsf{B}} } + \sugrarho^{-\frac13} a^{IJK} \V^{\mathsf{A}} {}_K \langle 0|_{\mathsf{A}}  \Bigr) \; ,
\end{align}
as well as 
\begin{align}
\langle 1|_{IJ}^K V^{-1} &=  e^{\sugrasigma} \sugrarho^{\frac{13}{9}} \biggl(  \V^{\mathsf{A}} {}_I \V^{\mathsf{B}} {}_J \V^{-1 K}{}_{\mathsf{C}}  \langle 1|_{{\mathsf{A}} {\mathsf{B}} }^{\mathsf{C}}  - \frac12 \sugrarho^{-\frac13}  a^{KPQ} \V^{\mathsf{A}} {}_I \V^{\mathsf{B}} {}_J  \V^{\mathsf{C}} {}_P \V^{\mathsf{D}} {}_Q \langle 2/3|_{{\mathsf{A}} {\mathsf{B}} {\mathsf{C}} {\mathsf{D}} }  \\
&\hspace{13mm} + \sugrarho^{-\frac23} \bigl( \delta^K_P b_{IJQ} + \delta^K_{[I} b_{J]PQ} \bigr) \Bigl( \V^{-1P}{}_{\mathsf{A}}  \V^{-1Q}{}_{\mathsf{B}}  \langle 1/3|^{{\mathsf{A}} {\mathsf{B}} } + \sugrarho^{-\frac13} a^{PQR} \V^{\mathsf{A}} {}_R \langle 0|_{\mathsf{A}}  \Bigr) \CR
&\hspace{13mm}  - \frac1{48} \sugrarho^{-\frac23} \varepsilon_{IJL_1\dots L_7} a^{KL_1L_2} a^{L_3L_4L_5} \V^{-1L_6}{}_{\mathsf{A}}  \V^{-1L_7}{}_{\mathsf{B}}  \langle 1/3 |^{{\mathsf{A}} {\mathsf{B}} }  \CR
&\hspace{13mm}+\sugrarho^{-1}  \Bigl( 2  \delta^K_{[I} h^L{}_{J]} -  2 h^K{}_{[I} \delta^L_{J]}    - \frac1{144}  \varepsilon_{IJP_1\dots P_7} a^{KP_1P_2} a^{P_3P_4P_5} a^{P_6P_7L} \Bigr)  \V^{{\mathsf{A}} }{}_L  \langle 0 |_{\mathsf{A}}   \biggr) \nonumber 
\end{align}
and
\begin{align}
\langle 4/3 |^{IJ} V^{-1} &= e^{\sugrasigma} \sugrarho^{\frac{16}{9}} \bigg( \V^{-1\, I}{}_{\mathsf{A}}  \V^{-1\, J}{}_{\mathsf{B}}  \langle 4/3|^{{\mathsf{A}} {\mathsf{B}} }  -   \sugrarho^{-1} h^{(I}{}_K \V^{-1\, J)}{}_{\mathsf{A}}  \V^{-1\, K}{}_{\mathsf{B}}  \langle 1/3|^{{\mathsf{A}} {\mathsf{B}} }  \nn\\
&\hspace{13mm} +\frac12 \sugrarho^{-1/3} a^{KL(I} \V^{-1 \, J)}{}_{\mathsf{A}}  \V^{\mathsf{B}} {}_K \V^{\mathsf{C}} {}_L  \langle 1|^{\mathsf{A}} _{{\mathsf{B}} {\mathsf{C}} } + \sugrarho^{-4/3} a^{KL(I} h^{J)}{}_L \V^{\mathsf{A}} {}_K \langle 0|_{\mathsf{A}}  \nn\\
&\hspace{13mm}- \frac18 \sugrarho^{-2/3} a^{KL(I} a^{J)PQ}\V^{\mathsf{A}} {}_K \V^{\mathsf{B}} {}_L \V^{\mathsf{C}} {}_P \V^{\mathsf{D}} {}_Q \langle 2/3|_{{\mathsf{A}} {\mathsf{B}} {\mathsf{C}} {\mathsf{D}} } \nn\\
&\hspace{13mm}-\frac1{288} \sugrarho^{-1} a^{K_1K_2(I} a^{J)K_3K_4} a^{K_5K_6K_7} \varepsilon_{K_1\ldots K_7 RS} \V^{-1\, R}{}_{\mathsf{A}}  \V^{-1\, S}{}_{\mathsf{B}}  \langle 1/3|^{{\mathsf{A}} {\mathsf{B}} } \nn\\
&\hspace{13mm} -\frac{1}{1152} \sugrarho^{-4/3} a^{K_1K_2(I} a^{J)K_3K_4} a^{K_5K_6K_7} a^{K_8K_9L} \varepsilon_{K_1\ldots K_9} \V^{\mathsf{A}} {}_L \langle 0|_{\mathsf{A}} \bigg)\,, \label{Vfourthird}
\end{align}
where use of~\eqref{eq:null9} was made repeatedly.
The (flattened) basis vectors are normalised such that
 \begin{align} 
 \label{eq:normbas}
 \langle 0|_{\mathsf{A}}  {}^{\mathsf{B}} |0\rangle &= \delta_{\mathsf{A}} ^{\mathsf{B}}  \; , & \langle 1/3|^{{\mathsf{A}} {\mathsf{B}} } {}_{{\mathsf{C}} {\mathsf{D}} }|1/3\rangle &= 2 \delta^{{\mathsf{A}} {\mathsf{B}} }_{{\mathsf{C}} {\mathsf{D}} }  \; , \nn\\
 \langle 2/3|_{{\mathsf{A}} {\mathsf{B}} {\mathsf{C}} {\mathsf{D}} }{}^{{\mathsf{E}} {\mathsf{F}} {\mathsf{G}} {\mathsf{H}} }|2/3\rangle &= 24\, \delta_{{\mathsf{A}} {\mathsf{B}} {\mathsf{C}} {\mathsf{D}}}^{{\mathsf{E}} {\mathsf{F}} {\mathsf{G}} {\mathsf{H}}}  \; , &
 \langle 1|_{{\mathsf{A}}{\mathsf{B}}}^{\mathsf{C}}  \, {}^{{\mathsf{E}}{\mathsf{F}}}_{\mathsf{G}} |1\rangle &= 2 \delta_{{\mathsf{A}}{\mathsf{B}}}^{{\mathsf{E}}{\mathsf{F}}} \delta^{\mathsf{C}}_{\mathsf{G}} + 4\delta_{{\mathsf{A}}{\mathsf{B}}}^{{\mathsf{C}}[{\mathsf{F}}} \delta^{{\mathsf{E}}]}_{\mathsf{G}}   \; , \nn\\
  \langle 4/3|^{{\mathsf{A}}{\mathsf{B}}} {} _{{\mathsf{C}}{\mathsf{D}}} |4/3\rangle &= \delta^{({\mathsf{A}}}_{({\mathsf{C}}} \delta^{{\mathsf{B}})}_{{\mathsf{D}})}  \; .
  \end{align}

\subsection{Inequivalent flows}
\label{sec:inequiv}

Using the decomposition of the basic representation, we can  discuss conjugacy of the various flowed algebras that were defined in Section~\ref{sec:flow}. As one of the main points will be comparing different units of flow $p$, we decorate the Virasoro generator by a label that keeps track of this and so write $\LL_0^\ord{p}$ and similarly for the other flowed generators $\T$ \textit{in this section only}.

In the case $p=0$ mod $3$, the algebra that commutes with $\LL_{0}^\ord{p}$ is again $\mf{gl}_1\oplus \mf{gl}_1\oplus \mf{e}_8$, composed of 
\begin{align}
\dK\; , \quad \LL_0^\ord{p}\; , \quad \T^{\ord{p}I}_{0\,\,\,\,\,\, J}\; , \quad \T_0^{\ord{p}IJK} \; , \quad \T^\ord{p}_{0 \,IJK}\; . 
\end{align}
Starting from the original vacuum of the basic module $\langle 0|$, one can built a state of eigenvalue $0$ with respect to $\LL_0^\ord{3}$ for $p=3$ 
as
\begin{align}
 \langle 0 |^\prime = \langle 0 | T_{1}^i T_{2}^j T_{1\, ij} \; . 
\end{align}
This state is in the highest weight representation of the ${\bf 147 250}_4$ of E$_8$ in~\eqref{eq:dere8}, and is therefore annihilated by all generators
\begin{align}
T_{-n-3 i}\; , \quad T_{-n-2}^{ij}\; , \quad T_{-n-1 ijk} \; , \quad T_{-n}^{\hspace{1.9mm} i}{}_j + \frac13  \delta_{n,0} \delta^i_j \dK\; , \quad T_{1-n}^{ijk} \; , \quad T_{2-n ij}\; , \quad T_{3-n}^i \; . 
\end{align}
for $n\geq 0$ and defines a vacuum state vector for the $\LL^\ord{3}_0$ decomposition of the basic module. One can construct the vacuum states of all $\LL^\ord{p}_0$ for $p=0$ mod $3$ using the same procedure, because  
\begin{align}
 \LL_0^\ord{p+3} = \LL_0^\ord{p}  + 3 \T_0^\ord{p}{}^i{}_i + 4 \dK \; , 
 \end{align}
for any $p$ and one can therefore obtain the spectral flowed subalgebra at $p+3$ from the one at $p$. Writing the vacuum $\langle 0|_{(q)}$ of the basic module of eigenvalue $0$ with respect to $\LL^\ord{p}_0$ for $p=3q$, one obtains by construction that 
\begin{align}
 \eta_{\alpha\beta}  \langle 0|_{(q)} \T^\alpha \otimes  \langle 0|_{(q)} \T^\beta = 0 \; ,  
\end{align}
and therefore that $ \langle 0|_{(q)} $ is in the E$_9$ orbit of $\langle 0 |_{(0)}$ for an element $g$ of the small Kac--Moody group  \cite{Peterson:1983}. The stabiliser of $\langle 0|_{(q)}$ determines the parabolic subgroup of Levi component $\text{GL}(1)\times \text{GL}(1) \times \text{E}_8$ and all the spectrally flowed $\mf{e}_8$ subalgebras for $p=0$ mod $3$ are therefore conjugate to each other in E$_9$. 

The same argument can be applied to case $p=1,2\mod 3$ using the level 3 module of weight $\Lambda_8$ in the labelling of Figure~\ref{fig:e9dynk}. Then one can further check that the cases $p=1$ and $p=-1$ are conjugate using that $\T_0^\ord{p}{}^i{}_i$ is conjugate to $-\T_0^\ord{p}{}^i{}_i$ in E$_8$. 

Therefore there are only two conjugacy classes, for $p=0$ mod $3$ and $p=\pm 1$ mod $3$. The cases $p=1$ and $p=2$ play a prominent role in our paper and they are related by an E$_9$ transformation (from the small Kac--Moody group).

\subsection{Reproducing physical Lagrangians}\label{app:Ltop to physical}

We give some details on the manipulations of \eqref{eq:E8coccycle as PP} that lead to the kinetic term for the $\mathrm{E}_8/(\mathrm{Spin}(16)/\ints_2)$ nonlinear sigma model.
We begin by noticing that we can rewrite \eqref{eq:E8coccycle as PP} by isolating $P^m$ for even or odd values of $m$ (also taking into account hermiticity).
We do so but then add up half of each such rewriting, thus finding (we hide an overall factor of $2\varrho$)
\allowdisplaybreaks
\begin{align}\label{eq:E8squaring first step}
&\sum_{n\in\ints} |n| \, P^n P^{-n-1}=\\\nonumber&
=\frac12P^1 P^0
+\frac12\sum_{n\ge1} ( P^{2n+1} - P^{2n-1} ) P^{-2n}
-\frac12\sum_{n\ge0} ( P^{-2n} - P^{-2n-2} ) P^{2n+1}
\\\nonumber&
=-\frac12P^0 \star P^0
+\frac12(P^1-\star P^0)P^0
+\frac12\sum_{n\ge1} ( P^{2n+1} - P^{2n-1} ) P^{-2n}
-\frac12\sum_{n\ge0} ( P^{-2n} - P^{-2n-2} ) P^{2n+1}
\intertext{%
In these and the following expressions we have hidden $\eta^{AB}$ as well as the E$_8$ indices on the currents, as they do not play any role in the computation. We remind the reader that these currents are spacetime one-forms and a wedge product is understood.
We now use twice a trick similar to what we did in the dilaton/central sector. In the first series, we add and subtract $\star P^{-2n}$ inside the parenthesis. In the second series, we add and subtract $\star P^{2n+1}$ to find}&
=-\frac12P^0 \star P^0
+\frac12\sum_{n\ge0} ( P^{2n+1} - \star P^{-2n} ) P^{-2n}
-\frac12\sum_{n\ge1} ( P^{2n-1} - \star P^{-2n} ) P^{-2n} \\&\qquad\nonumber
-\frac12\sum_{n\ge0}  ( P^{-2n} - \star P^{2n+1} )P^{2n+1}
+\frac12\sum_{n\ge0}  ( P^{-2n-2} - \star P^{2n+1} )P^{2n+1}
\intertext{In the second line, we add Hodge duals as done in the dilaton/central sector:}&
\nonumber
=-\frac12P^0 \star P^0
+\frac12\sum_{n\ge0} ( P^{2n+1} - \star P^{-2n} ) P^{-2n}
-\frac12\sum_{n\ge1} ( P^{2n-1} - \star P^{-2n} ) P^{-2n} \\\nonumber&\qquad
-\frac12\sum_{n\ge0} ( P^{2n+1} - \star P^{-2n} ) \star P^{2n+1} 
+\frac12\sum_{n\ge0} ( P^{2n+1} - \star P^{-2n-2} ) \star P^{2n+1}
\\&\label{e8 cocycle step 2}
=-\frac12P^0 \star P^0
\\&\qquad\nonumber
+\frac12\sum_{n\ge0} ( P^{2n+1} - \star P^{-2n} ) (P^{-2n}-\star P^{2n+1})
-\frac12\sum_{n\ge0} ( P^{2n+1} - \star P^{-2n-2} ) (P^{-2n-2}-\star P^{2n+1})\,,
\end{align}
so that the first term gives the physical kinetic term as in \eqref{eq:E8kinterm squaring}.

\bigskip

Let us now look at the axion sector in the SL(9) duality frame and reproduce \eqref{eq:axion topterm squaring}.
We start with the cocycle
\begin{align}
\label{axion cocycle}
&-2\varrho \sum_{n\in\ints}\left[
\Big(n-\frac23\Big) \frac16 \sQ^{n-2/3}{}^{\,\mathsf{ABC}} \sP^{-n-1/3}_{\mathsf{ABC}}
+\Big(n+\frac23\Big) \frac16 \sQ^{n+2/3}_{\mathsf{ABC}} \sP^{-n-5/3}{}^{\,\mathsf{ABC}}
\right]
\intertext{and in what follows, for brevity, we will hide the  $\mathsf{ABC}$ indices of the local SO(9)$_K$. From each term we extract the only term where both coefficients are along the negative modes and write the rest in terms of $n\ge1$:}\nonumber
\text{\eqref{axion cocycle}} &=
\frac19\varrho\, \sQ^{-1/3}\sP^{-2/3}
-\frac13\varrho \sum_{n=1}^\infty\Big(n-\frac13\Big)\sQ^{n-1/3}\sP^{-n-2/3}
+\frac13\varrho \sum_{n=1}^\infty\Big(n+\frac13\Big)\sQ^{-n-1/3}\sP^{n-2/3} 
\\\nonumber&\quad
+\frac29\varrho\, \sQ^{-2/3}P^{-1/3}
-\frac13\varrho \sum_{n=1}^\infty\Big(n-\frac23\Big)\sQ^{n-2/3}\sP^{-n-1/3}
+\frac13\varrho \sum_{n=1}^\infty\Big(n+\frac23\Big)\sQ^{-n-2/3}\sP^{n-1/3} \\[1ex]
&=\label{axion cocycle step 2}
-\frac1{36}\varrho\,\Omega^{-1/3}\Omega^{-2/3}
-\frac1{12}\varrho\sum_{n=0}^\infty \Big(
  \Omega^{-1/3-n}\,\Omega^{-4/3-n} + \Omega^{-2/3-n}\,\Omega^{-5/3-n}
\Big)
\end{align}
Now we want to reproduce squares of \eqref{eq:1/3twsd} and \eqref{eq:2/3twsd} in the series above.
To do so we follow the same procedure as for the E$_8$ case, but separately for the parts with weights shifted by $-1/3$ and $-2/3$, respectively.
To do so, let us define $X^n$ to correspond to either $\Omega^{-1/3+n}$ or $\Omega^{-2/3+n}$.
Then, in both cases the relevant term in the series above becomes
\begin{align}
&\hspace{-2em}\sum_{n=0}^\infty X^{-n} X^{-n-1} \\\nonumber
&=
-\frac12 X^{-1} X^0 
+\frac12 \sum_{n=1}^\infty (X^{-2n+1}-X^{-2n-1})X^{-2n}
+\frac12 \sum_{n=0}^\infty (X^{-2n}-X^{-2n-2})X^{-2n-1}\,.
\intertext{This is identical to (minus) the second line of \eqref{eq:E8squaring first step} if we rewrite that expression in terms of $\Omega$ (and rename $\Omega$ to $X$).
Of course, the objects we are dealing with here sit in the $\mathbf{84}$ and $\overline{\mathbf{84}}$ of SL(9) rather than the $\mathbf{248}$ of E$_8$, but this plays no role in the manipulations we are carrying out.
Twisted self-duality applies to $X$ here as it does to $\Omega$ there.
We thus reuse the end result:}
&=
\frac12X^0 \star X^0
\\&\qquad\nonumber
-\frac12\sum_{n\ge0} ( X^{-2n-1} - \star X^{-2n} ) (X^{-2n}-\star X^{-2n-1})
\\&\qquad\nonumber
+\frac12\sum_{n\ge0} ( X^{-2n-1} - \star X^{-2n-2} ) (X^{-2n-2}-\star X^{-2n-1}) \,.
\end{align}
Mapping back $X^n$ to $\Omega^{-1/3+n}$ or $\Omega^{-2/3+n}$ and hiding the squares of self-duality for brevity, we have
\begin{align}\nonumber
\text{\eqref{axion cocycle}} &=
-\frac1{18}\varrho\,\Omega^{-1/3}\Omega^{-2/3}
-\frac1{24}\varrho\,\Omega^{-1/3}\star\Omega^{-1/3}
-\frac1{24}\varrho\,\Omega^{-2/3}\star\Omega^{-2/3}
+\ldots\\\nonumber
&=
\frac1{18}\varrho\,\Omega^{-1/3}\Omega^{-2/3}
-\frac1{12}\varrho\,\Omega^{-1/3}\star\Omega^{-1/3}
+\frac1{24}\varrho(\Omega^{-1/3}-\star\Omega^{-2/3})(\star\Omega^{-1/3}-\Omega^{-2/3})
+\ldots\\
&=
\frac1{18}\varrho\,\Omega^{-1/3}\Omega^{-2/3}
-\frac1{12}\varrho\,\Omega^{-1/3}\star\Omega^{-1/3}
+\ldots
\end{align}
reproducing the physical Lagrangian \eqref{eq:axion topterm squaring} once the indices are reinstated.

\section{Details on the Weitzenb\"ock connection}
\label{app:Weitz}

In order to work out the Weitzenb\"ock connection~\eqref{eq:Weitz} and the embedding tensor components for the ansatz
\begin{align}
\label{eq:SO9ans2}
\cU^{-1} = \twistrho^{\LL_0}  e^{\twistsigma \dK} \mathsf{u}^{-1} 
\end{align}
we first note that the SL(9) matrix $\mathsf{u}$ acts on tensor generators written in components as
\begin{align}
\mathsf{u}\,  \T^I{}_J \mathsf{u}^{-1} = \mathsf{u}^{-1\,I}{}_K \mathsf{u}^{L}{}_J \T^K{}_L
\end{align}
and similarly for other tensors. 

With the ansatz~\eqref{eq:SO9ans2} and the solution~\eqref{eq:secsol} to the section constraint, we then get the trombone component of the embedding tensor to be of the simple form
\begin{align}
\label{eq:varth}
\langle \vartheta | = -\twistrho^{-1} \langle \partial | \cU^{-1} &= - \twistrho^{-1} \partial_{i} \left( \twistrho^{\frac{7}{9}} e^\twistsigma \mathsf{u}^{-1\, I}{}_K \mathsf{u}^{-1\, J}{}_L\right) \langle 1/3|^{0i} \,.
\end{align}

For the standard embedding tensor we work out the Maurer--Cartan derivative 
\begin{align}
\partial_i \cU \cU^{-1} = - \partial_i \mathsf{u}^I{}_K \mathsf{u}^{-1\, K}{}_J \T^J_{0\, I} - \partial_i\twistsigma \, \dK - \twistrho^{-1} \partial_i\twistrho \, \LL_0\,.
\end{align}

The expression~\eqref{eq:theta9} in the body of the paper is then computed as follows
\begin{align}
\label{eq:theta9app}
\langle \theta | &= \twistrho^{-\frac{2}{9}} e^\twistsigma \mathsf{u}^{-1\, 0}{}_K \mathsf{u}^{-1\, i}{}_L \partial_{i} \mathsf{u}^S{}_P \mathsf{u}^{-1\, P}{}_R \langle 1/3|^{KL} \T^R_{1\, S}  \nn\\
&\quad + \twistrho^{-\frac{11}{9}} e^\twistsigma   \mathsf{u}^{-1\, 0}{}_K \mathsf{u}^{-1\, i}{}_L\partial_{i} \twistrho \langle 1/3|^{KL} \LL_1 - \langle W^+ |\nn\\
&=\twistrho^{-\frac{2}{9}}  e^\twistsigma \mathsf{u}^{-1\, 0}{}_K \mathsf{u}^{-1\, i}{}_L \partial_{i} \mathsf{u}^S{}_P \mathsf{u}^{-1\, P}{}_R  \left( \langle 1/3|^{[KL}_{\phantom{[}} \T^{R]}_{1\, S} - \frac27  \langle 1/3|^{Q[K}_{\phantom{[}}  \T^{L\phantom{]}}_{1\, Q} \delta^{R]}_S \right)\nn\\
&\quad + \frac18  \twistrho^{-\frac{2}{9}}  e^\twistsigma \bigl(  \mathsf{u}^{-1\, 0}{}_K \partial_{i} \mathsf{u}^{-1\, i}{}_R - \mathsf{u}^{-1\, i}{}_K \partial_{i} \mathsf{u}^{-1\, 0}{}_R\bigr)  \langle 1/3|^{P(K}_{\phantom{]}} \T^{R)}_{1\, P} 
\nn\\
&\quad 
+\frac{9}{7}\twistrho^{-\frac{2}{9}}  e^\twistsigma \mathsf{u}^{-1\, [0}{}_K \partial_{i} \mathsf{u}^{-1\, i]}{}_R \langle 1/3|^{KR} \LL_1
+\twistrho^{-\frac{11}{9}} e^\twistsigma   \mathsf{u}^{-1\, 0}{}_K \mathsf{u}^{-1\, i}{}_L\partial_{i} \twistrho \langle 1/3|^{KL} \LL_1  - \langle W^+ |  \nn\\
&=\twistrho^{-\frac{2}{9}} e^\twistsigma \mathsf{u}^{-1\, 0}{}_K \mathsf{u}^{-1\, i}{}_L \partial_{i} \mathsf{u}^S{}_P \mathsf{u}^{-1\, P}{}_R  \left( \langle 1/3|^{[KL}_{\phantom{[}} \T^{R]}_{1\, S} - \frac27  \langle 1/3|^{Q[K}_{\phantom{[}}  \T^{L\phantom{]}}_{1\, Q} \delta^{R]}_S \right)\nn\\
&\quad +\frac18 \twistrho^{-\frac{2}{9}}  e^\twistsigma  \left(   \mathsf{u}^{-1\, 0}{}_K \partial_{i} \mathsf{u}^{-1\, i}{}_L-  \mathsf{u}^{-1\, i}{}_K \partial_{i} \mathsf{u}^{-1\, 0}{}_L - W^+_{00} \, \mathsf{u}^{-1\,0}{}_K \mathsf{u}^{-1\, 0}{}_L \right)  \langle 1/3|^{P(K}_{\phantom{]}} \T^{L)}_{1\, P} \nn\\
&\quad +\frac{9}{14} \twistrho^{-16/9} e^\twistsigma   \partial_{i} \left( \twistrho^{14/9}  \mathsf{u}^{-1\, 0}{}_K \mathsf{u}^{-1\, i}{}_L \right)\langle 1/3|^{KL} \LL_1  
\,,
\end{align}
where we used~\eqref{eq:shifrelt} and the  transformation
\begin{align}
\langle 1/3|^{IJ} \T^K_{0\, L} = 2 \delta_L^{[I} \langle 1/3|^{J]K} + \frac29 \delta^K_L \langle 1/3 |^{IJ}\,.
\end{align}
We have also specialised  $\langle w^+| = W^+_{00} \langle 4/3|^{00}$ and the terms in $\T^R_{1\, S}$ were rewritten using the decomposition into irreducible SL(9) representations 
\begin{align}
& \quad \langle 1/3|^{KL} \T^R_{1\, S} \nn\\*
&= \left[  \langle \tfrac13|^{KL} \T^R_{1\, S} -  \langle \tfrac13|^{[KL}_{\phantom{[}} \T^{R]}_{1\, S} + \frac14 \delta_S^{[K} \langle \tfrac13|^{L]P}_{\phantom{[}} \T^{R\phantom{]}}_{1\, P} -\frac1{40} \delta_S^R \langle \frac13|^{P[K}_{\phantom{[}} \T^{L]}_{1\, P} + \frac{11}{40} \langle \tfrac13|^{P[K}_{\phantom{[}} \T^{L\phantom{[}}_{1\, P}\delta^{R]}_S 
\right]_{\bf 2079}\nn\\*
&\quad + \left[  \langle \tfrac13|^{[KL}_{\phantom{[}} \T^{R]}_{1\, S} -\frac27  \langle \tfrac13|^{P[K}_{\phantom{[}}  \T^{L\phantom{]}}_{1\, P} \delta^{R]}_S \right]_{\bf 720} +\frac18 \left[  \delta_S^K \langle \tfrac13|^{P(L}_{\phantom{[}} \T^{R)}_{1\, P} -  \delta_S^L \langle \tfrac13|^{P(K}_{\phantom{[}} \T^{R)}_{1\, P} \right]_{\bf 45}\nn\\*
&\quad  + \frac9{70} \left[  \delta_S^K \langle \tfrac13|^{P[L}_{\phantom{[}} \T^{R]}_{1\, P} -  \delta_S^L \langle \tfrac13|^{P[K}_{\phantom{[}} \T^{R]}_{1\, P} +\frac29 \delta^R_S \langle \tfrac13|^{P[K}_{\phantom{]}} \T^{L]}_{1\, P}\right]_{\bf 36}\,,
\end{align}
where the subscripts are  written to highlight the irreducible representations associated to each projection. 
The module ${\bf 2079}=R(\lambda_1{+}\lambda_2{+}\lambda_8)$ is absent in the basic module due to 
\begin{align}
{\bf 45} \oplus {\bf 36} \oplus {\bf 720} = {\bf 45} \oplus\overline{\bf 9} \otimes {\bf 84}\,,
\end{align}
and~\eqref{eq:flowdec}, 
 and therefore does not appear in~\eqref{eq:theta9app}. For the ${\bf 36}$ we have used~\eqref{eq:36rel} when simplifying $\langle \theta|$, together with the fact that $\det \mathsf{u}=1$.


\section{Exceptional field theory conventions in eleven dimensions}
\label{SignConvention}
In order to fix the sign conventions for the exceptional field theory formulation of eleven dimensional supergravity, it is useful to compute the four-form field strength. Using the Kaluza--Klein ansatz \eqref{ThreeForm} for the three-form potential, one obtains for the four-form field strength 
\bea   F^{\scalebox{0.6}{11D}} \hspace{-2mm}&=& \hspace{-2mm} \frac{1}{24} f_{\Is\Js\Ks\Ls} ( \dd y^\Is +\widetilde{\langle 0|}{}^\Is |\mathcal{A}\rangle)  \wedge ( \dd y^\Js +\widetilde{\langle 0|}{}^\Js |\mathcal{A}\rangle) \wedge ( \dd y^\Ks +\widetilde{\langle 0|}{}^\Ks |\mathcal{A}\rangle) \wedge ( \dd y^\Ls +\widetilde{\langle 0|}{}^\Ls |\mathcal{A}\rangle)  \CR
  && + \frac16 D \alpha_{\Is\Js\Ks}  \wedge ( \dd y^\Is +\widetilde{\langle 0|}{}^\Is |\mathcal{A}\rangle)  \wedge ( \dd y^\Js +\widetilde{\langle 0|}{}^\Js |\mathcal{A}\rangle) \wedge ( \dd y^\Ks +\widetilde{\langle 0|}{}^\Ks |\mathcal{A}\rangle)  \CR
&&  + \frac12 \bigl( \mathcal{F}_{\Is\Js} + \alpha_{\Is\Js\Ks} \mathcal{F}^\Ks)   \wedge  ( \dd y^\Is +\widetilde{\langle 0|}{}^\Is |\mathcal{A}\rangle)  \wedge ( \dd y^\Js +\widetilde{\langle 0|}{}^\Js |\mathcal{A}\rangle)   \; ,   \eea
where
\bea f_{\Is\Js\Ks\Ls}  \hspace{-2mm}&=& \hspace{-2mm}  4 \partial_{[\Is} \alpha_{\Js\Ks\Ls]} \CR
D \alpha_{\Is\Js\Ks}  \hspace{-2mm}&=& \hspace{-2mm}  \dd \alpha_{\Is\Js\Ks} - 4  \widetilde{\langle 0|}{}^\Ls |\mathcal{A}\rangle \partial_{[\Ls} \alpha_{\Is\Js\Ks]} - 3 \partial_{[\Is} \bigl(  \widetilde{\langle 0|}{}^\Ls |\mathcal{A}\rangle \alpha_{\Js\Ks]\Ls} \bigr) -3 \partial_{[\Is} \widetilde{\langle 1/3|}{}_{\Js\Ks]}  |\mathcal{A}\rangle  \CR
\mathcal{F}_{\Is\Js}    \hspace{-2mm}&=& \hspace{-2mm}  \dd   \widetilde{\langle 1/3|}{}_{\Is\Js}  |\mathcal{A}\rangle - 3 \widetilde{\langle 0|}{}^\Ks |\mathcal{A}\rangle \wedge  \partial_{[\Is}   \widetilde{\langle 1/3|}{}_{\Js\Ks]}  |\mathcal{A}\rangle +2 \partial_{[\Is} \bigl( \widetilde{\langle 0|}{}^\Ks |\mathcal{A}\rangle \wedge     \widetilde{\langle 1/3|}{}_{\Js]\Ks}  |\mathcal{A}\rangle\bigr)\CR
 && + 4\partial_{[\Is}  \widetilde{\langle 0|}{}^\Ks  | \mathcal{C} \rangle \widetilde{\langle 1/3|}{}_{\Js]\Ks}  |\mathcal{C}\rangle  \CR
 \mathcal{F}^\Is  \hspace{-2mm}&=& \hspace{-2mm}  \dd \widetilde{\langle 0|}{}^\Is |\mathcal{A}\rangle -\widetilde{\langle 0|}{}^\Js |\mathcal{A}\rangle \partial_\Js \widetilde{\langle 0|}{}^\Is |\mathcal{A}\rangle\; .  \eea
The matching of the Kaluza--Klein ansatz with the exceptional field theory parametrisation is fixed such that the covariant derivative and the field strengths are compatible. One computes using \eqref{Vtilde11D} that the covariant derivative in exceptional field theory gives \footnote{Using $D \mathcal{V} =\dd \mathcal{V} - \widetilde{\langle 0|}{}^{\Is} |\mathcal{A}\rangle \partial_{\Is} \mathcal{V} - \eta_{\alpha\beta}  \widetilde{\langle 0|}{}^{\Is}  T^\alpha \partial_{\Is} |\mathcal{A}\rangle \mathcal{V} T^\beta + h \mathcal{V}$, where $h$ is the local $K(\mf{e}_9)$ compensating transformation.}
\be D\alpha_{\Is\Js\Ks} =  \dd \alpha_{\Is\Js\Ks} - \widetilde{\langle 0|}{}^{\Ls} |\mathcal{A}\rangle \partial_{\Ls} \alpha_{\Is\Js\Ks} +3 \partial_{\tilde{P}}  \widetilde{\langle 0|}{}^{\tilde{P} }  \T_0^{\; \Ls}{}_{[\Is} |\mathcal{A}\rangle\alpha_{\Js\Ks]\Ls} - \frac13 \partial_{\Ls}  \widetilde{\langle 0|}{}^{\Ls}  |\mathcal{A}\rangle\alpha_{\Is\Js\Ks}- \partial_{\Ls}  \widetilde{\langle 0|}{}^{\Ls}  \T_{\frac13\, \Is\Js\Ks } |\mathcal{A}\rangle \ee
Note in particular that this determines the sign of $\alpha_{\Is\Js\Ks}$ in  \eqref{Vtilde11D} for a fixed sign of $ \widetilde{\langle 1/3|}{}_{\Is\Js} |\mathcal{A}\rangle$. 
The two-form ansatz is justified by checking that it matches the exceptional contribution in  the field strength
 \bea \mathcal{F}_{\Is\Js} \hspace{-2mm} &=& \hspace{-2mm}  \widetilde{\langle 1/3|}{}_{\Is\Js} \Bigl( \dd |\mathcal{A}\rangle - \tfrac12 \widetilde{\langle 0|}{}^\Ks |\mathcal{A}\rangle \partial_\Ks  |\mathcal{A}\rangle  + \tfrac12 \eta_{\alpha\beta} \widetilde{\langle 0|}{}^\Ks  T^\alpha  \partial_\Ks |\mathcal{A}\rangle  T^\beta  |\mathcal{A}\rangle + \tfrac12  \widetilde{\langle 0|}{}^\Ks \partial_\Ks |\mathcal{A}\rangle   |\mathcal{A}\rangle  \CR
 &&   \eta_{\alpha\beta} \partial_\Ks \widetilde{\langle 0|}{}^\Ks T^\alpha | \mathcal{C} \rangle   T^\beta |\mathcal{C}\rangle + \eta_{\alpha\beta}  \widetilde{\langle 0|}{}^\Ks T^\alpha | \mathcal{C}_{[1}  \rangle  T^\beta |\mathcal{C}_{2] \Ks}\rangle + 2  \widetilde{\langle 0|}{}^\Ks  | \mathcal{C}_{[1}  \rangle   |\mathcal{C}_{2] \Ks}\rangle   \CR
&& + \eta_{-1\, \alpha\beta}  \widetilde{\langle 0|}{}^\Ks T^\alpha | \mathcal{C}^+_{1}  \rangle  T^\beta |\mathcal{C}^+_{2 \Ks}\rangle   \Bigr) \CR 
&=& \dd   \widetilde{\langle 1/3|}{}_{\Is\Js}  |\mathcal{A}\rangle - 3 \widetilde{\langle 0|}{}^\Ks |\mathcal{A}\rangle \wedge  \partial_{[\Is}   \widetilde{\langle 1/3|}{}_{\Js\Ks]}  |\mathcal{A}\rangle +2 \partial_{[\Is} \widetilde{\langle 0|}{}^\Ks |\mathcal{A}\rangle \wedge     \widetilde{\langle 1/3|}{}_{\Js]\Ks}  |\mathcal{A}\rangle \CR
 && + \partial_{[\Is} \Bigl( 4 \widetilde{\langle 0|}{}^\Ks  | \mathcal{C} \rangle \widetilde{\langle 1/3|}{}_{\Js]\Ks}  |\mathcal{C}\rangle + \widetilde{\langle 0|}{}^\Ks  | \mathcal{A} \rangle \widetilde{\langle 1/3|}{}_{\Js]\Ks}  |\mathcal{A}\rangle \Bigr) \; ,  \eea
 and
 \bea \mathcal{F}^{\Is} \hspace{-2mm} &=& \hspace{-2mm}  \widetilde{\langle 0|}{}^{\Is} \Bigl( \dd |\mathcal{A}\rangle - \tfrac12 \widetilde{\langle 0|}{}^\Ks |\mathcal{A}\rangle \partial_\Ks  |\mathcal{A}\rangle  + \tfrac12 \eta_{\alpha\beta} \widetilde{\langle 0|}{}^\Ks  T^\alpha  \partial_\Ks |\mathcal{A}\rangle  T^\beta  |\mathcal{A}\rangle + \tfrac12  \widetilde{\langle 0|}{}^\Ks \partial_\Ks |\mathcal{A}\rangle   |\mathcal{A}\rangle  \CR
 &&   \eta_{\alpha\beta} \partial_\Ks \widetilde{\langle 0|}{}^\Ks T^\alpha | \mathcal{C} \rangle   T^\beta |\mathcal{C}\rangle + \eta_{\alpha\beta}  \widetilde{\langle 0|}{}^\Ks T^\alpha | \mathcal{C}_{[1}  \rangle  T^\beta |\mathcal{C}_{2] \Ks}\rangle + 2  \widetilde{\langle 0|}{}^\Ks  | \mathcal{C}_{[1}  \rangle   |\mathcal{C}_{2] \Ks}\rangle   \CR
&& + \eta_{-1\, \alpha\beta}  \widetilde{\langle 0|}{}^\Ks T^\alpha | \mathcal{C}^+_{1}  \rangle  T^\beta |\mathcal{C}^+_{2 \Ks}\rangle   \Bigr) \CR 
&=&  \dd \widetilde{\langle 0|}{}^\Is |\mathcal{A}\rangle -\widetilde{\langle 0|}{}^\Js |\mathcal{A}\rangle \partial_\Js \widetilde{\langle 0|}{}^\Is |\mathcal{A}\rangle\; .    \eea
Note that there is here a  redefinition of the 2-form.


\section{Gauge invariance and uplift formul\ae}
\label{sec:gaugeinv}

In order to understand the dependence of the uplift ansatz on the pure gauge fields $b_{IJK}$ and $\delta_{K(I} h^K{}_{J)}$, it is useful to consider the corresponding gauge transformations in eleven-dimensional supergravity and gauged supergravity. For this purpose let us recall the gauge transformation in gauged supergravity 
\bea 
\delta | A\rangle \hspace{-2mm} &=& \hspace{-2mm}  \dd |\lambda\rangle + \eta_{-1\alpha\beta} \langle \theta | \T^\alpha |\lambda\rangle \T^\beta |A\rangle + \eta_{-1\alpha\beta} \langle \theta | \T^\alpha \otimes T^\beta  |\hspace{-0.5mm}| \Sigma \rangle\hspace{-1mm}\rangle \; , \label{GaugeTExplicit}  \\*
\delta   |\hspace{-0.5mm}| C \rangle\hspace{-1mm}\rangle  \hspace{-2mm} &=& \hspace{-2mm}  \eta_{-1\alpha\beta} \langle \theta | \T^\alpha | \lambda\rangle \bigl( \mathds{1}\otimes \T^\beta + \T^\beta\otimes \mathds{1}\bigr)   |\hspace{-0.5mm}| C \rangle\hspace{-1mm}\rangle  + \frac{1}{2} \dd |\lambda\rangle \otimes \wedge |A\rangle - \frac{1}{2}  |A\rangle \otimes \wedge \dd  |\lambda \rangle \CR
&&  + \dd  |\hspace{-0.5mm}| \Sigma \rangle\hspace{-1mm}\rangle  - \frac{1}{2} \eta_{-1\alpha\beta} |A\rangle \otimes   \wedge \langle \theta | \T^\alpha \otimes T^\beta  |\hspace{-0.5mm}| \Sigma \rangle\hspace{-1mm}\rangle + \frac12  \eta_{-1\alpha\beta} \langle \theta | \T^\alpha \otimes T^\beta  |\hspace{-0.5mm}| \Sigma \rangle\hspace{-1mm}\rangle \otimes \wedge |A\rangle \; , \nn 
\eea
where both the two-form $|\hspace{-0.5mm}| C \rangle\hspace{-1mm}\rangle$ and one-form gauge parameter $|\hspace{-0.5mm}| \Sigma \rangle\hspace{-1mm}\rangle$ are in the symmetric tensor product of two copies of the basic module.  One can redefine $|\hspace{-0.5mm}| \Sigma \rangle\hspace{-1mm}\rangle$ such that the gauge transformation of the gauge field becomes a covariant derivative, but both forms will be  useful in this  section.

The following gauge transformation of the three-form in eleven dimensions 
\bea \delta A^{\scalebox{0.6}{11D}} \hspace{-2mm} &=& \hspace{-2mm} \dd \Bigl( \frac \coup 2 \lambda^K_{IJ} Y_K \partial_i Y^I \partial_j Y^J ( {\rm d}y^i +\mathcal{A}^i ) \wedge  ( {\rm d}y^j + \mathcal{A}^j )\Bigr)  \CR
\hspace{-2mm} &=& \hspace{-2mm} \frac \coup 2 \delta_{L[K}  \lambda^L_{IJ]} \partial_i Y^I \partial_j Y^J \partial_k Y^K( {\rm d}y^i + \mathcal{A}^i ) \wedge  ( {\rm d}y^j + \mathcal{A}^j ) \wedge ( {\rm d}y^k + \mathcal{A}^k )  \CR
&& +\frac \coup2 Y^K \partial_i Y^I \partial_j Y^J D \lambda_{IJ}^K  \wedge  ( {\rm d}y^i + \mathcal{A}^i ) \wedge  ( {\rm d}y^j +\mathcal{A}^j )\CR
&& -\coup^2 \lambda_{IJ}^K \langle 1/3|^{IL} |F\rangle   Y_K Y_L \partial_i Y^J \wedge  ( {\rm d}y^i +\mathcal{A}^i )    \eea
with
\bea D \lambda_{IJ}^K &=&\dd\lambda_{IJ}^K - \coup \delta_{PQ} \langle 1/3|^{KP} |A\rangle \lambda^Q_{IJ} +2 \coup \delta_{P[I} \langle 1/3|^{PQ} |A\rangle \lambda^K_{J]Q} \; , \CR
\langle 1/3|^{IJ} |F\rangle &=&  \langle 1/3|^{IJ} |\dd A\rangle+\coup \delta_{KL}  \langle 1/3|^{IK} |A\rangle \wedge \langle 1/3|^{JL} |A\rangle \; ,  \eea
is equivalent to the gauged supergravity  gauge transformation defined as a covariant derivative 
\bea \delta b_{IJK} &=& -3 \Theta_{L[I} \lambda_{JK]}^L \CR
\delta \langle 1|^K_{IJ}  |A\rangle  &=& \dd \lambda^K_{IJ} - \Theta_{PQ} \langle 1/3|^{KP} |A\rangle \lambda_{IJ}^Q + 2 \Theta_{P[I} \langle 1/3|^{PQ} |A\rangle \lambda_{J]Q}^K   \eea
of parameter  
\be \lambda^{K}_{IJ} = \langle 1|^K_{IJ} |\lambda\rangle \; . \ee
Note moreover that the four-form field strength in eleven dimensions only depends on the field $b_{IJK}$ through its covariant derivative and the linear combination 
\be \langle 1|^K_{IJ} |F\rangle - b_{IJL} \langle 1/3|^{KL} |F\rangle \ee
for which the right-hand side of \eqref{F1dual} and \eqref{FSO9duality} does not depend on the field $b_{IJK}$, as one sees in \eqref{F1DualityRight}.

We find therefore that $b_{IJK}$ can consistently be gauged away both in eleven-dimensional supergravity and gauged supergravity.

One similarly exhibits that $h^I{}_J$ only appears non-trivially in the gauged supergravity Lagrangian through it antisymmetric component $\Theta_{K[I} h^K{}_{J]}$, while its symmetric component is pure gauge. To see this, note that one can use a diffeomorphism along the circle coordinate 
\be \y^9 \rightarrow y^9 - \coup Y_I Y_J \xi^{IJ}(x)  \label{Diffxi} \ee
for a symmetric tensor $\xi^{IJ}$ function of the external coordinates. This diffeomorphism only affects the fibre one-form as
\bea  {\rm d}y^9 + \mathcal{A}^{9} + K_i ({\rm d}y^i+\mathcal{A}^i)  \hspace{-2mm} &=& \hspace{-2mm} {\rm d}y^9 + \coup Y_I Y_J \langle 4/3|^{IJ} |A\rangle + K_i \bigl( {\rm d}y^i +\coup Y_I \mathring g^{ij} \partial_j Y_J \langle 1/3|^{IJ} |A\rangle \bigr) \CR
\hspace{-2mm} &\rightarrow & \hspace{-2mm} {\rm d}y^9 + \coup Y_I Y_J \bigl( \langle 4/3|^{IJ} |A\rangle -\dd \xi^{IJ} -2 \coup \delta_{KL}  \langle 1/3|^{K(I} |A\rangle  \xi^{J)L} \bigr) \CR
&& + \bigl( K_i  - 2 \coup Y_I \partial_i Y_J  \xi^{IJ} \bigr) \bigl( {\rm d}y^i +\coup Y_K \mathring g^{ij} \partial_j Y_L \langle 1/3|^{KL} |A\rangle \bigr) \; .  \eea
Using the metric ansatz one obtains
\bea 
&& \rho^{-\frac{8}{9}}   e^{2\varsigma} G^{ij} \bigl( K_j - 2 \coup Y_I \partial_j Y_J  \xi^{IJ} \bigr)\nn\\
&=& -\coup^2  \sugrarho^{-\frac{16}{9}} (\det \mathring g)^{\frac{1}{9}} Y_{I}  Y_J Y_{K} \mathring g^{ij} \partial_j Y_L\bigl(  \langle 4/3|^{IJ} M^{-1} \, {}^{KL}|1/3\rangle+2 \coup \delta_{PQ}  \xi^{P(I} \langle 1/3|^{J)Q} M^{-1} \, {}^{KL}|1/3\rangle \bigr) \nonumber \eea
such that this diffeomorphism is equivalent to the gauge transformation 
\bea \delta  \langle 4/3|^{IJ} |A\rangle &=& \dd \xi^{IJ} +2 \coup \delta_{KL}  \langle 1/3|^{K(I} |A\rangle  \xi^{J)L}\; ,  \CR
\delta  h^I{}_J &=&  2 \coup \delta_{JK} \xi^{IK}  - \frac{2}{9} \coup \delta_J^I \delta_{KL} \xi^{KL}\; . \eea
In gauge supergravity one can identify
\be \xi^{IJ} = \langle 4/3|^{IJ} |\lambda\rangle\; .  \ee
The trace component is not relevant to this discussion, therefore we assume $\Theta_{IJ} \xi^{IJ} = 0$ to simplify expressions. One then obtains from \eqref{GaugeTExplicit} the gauge transformation of the following fields as
\bea \delta h^I{}_J &=& 2 \Theta_{JK} \xi^{IK}\; ,  \CR
\delta \langle 4/3|^{IJ} |A\rangle &=& \dd\xi^{IJ} + 2 \Theta_{KL} \langle 1/3|^{K(I} |A\rangle \xi^{J)L} \; , \CR
\delta \langle 1|^K_{IJ}|A\rangle  &=& -4\Theta_{L[I} \langle 0|_{J]} |A\rangle \xi^{KL} + 4 \delta^K_{[I} \Theta_{J]P} \langle 0|_Q |A\rangle \xi^{PQ} \; .  \eea

With this transformation of $h^I{}_J$, one checks that 
\bea \delta  \langle 4/3|^{IJ} M^{-1} \, {}^{KL}|1/3\rangle&=& - 2\Theta_{PQ} \xi^{P(I}  \langle 1/3|^{J)Q} M^{-1} \, {}^{KL}|1/3\rangle \\
\delta  \langle 1|^{P}_{IJ} M^{-1} \, {}^{KL}|1/3\rangle&=&  -4  \xi^{PQ}  \Theta_{Q[I}  \langle 0|_{J]}  M^{-1} \, {}^{KL}|1/3\rangle+ 4   \delta^P_{[I} \Theta_{J]Q} \xi^{QR} \langle 0|_{R}  M^{-1} \, {}^{KL}|1/3\rangle  \nonumber \eea
such that the three-form component transforms under the associated diffeomorphism
\be \delta \mathcal{A}_{ij} = 4 \coup  \xi^{IJ} Y_I \partial_{[i} Y_J \mathcal{A}_{9 j]} \; , \quad \delta \alpha_{ijk} = 6  \coup \xi^{IJ} Y_I \partial_{[i} Y_J \alpha_{9 jk]}\; , \ee
 using \eqref{alphaIJ} and \eqref{alphaIJK}.

Let us finally note that this  gauge transformation acts on the field strength 
\be \delta \langle 4/3|^{IJ} |F\rangle =   2  \Theta_{KL} \langle 1/3|^{K(I} |F\rangle \xi^{J)L} \ee
consistently with the property that the derivative of the potential in the right-hand side of \eqref{FKKduality} is not gauge invariant, but transforms as 
\be \delta \frac{\partial V_{\rm gsugra}}{\partial \Theta^{IJ}} =   2  \Theta_{KL} \langle 1/3|^{K(I} M^{-1} |\theta \rangle  \xi^{J)L} \ee
so that \eqref{FKKduality} transforms into the Yang--Mills equation \eqref{FSO9duality}.

This completes the proof of equivalence between the diffeomorphism \eqref{Diffxi} and the gauged supergravity gauged transformation of parameter $\xi^{IJ}$. One can therefore gauge fix $\Theta_{K(I} h^K{}_{J)}$ to any convenient value in the equations of motion to determine the eleven-dimensional fields.

\section{Embedding tensors with uplift} 
\label{app:uplift}

It would be highly desirable to be able to classify the most general consistent truncations of ten- and eleven-dimensional maximal supergravity to gauged maximal supergravity in two dimensions. 
We may assume that all such truncations are necessarily generalised Scherk--Schwarz reductions, as seems plausible from the requirement that maximal supersymmetry must be preserved, see for instance \cite{Lee:2014mla,Cassani:2019vcl}.
Then, the problem can be roughly divided into two objectives: to classify all inequivalent embedding tensors of the two-dimensional theory admitting a gSS uplift and to explicitly identify the internal space and twist matrix for each case.

These are however extremely difficult tasks in general, that have not yet been completed for truncations to $D\ge3$ maximal supergravities.
Significant progress has been made in recent years for E$_{n}$ ExFTs with $n\le7$ \cite{duBosque:2017dfc,Inverso:2017lrz,Bugden:2021wxg,Bugden:2021nwl,Hulik:2022oyc}.
Necessary and sufficient constraints for an embedding tensor to admit an uplift have been identified \cite{Inverso:2017lrz,Bugden:2021wxg,Bugden:2021nwl,Hulik:2022oyc}, and a general construction procedure for the twist matrix -- assuming such constraints are satisfied -- was determined~\cite{Inverso:2017lrz}.
The classification of solutions of such constraints, up to duality orbits, is at the time of this writing an unsolved problem.
Duality covariant, necessary conditions for the existence of an uplift of a $D=3$ gauged maximal supergravity have recently been presented in \cite{Eloy:2023zzh} and  analogous set of necessary conditions for $D=2$ is presented in the companion paper~\cite{gss} (see equations (3.71), (3.72) there).

In this appendix we take a complementary point of view.
We impose that the embedding tensor must originate from a twist matrix satisfying the section constraint.
By fixing a solution of the section constraint, we then find which entries within the embedding tensor can actually be generated by projecting a putative Weitzenb\"ock connection through \eqref{eq:theta from W}.
A similar approach was recently taken in \cite{Hassler:2022egz} for $D\ge4$ supergravities.
We will show that any Lagrangian embedding tensor $\bra\theta$ admitting an uplift is only parametrised by \emph{finitely} many components, which we identify in equations~\eqref{eq:uplifttheta11d} and~\eqref{eq:upliftthetaIIB} below for uplifts to eleven-dimensional and IIB supergravities, respectively.
Notice that the conditions found in this way break the exceptional group to a parabolic subgroup (the one preserving the fixed choice of section) and therefore an embedding tensor with uplift is only required to match the ones identified with this procedure {up to the action of a rigid E$_9$ element}. 
We do not prove whether a twist matrix actually exists for the embedding tensors parametrised by~\eqref{eq:uplifttheta11d} and~\eqref{eq:upliftthetaIIB}.
We only consider reductions from either eleven-dimensional or type IIB supergravity, excluding for example the case of massive type IIA, since a Romans mass deformation of E$_9$ ExFT analogous to \cite{Ciceri:2016dmd} is not yet available.

\medskip

We will use a basis appropriate to the chosen solution of the section constraint.
For instance, for eleven-dimensional supergravity we use the basis \eqref{eq:bas1} and internal derivatives take the form \eqref{eq:secsol}.
For short, only in this appendix we drop all the tildes introduced in \eqref{eq:bas1} to distinguish the $p=1$ flowed basis from the $p=2$ one in \eqref{eq:bas2}.
We do the same for the generators $\T^\alpha$, which we always assume to be set in a basis adapted to the choice of section constraint (i.e. the $p=1$ flowed SL(9) basis for the eleven-dimensional supergravity section). Therefore in this section we shall write 
\be \langle \partial | = \langle 0 |^I \partial_I \ee
where $I=1$ to $9$ for eleven-dimensional supergravity, and $I=1$ to $8$ for type IIB supergravity. We assume that the appropriate basis is used also for type IIB supergravity, such that 
\be \langle 0|^I \LL_{-n} = 0 \; , \qquad \forall n \ge 1 \; . \label{L0Section} \ee
The IIB basis will be further described below.

\medskip

The approach followed in this appendix is made possible by an observation on the general form of the twist matrix.
One first notices that the internal space must be a homogeneous space $G/H$ where $G$ is the gauged supergravity gauge group and $H$ some subgroup \cite{Grana:2008yw}. 
Notice that in $D=2$ both $G$ and $H$ are infinite-dimensional. 
Following the analysis of \cite{Inverso:2017lrz}, the internal vectors $\bra{0}^Ir^{-1}\cU^{-1} = \bra{k^I}$ generate the transitive action of $G$.
The ancillary gauge parameters appearing in E$_8$ and E$_9$ generalised diffeomorphisms do not affect this observation, hence we can carry over any conclusions from \cite{Inverso:2017lrz} which only rely on this observation.
In particular, it implies that the twist matrix always decomposes as
\be\label{eq:twistdeco} 
\mathcal{U} = L \mathcal{E}^{-1} 
\ee
where $L(y) \in G$ is the coset representative of $G/H$, while $\mathcal{E}$ belongs to the parabolic subgroup preserving the choice of solution of the section constraints, i.e. only includes generators of non-positive mode number with respect to the Virasoro generator $\LL_{0}$ satisfying \eqref{L0Section}.
Explicitly,
\be
\bra0^I\cE = g^I{}_J \bra0^J\,,
\ee
with $g^I{}_J$ a GL$(d)$ element.
Because the embedding tensor $\langle \theta|$  is gauge invariant, one then has
\be \langle \theta|\mathcal{U} = \langle \theta | \mathcal{E}^{-1} \; . \label{thetaU=thetaE} \ee
This is important because it will be easier to constrain the components that can be non-vanishing in $ \langle \theta|\mathcal{U}$ and the action of  $\mathcal{E}$ preserves the highest possible $\LL_0$ degree. 

\medskip

It will be convenient to define the conjugate Weitzenb\"ock connection 
\be \langle \widetilde{\mathcal{W}}_\alpha | \otimes \T^\alpha = \langle \mathcal{W}_\alpha | r \mathcal{U} \otimes \mathcal{U}^{-1} \T^\alpha \mathcal{U} =  \langle e^M | \otimes \mathcal{U}^{-1} \partial_M \mathcal{U}\; . \ee
We shall assume that $\mathcal{U} \in {\rm E}_9$. The most general case can be analysed similarly, but requires slightly heavier notation, so we shall refrain from writing it. 
With this definition 
\be \langle \vartheta | =  \langle \widetilde{\mathcal{W}}_\alpha | \T^\alpha r^{-1} \mathcal{U}^{-1} \; , \qquad  \langle \theta | =  -\langle \widetilde{\mathcal{W}}_\alpha | \mathsf{S}_1(\T^\alpha)  r^{-2} \mathcal{U}^{-1}- \langle \widetilde{W}^+ | r^{-2} \mathcal{U}^{-1} \; , \ee
where 
\be  \langle \widetilde{W}^+ |  = r^{-1} \langle {W}^+ |- \omega_{-1}^\alpha(\mathcal{U}) \langle \widetilde{\mathcal{W}}_\alpha |\; .  \ee 
Because $r^{-1} \mathcal{U}^{-1}$ is invertible, the condition $\langle \vartheta|=0$ implies 
\be \langle \widetilde{\mathcal{W}}_\alpha | \T^\alpha  = 0 \ee
pointwise. As we shall see, this equation severely constrains the possible non-zero components of $\langle \theta |r^2 \mathcal{U}$ and in turn $ \langle \theta| $.

\subsection{Eleven-dimensional supergravity}

Let us start with eleven-dimensional supergravity. One uses therefore the GL$(9)$  decomposition,
\begin{align}
\overline{R(\Lambda_0)_{-1}} &= {\bf 9}_{\frac{4}{9}} \oplus \overline{\bf 36}_{\frac{7}{9}} \oplus {\bf 126}_{\frac{10}9} \oplus ({\bf 36} \otimes \overline{\bf 9})_{\frac{13}{9}} \oplus \left( {\bf 9}\otimes \overline{\bf 84}   \oplus \overline{\bf 45} \right)_\frac{16}{9} \nn\\
&\quad \oplus ({\bf 9} \otimes {\bf 84} \oplus \overline{\bf 1008} )_{\frac{19}{9}} \oplus \ldots,
\end{align}
and we recall that we drop for short all tildes in \eqref{eq:bas1} in this appendix. The conjugate Weitzenb\"ock connection then takes the form
\begin{multline} \langle  \widetilde{\mathcal{W}}_{\alpha} | \otimes \T^\alpha  =  \langle 0|^I \otimes \sum_{n} \Bigl(  \widetilde{W}^\ord{n}_{I}{}^J{}_K \T_n^K{}_J +\tfrac16 \widetilde{W}^\ord{n+\frac13}_{I}{}^{JKL} \T_{n+\frac13\, JKL} + \tfrac16 \widetilde{W}^\ord{n+\frac23}_{I; JKL} \T_{n+\frac23}^{JKL} \Bigr) \\
+ \widetilde{W}_I^0  \langle 0|^I \otimes \LL_0  +\widetilde{W}_I^\dK  \langle 0|^I \otimes \dK\; ,  \end{multline}
where the index $I$ labels the derivative along the eleven coordinates, while the adjoint indices are labeled with the indices $JKL$. The condition that there is no trombone gives therefore 
\bea 0 \hspace{-2mm} &\overset{!}{=}&\hspace{-2mm} \langle  \widetilde{\mathcal{W}}_{\alpha} | \T^\alpha \CR
\hspace{-2mm}& =&\hspace{-2mm} \langle 0|^I \bigl( \tfrac49 \widetilde{W}_I^0 +\widetilde{W}_I^\dK - \widetilde{W}^\ord{n}_{J}{}^J{}_I \bigr) + \frac12 \langle 1/3|_{IJ}  \widetilde{W}_K^\ord{\frac13}{}^{IJK} + \frac16  \langle 2/3|^{IJKL}  \widetilde{W}^\ord{\frac23}_{I; JKL} + \langle 1|^{IJ}_K \widetilde{W}^\ord{1}_{I}{}^K{}_J \T_n^K{}_J \CR
&& +\sum_{n\ge 1} \langle 0|^I \Bigl(   \tfrac16 \widetilde{W}^\ord{n+\frac13}_{I}{}^{JKL} \T_{n+\frac13\, JKL} + \tfrac16 \widetilde{W}^\ord{n+\frac23}_{I; JKL} \T_{n+\frac23}^{JKL} + \widetilde{W}^\ord{n+1}_{I}{}^J{}_K \T_n^K{}_J  \Bigr)  \;  .  \eea
This equation must be true for each basis element separately, and one finds that it sets to zero all the components of $  \widetilde{\mathcal{W}}_{\alpha}$ with $n\ge 4/3$. To prove this we compute the relations 
\bea \langle 0 |^I \T_{\frac13 + n\, JKL} \T_{-\frac13 -n}^{PQR}  &=& 18 \delta^{I[PQ}_{JKL} \langle 0|^{R]} + 6 n \delta_{JKL}^{PQR} \langle 0 |^I \CR
\langle 0 |^I \T_{\frac23 + n}^{JKL} \T_{-\frac23 -n\, PQR} &=& - 24 \delta_{PQR}^{[IJK} \langle 0 |^{L]} + 6 n \delta_{PQR}^{JKL} \langle 0|^I \CR
\langle 0 |^I \T_{1+ n}^{\hspace{4mm} J}{}_K \T_{-1 -n}^{\hspace{5mm}  P}{}_Q &=& - 2 \delta_K^P \delta_Q^{[I} \langle 0 |^{J]} + \delta_K^I \delta_Q^J \langle 0 |^P + n \delta_K^P \delta_Q^J \langle 0 |^I - \frac{n+1}{9} \delta_K^J \delta_Q^P \langle 0 |^I \; .  \eea
One finds from these formulas that none of the components of $ \langle 0 |^I \T_{\frac13 + n\, JKL}$, $\langle 0 |^I \T_{\frac23 + n}^{JKL} $ and $\langle 0 |^I \T_{1+ n}^{\hspace{4mm} J}{}_K$ vanish for $n\ge 1$, because their norm square is strictly positive. We conclude therefore that 
\be \widetilde{W}^\ord{n+\frac13}_{I}{}^{JKL} = 0 \; , \quad  \widetilde{W}^\ord{n+\frac23}_{I; JKL} \; , \quad \widetilde{W}^\ord{n+1}_{I}{}^J{}_K = 0 \; , \quad \forall n\ge 1 \ee
and
\be \widetilde{W}^\ord{\frac13}_K{}^{IJK} = 0 \; , \quad \widetilde{W}^\ord{\frac23}_{[I;JKL]}=0\; , \quad \widetilde{W}^\ord{1}_{[I\hspace{4mm} K]}{}^{\hspace{-5.5mm}J\hspace{3mm} }=0\; ,\quad \widetilde{W}_I^\dK=\widetilde{W}^\ord{0}_J{}^J{}_I - \tfrac49 \widetilde{W}^0_{I} \; .  \ee

From these constraints one gets immediately that $ \langle \widetilde{\mathcal{W}}_\alpha | \mathsf{S}_{-n}(\T^\alpha)   = 0 $ for $n\ge 1$ and that the embedding tensor can be written as
\begin{multline}  \langle \theta |r^2 \mathcal{U}  = \langle 0|^I \widetilde{\Theta}^\ord{\frac{4}{9}}_I + \frac12 \langle 1/3|_{IJ} \widetilde{\Theta}^{\ord{\frac{7}{9}} IJ} + \frac1{24} \langle 2/3|^{IJKL} \widetilde{\Theta}^\ord{\frac{10}{9}}_{IJKL} +\frac12  \langle 1|^{IJ}_K \widetilde{\Theta}_{IJ}^{\ord{\frac{13}{9}} K}\\ - \frac16 \langle 4/3|^L_{IJK}   \widetilde{W}^\ord{\frac13}_{L}{}^{IJK}- \frac16  \langle 5/3|^{I,JKL}   \widetilde{W}^\ord{\frac23}_{I;JKL} - \langle 2|^{IJ}_K \widetilde{W}^{\ord{1} K}_{I\hspace{5mm} J} \end{multline}
where the first four components are a priori generic while one defines $\langle 4/3|^L_{IJK} $, $\langle 5/3|^{I,JKL}$ and $\langle 2|^{IJ}_K$ in the corresponding irreducible SL(9) representations  ${\bf 720}$, $\overline{\bf 630}$ and $\overline{\bf 396}$, i.e. 
\be  \langle 2|^{IJ}_K =  \langle 2|^{JI}_K \; , \quad \langle 2|^{IJ}_J = 0 \; , \qquad \langle 5/3|^{[I,JKL]}= 0 \; , \qquad \langle 4/3|^K_{IJK} = 0 \; .  \ee
One checks that this structure is preserved by the action of the parabolic subgroup of negative $\LL_0$ degree. Indeed the two other irreducible representations $\overline{\bf 126}$ and ${\bf 1008}$ of degree $\frac{19}{9}$ cannot be obtained from the degree $\frac{22}{9}$ element in the $\overline{\bf 396}$, and the two other irreducible representations ${\bf 45}$ and ${\bf 36}$ of degree  $\frac{16}{9}$ cannot be obtained from the degree $\frac{19}{9}$ element in the $\overline{\bf 630}$ or the degree $\frac{22}{9}$ element in the $\overline{\bf 396}$. We conclude therefore from \eqref{thetaU=thetaE} that the embedding tensor admits the same expansion 
\begin{multline}  \langle \theta |  = \langle 0|^I {\Theta}^\ord{\frac{4}{9}}_I + \frac12 \langle 1/3|_{IJ} {\Theta}^{\ord{\frac{7}{9}} IJ} + \frac1{24} \langle 2/3|^{IJKL} {\Theta}^\ord{\frac{10}{9}}_{IJKL} +\frac12  \langle 1|^{IJ}_K {\Theta}_{IJ}^{\ord{\frac{13}{9}} K}\\ +\frac16 \langle 4/3|^L_{IJK}   {\Theta}^\ord{\frac{16}{9}}_{L}{}^{IJK}+ \frac16  \langle 5/3|^{I,JKL}   {\Theta}^\ord{\frac{19}{9}}_{I,JKL} + \langle 2|^{IJ}_K {\Theta}^{\ord{\frac{22}{9}} K}_{IJ} \label{eq:uplifttheta11d} \end{multline}
where the last three components are in the corresponding SL(9) irreducible representations.

To further constrain the components of the embedding tensor, we can use the quadratic constraint~\eqref{QuadraCons}.
For short we introduce the notation 
\be  {\Theta}^K_{IJ} \equiv {\Theta}^{\ord{\frac{22}{9}} K}_{IJ}  \; , \ee
for the component of main interest for us. The first components of the quadratic constraint give\footnote{Where we use $ \langle 2|^{IJ}_K \T_{-1}^P{}_Q = \delta_Q^{(I} \langle 1|^{J)P}_K - \tfrac1{10} \delta_K^{(I} \langle 1|^{J)P}_Q + \tfrac1{10} \delta_K^{(I} \delta^{J)}_Q  \langle 1|^{PL}_L \; .  $}
\be {\Theta}^P_{IJ} {\Theta}^Q_{KL} \Bigl(2 \langle 1 |_P^{IK} \otimes \langle 2|_Q^{LJ} - \delta_Q^I \langle 1 |_P^{JR} \otimes \langle 2|_R^{KL}  \Bigr) = 0 \; ,  \ee
and 
\be {\Theta}^P_{IJ} {\Theta}^Q_{KL} \Bigl(2\langle 4/3|_{PQR}^{I} \otimes \langle 5/3|^{K,LJR} - \delta_Q^I \langle 4/3|_{PRS}^{J} \otimes \langle 5/3|^{K,LRS} \Bigr)   = 0 \; , \ee
which imply the two equations 
\be {\Theta}^{[P}_{IJ} {\Theta}^{Q]}_{KL}= 0 \; , \qquad {\Theta}^P_{IJ} {\Theta}^Q_{KP} = 0 \; . \label{QuadraticTheta21}  
\ee
The first equation gives that ${\Theta}^P_{IJ}$ factorises in $u^P {\Theta}_{IJ}$  and one gets 
\be {\Theta}^P_{IJ} = u^P {\Theta}_{IJ} \; , \qquad u^J {\Theta}_{IJ} = 0 \; , \ee
such that $u^I$ defines a specific direction in SL$(9)$ and ${\Theta}_{IJ}$ is a symmetric tensor in the orthogonal subspace. Without loss of generality we can always choose coordinates such that $i=1$ to $8$ and
\be {\Theta}^k_{IJ} = 0\; , \qquad {\Theta}^K_{9J} = 0 \; ,  \ee
and the only non-vanishing components are ${\Theta}^9_{ij}$, which defines a symmetric matrix of rank $8-r\le 8$. 

The next constraint we get from \eqref{QuadraCons} is 
\be {\Theta}^\ord{\frac{19}{9}}_{K,L_1L_2L_3} {\Theta}^P_{IJ} \bigl( 2 \langle 2/3|^{L_1L_2L_3I} \otimes \langle 2|^{JK} - \delta_P^K \langle 2/3|^{L_1L_2L_3Q} \otimes \langle 2|^{IJ}_Q \bigr) = 0 \ee
which implies the two equations 
\be {\Theta}^\ord{\frac{19}{9}}_{K),[L_1L_2L_3} {\Theta}^P_{I](J}= 0 \; , \qquad {\Theta}^\ord{\frac{19}{9}}_{P,L_1L_2L_3} {\Theta}^P_{IJ} = 0 \; . \ee
One finds therefore that ${\Theta}^\ord{\frac{19}{9}}_{I,JKL}$ is orthogonal to the vector $u^I$ on its first index. One can write the general solution as the sum of two terms
\be {\Theta}^\ord{\frac{19}{9}}_{I,JKL} = {\Theta}_{I[J}^P \Lambda_{KL]P}  + {\Theta}_{I,JKL} \ee
where $\Lambda_{IJK}$ is an arbitrary antisymmetric tensor and ${\Theta}_{I,JKL}$ satisfies 
\be u^I {\Theta}_{I,JKL}= 0 \; , \quad  {\Theta}_{K,[L_1L_2L_3} {\Theta}^P_{L_4]J} = 0 \; . \ee
The term in $\Lambda_{IJK}$ can be absorbed in a E$_9$ transformation and can therefore be disregarded.  There is a non-trivial solution ${\Theta}_{I,JKL}$ to this equation if and only if ${\Theta}^9_{ij}$ has rank at most three, i.e. $r\ge 5$. To describe the solution it is convenient to split the indices $i=1$ to $8$, to $a=1$ to $8-r$ and $\hat{a}=8-r+1$ to $8$, such that ${\Theta}^9_{ab} $ is non-degenerate and the other components ${\Theta}^9_{a\hat{b}} = {\Theta}^9_{\hat{a}\hat{b}}=0 $.

$\bullet$ $r=5$ : The solution has ${\Theta}_{i,123}$ arbitrary and the other components vanish.

$\bullet$ $r=6$ : The solution has ${\Theta}_{i,12L}$ arbitrary and the other components vanish.

$\bullet$ $r=7$ : The solution has ${\Theta}_{i,1KL}$ arbitrary and the other components vanish.

$\bullet$ $r=8$ : ${\Theta}^K_{IJ}=0$ and there is no constraint in these components.

\medskip

At the next orders, the constraints become more and more complicated and we will not give the full solution. If one assumes that $\Theta_{ij}$ is maximal rank, one finds the unique solution 
 \be \langle \theta |  = \langle 0|^9 {\Theta}_{00} +   \langle 1|^{i9}_9 {\Theta}_{0i}+  \langle 2|^{ij}_9 {\Theta}_{ij} \; , \ee
 corresponding to the CSO$(p,q,r)$ gaugings discussed in this paper.

\subsection{Type IIB supergravity}

The basis appropriate to make the type IIB section constraint manifest corresponds to the  grading 
\be \mathfrak{e}_9 = \bigoplus_n \Bigl( ( \mf{sl}_2 \oplus \mf{sl}_8)_n \oplus ({\bf 2},\overline{\bf 28})_{\frac14+n} \oplus ({\bf 1},{\bf 70})_{\frac12+n} \oplus ({\bf 2},{\bf 28})_{\frac34+n} \bigr) \oplus \langle \dK,\LL_0\rangle\; .  \label{SL2SL8grad} \ee
This decomposition can be obtained by spectral flow. Starting from the associated $\mathds{Z}_4$ graded decomposition of $\mf{e}_8$, one further decomposes $\mf{sl}_8$ into $\mf{gl}_7$ with the $\mf{sl}_8$ generators $T^I{}_J$ splitting into  
\be T^i{}_j\; , \quad T^0{}_j\; , \quad T^i{}_0 \; , \quad T^0{}_0 = - T^k{}_k \; , \ee
for $I=0$ to $7$ and $i=1$ to $7$. The spectrally flowed Virasoro generators are then 
\be 
\LL^\ord{p}_n = L_n + p T_n^k{}_k + \frac{7p^2}{16} \delta_{n,0} \dK \; , 
\ee
and the corresponding $\hat{\mf{e}}_8$ generators are 
\be 
\T_n^i{}_j = T_n^i{}_j + \frac{p^2}{8} \delta^i_j \dK  \; , \quad \T_n^i{}_0 = T_{n+p}^i{}_0 \; , \quad \T_n^0{}_j = T_{n-p}^0{}_j \; . 
\ee
One finds that the Virasoro generator $\LL_0$ defines the grading \eqref{SL2SL8grad} for $p=1$ mod 4. For $p=3$ mod 4 one gets the same graded decomposition with the conjugate representations. We shall use the $p=1$ basis for the uplift to ten dimensions.  

One finds then the corresponding decomposition of the basic module
\bea 
\overline{R(\Lambda_0)_{-1}} \hspace{-2mm} &=& \hspace{-2mm}  ({\bf1},{\bf 8})_{\frac7{16}} \oplus ( {\bf 2},\overline{\bf 8})_{\frac{11}{16}} \oplus ({\bf1},\overline{\bf 56})_{\frac{15}{16}} \oplus ( {\bf 2},{{\bf 56}})_{\frac{19}{16}} \oplus \bigl( ({\bf1},\overline{\bf 8}\otimes  {\bf 28})\oplus ( {\bf 3},{\bf 8} )\bigr)_{\frac{23}{16}}  \CR
&& \oplus \bigl( ( {\bf 2},{\bf 8}\otimes \overline{\bf 28} ) \oplus  ({\bf 2},\overline{\bf 8})\bigr)_{\frac{27}{16}} \oplus \bigl( ({\bf 3},\overline{\bf 56}) \oplus ({\bf1},{\bf 8}\otimes {\bf 70})\oplus ({\bf1},\overline{\bf 168})\bigr)_{\frac{31}{16}} \CR
&&\oplus \bigl( ({\bf 2},\overline{\bf 8}\otimes {\bf 70}) \oplus ({\bf 2},{\bf 56}) \oplus ({\bf 2},{\bf 168})\bigr)_{\frac{35}{16}} + \dots 
\eea
where the $({\bf1},{\bf 8})_{7/{16}}$ corresponds to the  derivatives in the eight internal coordinates.

As in the preceding section, one writes the conjugate Weitzenb\"ock connection in the appropriate basis as 
\begin{multline}  \langle \widetilde{\mathcal{W}}_{\alpha} | \otimes \T^\alpha    =  \langle 0|^I \otimes \sum_{n} \Bigl(  \widetilde{W}^\ord{n}_{I}{}^J{}_K \T_n^K{}_J +\widetilde{W}^\ord{n}_{I}{}^\alpha{}_\beta \T_n^\beta{}_\alpha +\tfrac12 \widetilde{W}^\ord{n+\frac14}_{I}{}^{JK}_\alpha \T_{n+\frac12\, JK}^{\hspace{8mm} \alpha} \\
 + \frac{1}{24} \widetilde{W}^{\ord{n+\frac12}}_{I;JKLP}  \T_{n + \frac12}^{JKLP} + \tfrac12 \widetilde{W}^{\ord{n+\frac34}\; \alpha}_{I; JK} \T_{n+\frac14\, \alpha }^{JK} \Bigr) + \widetilde{W}_I^0  \langle 0|^I \otimes \LL_0  +\widetilde{W}_I^\dK  \langle 0|^I \otimes \dK \end{multline}
 where $I,J,K$ are the SL(8) indices and  $\alpha,\beta$ (on the right-hand side) are the SL(2) indices. There should not be any confusion with the index $\alpha$ of $\mf{e}_9$ on the left-hand side.

The decomposition of the basic module includes
\bea \overline{R(\Lambda_0)_{-1}}\hspace{-2mm} &\supset&\hspace{-2mm} ({\bf1},{\bf 8})_{\frac{7}{16}} \oplus ( {\bf 2},\overline{\bf 8})_{\frac{11}{16}} \oplus ({\bf1},\overline{\bf 56})_{\frac{15}{16}} \oplus ( {\bf 2},{{\bf 56}})_{\frac{19}{16}} \oplus \bigl( ({\bf1},\overline{\bf 8}\otimes  {\bf 28})\oplus ( {\bf 3},{\bf 8} )\bigr)_{\frac{23}{16}}\\ 
&& \hspace{0mm} \oplus  \bigoplus_{n=0}^\infty \Bigl[ \bigl( ({\bf1},{\bf 8})\otimes ({\bf 2},\overline{\bf 28})\bigr)_{\frac{27}{16}+n} \oplus  \bigl(({\bf1}, {\bf 8})\otimes ({\bf1},{\bf 70}) \bigr)_{\frac{31}{16}+n} \oplus  \bigl( ({\bf1},{\bf 8})\otimes ({\bf 2},{\bf 28})\bigr)_{\frac{35}{16}+n} \CR
&& \hspace{9mm}\oplus  \bigl( ({\bf1},{\bf 8}\otimes {\bf 63})\oplus ({\bf 3},{{\bf 8}}) \bigr)_{\frac{39}{16}+n} \Bigr] \nonumber \; . 
\eea
From this decomposition and the condition that $\langle \vartheta|=0$ one then concludes that all the components of the Weitzenb\"ock connection of $\LL_0$ degree greater or equal to $\frac{5}{4}$ vanish, because the projection to the basic module does not project these components to  smaller representations. One moreover gets that 
\be \widetilde{W}^\ord{\frac14}_J{}^{IJ}_\alpha = 0 \; , \quad \widetilde{W}^\ord{\frac12}_{[I;JKLP]}=0\; , \quad  \widetilde{W}^{\ord{\frac34}\hspace{1mm} \alpha}_{[I;JK]} = 0 \; , \quad \widetilde{W}^\ord{1}_{[I\hspace{4mm} K]}{}^{\hspace{-5.5mm}J\hspace{3mm} }=0\; , \quad \widetilde{W}^\ord{1}_{I}{}^\alpha{}_\beta = 0 \; , \ee
and $\widetilde{W}_I^\dK$ is determined. From these constraints one gets immediately that $ \langle \widetilde{W}_\alpha| \sS_{-n}(\T^\alpha) = 0 $ for $n\ge 1$ and that the embedding tensor can be written as
\begin{multline}  
\langle \theta |r^2 \mathcal{U}  = \langle 0|^I \widetilde{\Theta}^\ord{\frac{7}{16}}_I + \langle 1/4|_{I}^\alpha \widetilde{\Theta}^{\ord{\frac{11}{16}} I}_{\hspace{5mm} \alpha} + \frac1{6} \langle 1/2|_{IJK} \widetilde{\Theta}^{\ord{\frac{15}{16}} IJK} +\frac16 \langle 3/4|^{IJK}_\alpha \widetilde{\Theta}^{\ord{\frac{19}{16}} \; \alpha }_{IJK} \\ +\frac12  \langle 1|^{IJ}_K \widetilde{\Theta}_{IJ}^{\ord{\frac{23}{16}} K}+ \langle 1|^{I}_{\alpha\beta} \widetilde{\Theta}_{I}^{\ord{\frac{23}{16}} \alpha\beta} 
 - \frac12 \langle 5/4|^{K\, \alpha}_{IJ}   \widetilde{W}^\ord{\frac14}_K{}^{IJ}_\alpha \\ - \frac1{24}  \langle 3/2|^{I,JKLP}  \widetilde{W}^\ord{\frac12}_{I;JKLP} - \frac12 \langle 7/4|^{I, JK}_{\hspace{2mm} \alpha }   \widetilde{W}^{\ord{\frac34}\hspace{2mm} \alpha}_{I;JK}  - \langle 2|^{IJ}_K \widetilde{W}^{\ord{1} K}_{I\hspace{5mm} J} \; .
\end{multline}
where the basis elements $ \langle 5/4|^{K\, \alpha}_{IJ}  $,  $\langle 3/2|^{I,JKLP} $, $\langle 7/4|^{I, JK}_{\hspace{2mm} \alpha }  $ and $\langle 2|^{IJ}_K $ are in the corresponding irreducible representations.

One can now use \eqref{thetaU=thetaE} and check that all the generators of negative $\LL_0$ degree preserve this form such that the embedding tensor decomposes as well as
\begin{multline} 
\langle \theta | = \langle 0|^I {\Theta}^\ord{\frac{7}{16}}_I + \langle 1/4|_{I}^\alpha {\Theta}^{\ord{\frac{11}{16}} I}_{\hspace{5mm} \alpha} + \frac1{6} \langle 1/2|_{IJK} \widetilde{\Theta}^{\ord{\frac{15}{16}} IJK} +\frac16 \langle 3/4|^{IJK}_\alpha \widetilde{\Theta}^{\ord{\frac{19}{16}} \; \alpha }_{IJK}  \\* +\frac12  \langle 1|^{IJ}_K \Theta_{IJ}^{\ord{\frac{23}{16}} K}+ \langle 1|^{I}_{\alpha\beta} \Theta_{I}^{\ord{\frac{23}{16}} \alpha\beta} 
 + \frac12 \langle 5/4|^{K\, \alpha}_{IJ}   {\Theta}^\ord{\frac{27}{16}}_K{}^{IJ}_\alpha \\* 
 + \frac1{24}  \langle 3/2|^{I,JKLP}  {\Theta}^\ord{\frac{31}{16}}_{I,JKLP} + \frac12 \langle 7/4|^{I, JK}_{\hspace{2mm} \alpha }   {\Theta}^{\ord{\frac{35}{16}} \alpha}_{\quad I,JK}  + \langle 2|^{IJ}_K \Theta^{\ord{\frac{39}{16}} K}_{\quad IJ} \; . \label{eq:upliftthetaIIB}
\end{multline}

As for the eleven-dimensional case, the highest degree component is a vector valued symmetric tensor $\Theta^{\ord{\frac{39}{16}} K}_{IJ}$, and one expects the same constraint \eqref{QuadraticTheta21} to follow from the quadratic constraint \eqref{QuadraCons}. There is again a solution 
 \be \langle \theta |  = \langle 0|^8 {\Theta}_{00} +   \langle 1|^{i8}_8 {\Theta}_{0i}+  \langle 2|^{ij}_8 {\Theta}_{ij} \; , \label{SL8EmbedTens}\ee
to the quadratic constraint that corresponds to CSO$(p,q,r)$ gaugings obtained by reduction of type IIB supergravity on a circle times $S^7$ or other hyperboloids and contractions thereof. 

To understand this particular example it is useful to consider the $p=2$ spectral flowed basis in which
\be \mathfrak{e}_9 = \bigoplus_n \Bigl( ( \mf{sl}_2 \oplus \mf{e}_7)_n \oplus ({\bf 2},{\bf 56})_{\frac12+n} \bigr) \oplus \langle \dK,\LL_0\rangle \ee
and the basic module decomposes as 
\be
\overline{R(\Lambda_0)_{-1}} = ({\bf 2},{\bf1})_{\frac{1}{4}} \oplus ({\bf1},{\bf 56})_{\frac{3}{4}} \oplus \bigl( ({\bf 2},  {\bf 133})\oplus  ({\bf 2},{\bf1}) )\bigr)_{\frac{5}{4}} \oplus  \bigl( ({\bf1}, {\bf 56}\oplus {\bf 912} ) \oplus ( {\bf 3},{\bf 56}) \bigr)_{\frac{7}{4}} \oplus \dots 
\ee
One finds therefore that the symmetric tensor \eqref{SL8EmbedTens} of SL(8)  sits inside the ${\bf 912}_{7/4} $ and is a solution to the quadratic constraint by embedding in $\mf{e}_7$. This consistent truncation is T-dual to the reduction of eleven-dimensional supergravity on $S^7$ further reduced over a torus $T^2$, and is therefore already known to be a consistent truncation \cite{deWit:1984nz,Nicolai:2011cy,Godazgar:2013nma}.


\providecommand{\href}[2]{#2}\begingroup\raggedright\endgroup

\end{document}